\documentclass[12pt]{article}
\usepackage{latexsym}
\newcommand{\ft}[2]{{\textstyle\frac{#1}{#2}}}
\def\tilde{\widetilde}
\def\l{\lambda}
\def\1bar{1\hskip -.275cm -}
\def\2bar{2\hskip -.275cm -}
\def\3bar{3\hskip -.275cm -}
\begin{document}
\begin{titlepage}
\begin{flushright}
Preprint DFTT 99/47\\
hep-th/9909137 \\
September 1999\\
\end{flushright}
\vskip 2cm
\begin{center}
{{\Large \bf  The complete ${{\cal N}=3}$ Kaluza Klein spectrum \\ 
of $11 D$ supergravity on $AdS_4\times N^{010}$
}}\\
\vfill
{\large   Piet Termonia } \\
\vfill
{ \sl  Dipartimento di Fisica Teorica, Universit\'a di Torino, via P.
Giuria 1,
I-10125 Torino, \\
 Istituto Nazionale di Fisica Nucleare (INFN) - Sezione di Torino,
Italy \\}
\end{center}
\vfill
\begin{abstract}
We derive the invariant operators of the
zero-form, the one-form, the two-form and the
spinor 
from which the mass spectrum
of Kaluza Klein of eleven-dimensional supergravity on 
$AdS_4\times N^{010}$ 
can be derived 
by means of harmonic analysis.
We calculate their eigenvalues for all representations
of $SU(3)\times SO(3)$.  
We show that the information contained in these operators
is sufficient to reconstruct the complete ${\cal N}=3$
supersymmetry content of the compactified theory.
We find the ${\cal N}=3$
massless graviton multiplet, the Betti multiplet and the
$SU(3)$ Killing vector multiplet.
\end{abstract}
\vspace{2mm} \vfill \hrule width 3.cm
{\footnotesize
 Supported by   EEC  under TMR contract
 ERBFMRX-CT96-0045}
\end{titlepage}
\section{Introduction}
In order to challenge Maldacena's conjecture
\cite{maldacena} in all kinds of circumstances there
is a strong need for solved gauge theories on AdS manifolds. 
Indeed, they provide one of the two comparison terms in
the anti-de Sitter/conformal field theory 
correspondence \cite{klebwit}. 
A good deal of information to be checked
lies already in the mass spectrum and the 
supersymmetry multiplet structure of such theories.
\par 
A class of eleven-dimensional supergravity \cite{CremmerJuliaScherk}
solutions which may serve as
compactified supersymmetric vacuum backgrounds is given by the
Freund-Rubin solutions \cite{freurub}. 
They have the form $AdS_4 \times M$, where $M$
is a compact seven-dimensional Einstein manifold.
A particular convenient number of manifolds to use for such
Kaluza Klein compactifications are the $G/H$ coset spaces. 
Part of their
attraction comes from the fact that it is immediate
to read off their isometry groups.
Consequently one knows the gauge symmetries of
the fundamental particle interactions.
Moreover, it is clear how to calculate the mass spectrum of the
four-dimensional theory. One can use 
harmonic analysis. As was shown in 
\cite{Salamstrathdee, castdauriafre} and in a series of papers
\cite{noi321}, this can be done by
exhibiting the group structure of the constituents of the 
coset rather than laboriously solving the 
differential equations.
Hence the calculation of the mass spectrum gets reduced
to a group-theoretical investigation of the coset and the
final masses are found upon calculating the eigenvalues
of some discrete operators. This is food for computers.
In this paper we outline the necessary steps to be performed
to reach this goal.
Even in the compactifications where the calculations 
become too gigantic the masses can still successfully be 
calculated in this way.
\par
The complete list of the
seven-dimensional coset spaces that may serve to compactify 
eleven-dimensional supergravity to four dimensions is known
\cite{castromwar}.
Also the number of the 
remaining supersymmetries and their geometries have been
intensively studied in the past. It turns out
to contain some non-extremal supersymmetric
cases that present themselves as promising 
candidates for the anti-de Sitter/conformal field theory check.
This as opposed to Kaluza Klein on the seven-sphere
which is related to the extremal ${\cal N}=8$ supergravity
theory \cite{dewitnicolai}.
There the spectrum can be derived from
the short unitary irreducible representation of $Osp(8\vert 4)$
with highest spin two, see \cite{gunawar}.
From the perspective of the three-dimensional superconformal 
theory this means that all the composite primary operators 
have conformal weight equal to their naive dimensions.
Hence no anomalous dimensions are generated.  
One of such non-trivial compactifications is the
one on $AdS \times {M}^{111}$, where $M^{111}$ is one of
Witten's $M^{pqr}$ spaces \cite{kkwitten, noi321}. 
For this manifold the
complete spectrum has been calculated and arranged in ${\cal N}=2$
multiplets, see \cite{multanna} and 
\cite{m111}. This spectrum provides some ideal material for the
anti-de Sitter/conformal field theory check
which has already successfully been done \cite{adscftcheckers}. 
\par
In many cases of $G/H$ compactifications the isometry group can be read
off directly, being the group $G$. If that is true, then the superisometry
group is simply the supergroup $Osp(4\vert {\cal N})\times G^\prime$, where
$G=G^\prime \otimes SO({\cal N})$ and $\cal N$ is determined by the number of
Killing spinors that are allowed on the manifold. Yet this becomes slightly
more complicated when the normalizer $N$ of $H$ is non zero. Then the
isometry group becomes $G\times N$ in stead of $G$.
This is for instance the case for eleven-dimensional 
supergravity on a background of
\begin{eqnarray}
AdS_4 \times N^{010}                \,,
\label{AdSN010intro}
\end{eqnarray}
with
\begin{eqnarray}
N^{010} \equiv \frac{SU(3)}{U(1)}\,.
\label{N010U1}
\end{eqnarray}
which was introduced in the paper \cite{castrom}.
It has been proven to yield a ${\cal N}=3$ 
background\footnote{Actually there are many different manifolds
of this form due to the present freedom in the choice of the vielbeins
on this manifolds. However, as shown in \cite{castrom, cast} there
is a particular choice for which ${\cal N}=3$. } 
where the $SO(3)$ R-symmetry group is the normalizer of the $U(1)$  in the
denominator. Here the identification of the resulting $SO(3)$ 
multiplet spectrum is not a straightforward exercise,
as can be seen in \cite{cast}
if one takes the description (\ref{N010U1}) for the manifold. 
Still as L. Castellani and L. J. Romans showed in \cite{castrom},
one can elegantly circumvent this difficulty by taking a
description  which has  the normalizer taken into account from the very
start. In particular, as they suggested in \cite{castrom, cast}, one
can exhibit the fact that the ${\cal N}=3$ supersymmetric $N^{010}$ is equal
to
\begin{eqnarray}
\frac{SU(3)\times SU(2)}{SU(2)\times U(1)}
\label{N010SU2}
\end{eqnarray}
where the $SU(2)$ in the denominator is diagonally embedded in
$SU(3)\times SU(2)$. In this way the $SO(3)\approx SU(2)$ is everywhere
explicitly present.
\par
The harmonic analysis on $N^{010}$ has been done partly
by Castellani in \cite{cast}, using the formulation (\ref{N010U1}).
Yet due to lack of computing power his calculation 
did not cover the complete spectrum.
It is the scope of this paper to extend the technique of harmonic
analysis to the alternative formulation (\ref{N010SU2})
and derive really the complete spectrum.
This choice is motivated by the fact that the field components
will then automatically be organized in irreducible representations
of $SO(3)$. This will make the matching 
of the spectrum with ${\cal N}=3$ multiplets possible. 
We present here a sufficient part of the
mass operator eigenvalues that contains enough information
to reconstruct the complete spectrum by using the supersymmetry
mass relations \cite{noi321}.
This will be used in a subsequent paper \cite{n010}
to derive the complete multiplet structure of the ${\cal N}=3$ $N^{010}$
Freund-Rubin compactifications.
\par
The importance of the spectrum of eleven-dimensional 
supergravity on (\ref{AdSN010intro}) lies in the fact that
an anti-de Sitter/conformal field theory 
comparison in this background is fairly facilitated
by the fact that for the four-dimensional theory the structure 
of the effective Lagrangian is already known to be
\cite{n3},
\begin{eqnarray}
\frac{SU(3,8)}{SU(3)\times SU(8)\times U(1)} \,.
\end{eqnarray}
Still, as becomes clear from the work presented in this
paper, the mass spectrum of the ${\cal N}=3$ theory
is going to be enough non trivial to
provide another powerful check of the 
anti-de Sitter/conformal field theory duality.
\par
The spectrum that we obtain here has to be
organized in ${\cal N}=3$ multiplets.
A systematic classification of all the
possible short ${\cal N}=3$ multiplets and their superspace structure
does not exist in the literature. 
Only the short vector multiplet in table \ref{vectormultiplet}
\begin{table}
\begin{tabular}{||c|c|c|c|c|c|c||}
\hline
spin &
energy &
isospin &
mass &
name &
mass &
name \\
\hline
\hline
$1$      & $J+1$      & $J-1$   & $16E_0(E_0-1)$  & $A/W$ 
                                & $16E_0(E_0-1)$  & $Z$    \\
$\ft12$  & $J+\ft32$  & $J-1$   & $-4E_0$  &  $\lambda_{L,T}$
                                & $4E_0$ &  $\lambda_T$  \\
$\ft12$  & $J+\ft32$  & $J-2$   & $-4E_0$  &  $\lambda_{L,T}$
                                & $4E_0$  &   $\lambda_T$  \\ 
$\ft12$  & $J+\ft12$  & $J$     & $4E_0-4$  & $\lambda_{L,T}$ 
                                & $-4E_0+4$  & $\lambda_T$   \\ 
$\ft12$  & $J+\ft12$  & $J-1$   & $4E_0-4$  &  $\lambda_{L,T}$
                                & $-4E_0+4$  &  $\lambda_T$    \\ 
$0$      & $J+2$      & $J-2$   & $16E_0(E_0+1)$  & $\phi,S/W$  
                                & $16E_0(E_0+1)$  & $\pi$   \\ 
$0$      & $J+1$      & $J$     & $16E_0(E_0-1)$  &   $\pi$
                                & $16E_0(E_0-1)$ &   $\phi$  \\ 
$0$      & $J+1$      & $J-1$   & $16E_0(E_0-1)$  &   $\pi$
                                & $16E_0(E_0-1)$ &  $\phi$ \\ 
$0$      & $J+1$      & $J-2$   & $16E_0(E_0-1)$  &  $\pi$ 
                                & $16E_0(E_0-1)$  &  $\phi$ \\ 
$0$      & $E_0=J$    & $J$     & $16(E_0-2)(E_0-1)$  & $\phi,S/W$  
                                & $16(E_0-2)(E_0-1)$ &  $\pi$ \\ 
\hline
\end{tabular}
\caption{The states of the $Osp(3|4)$ vector multiplet representation
organized in representations of $SO(2)\times SU(2)_{\rm spin}$ and 
$SO(3)_{\rm isospin}$ as they appear in Kaluza Klein
of eleven-dimensional supergravity. The unitary bound for the ground state 
is satisfied $E_0=J$. The names of the fields are chosen as in \cite{noi321}.}
\label{vectormultiplet}
\end{table}
has been found explicitly in \cite{freedmannicolai}.
As in the case of the $AdS \times M^{111}$ Kaluza Klein
compactification the matching of the mass spectrum
will teach us a lot if not everything about the existence 
of other short multiplets. Following the arguments of
\cite{adscftcheckers} it turns out to be necessary to
decompose the resulting ${\cal N}=3$ multiplets in
${\cal N}=2$ multiplets.
Upon doing so the 
states of the resulting ${\cal N}=3$ 
multiplets can then be recognized as 
coming from the on-shell field components of 
the superfields on the ${\cal N}=2$ superspace that was introduced 
in \cite{susp}.
Their superfield constraints are then straightforwardly
read off.
We postpone this issue to a future publication.
\par
The work that we present here fits in a much wider
project that is currently being carried out by the
Torino group. The final scope of this $AdS\times N^{010}$
spectrum is to check the anti-de Sitter/conformal
field theory correspondence as it has been  
carried out \cite{adscftcheckers} for $AdS \times M^{111}$.
It will be done \cite{Q111} 
also for eleven-dimensional supergravity on $AdS \times Q^{ppp}$.
\par
This paper is organized as follows.
We start with a short description of the geometry of $N^{010}$.
We thereby restrict ourselves to the essentials that we will need
further on. We leave a more rigorous treatment to a future 
publication \cite{castn010}. 
Then we will repeat the standard concepts of
harmonic analysis and explain how they can be applied to
eleven-dimensional supergravity on $AdS_4 \times N^{010}$. Using these
techniques we will then compute the zero-form operator $M_{(0)^3}$,
the one-form operator
$M_{(0)^2 (1)}$, 
the two-form operator $M_{(0)(1)^2}$
and the spinor operator $M_{(1/2)^3}$,
the notation 
of these operators being the same as in
\cite{castdauriafre, noi321}.
We will then  
present their eigenvalues.
We conclude by arguing that
the information  obtained in this paper is sufficient to
calculate the complete spectrum. We will do so by means 
of a concrete example: the massless graviton multiplet.
Moreover, in this way we will prove that the remaining spectrum is indeed
${\cal N}=3$ supersymmetric. Finally, we will identify the
series of irreducible representations of $SU(3)\times SU(2)$
with a common field content. For these series we will 
list the eigenvalues that are present. From these eigenvalues
the masses are then obtained by applying the mass formulas
from \cite{noi321}. 
\section{The geometry of $N^{010}$}
In this section we introduce the essential 
concepts\footnote{The material in this section 
was explained to me by Leonardo 
Castellani. I am indebted  to him for this.
He also calculated the rescalings at an early
stage of this research.
}
of the geometry of $N^{010}$. We restrict ourselves 
to the elements that serve our purposes.
We leave a more elaborated treatment to 
\cite{castn010}.
We fix some freedom in the choice of the vielbeins.
This will ensure that the compactified theory on (\ref{AdSN010intro})
will have ${\cal N}=3$ supersymmetry.
\par
The $N^{010}$ coset spaces are special manifolds of the
class of $N^{pqr}$ coset spaces,
\begin{eqnarray}
\frac{G}{H}=\frac{SU(3)\times U(1)}{U(1)\times U(1)}
\end{eqnarray}
where the integer numbers $p,q,r$ specify the way in which the
two $U(1)$ generators $M$ and $N$ of $H$ are embedded in $G$,
\begin{eqnarray}
  \label{eq:pqr}
  M &=& -\frac{\sqrt{2}}{RQ} 
    \left(
        \ft{i}{2} rp \sqrt{3} \lambda_8
      + \ft{i}{2} rq \lambda_3
      - \ft{i}{2} (3 p^2 + q^2) Y
    \right) \,,
    \nonumber \\
  N &=& -\frac{1}{Q}
    \left(
      - \ft{i}{2} q \lambda_8
      + \ft{i}{2} p \sqrt{3} \lambda_3
    \right) 
    \,,
\end{eqnarray}
with
\begin{eqnarray}
  \label{eq:RQTr}
  R=\sqrt{3 p^2 + q^2 + 2 r^2} \,, 
  \qquad
  Q=\sqrt{3 p^2 + q^2} \,,
\end{eqnarray}
where $Y$ is the $U(1)$ generator in $SU(3) \times U(1)$
and the $SU(3)$ generators $\lambda$ are given in appendix
\ref{conventions}.
We do not get
into details in this paper but for a detailed description
of the definition, the geometry and the properties of these
spaces we refer the reader to the literature 
\cite{castrom, cast, castn010}.
These spaces are seven dimensional and can be used to make a
Freund-Rubin background for eleven-dimensional supergravity,
\begin{eqnarray}
AdS \times N^{010} \,.
\end{eqnarray}
They are solutions of the field equations of eleven-dimensional
supergravity.
\par
In this paper we restrict our attention to the case of $N^{010}$.
It has been shown that among these spaces there exists a class
of cosets,
\begin{eqnarray}
\frac{SU(3)}{U(1)}
\end{eqnarray}
for $p=0, r=0$ which has $SU(3)$ holonomy and
yields an ${\cal N}=3$ supergravity. In fact, without
loss of generality one can put $q=1$ in this case.
The normalizer of $U(1)$ in $SU(3)$ is $U(1)\times SO(3)$.
So the actual isometry is $SU(3)\times SO(3)$ and
the resulting isometry supergroup of the $AdS_4 \times N^{010}$
becomes $Osp(4\vert 3)\times SU(3)$. So the extra $SO(3)$  in the
normalizer becomes the R symmetry of the supergroup.
\par
In this paper we derive the mass spectrum of the Kaluza Klein
theory on $AdS_4\times N^{010}$ and to this end it is more
convenient to use the alternative description which was proposed
by Castellani and Romans,
where the $SO(3)\approx SU(2)$ is already  present form the start. Hence
we use
\begin{eqnarray}
N^{010}=\frac{SU(3)\times SU(2)}{SU(2)\times U(1)}
\end{eqnarray}
For the generators of $SU(3)$ we take $\ft{i}{2} \lambda$
and for $SU(2)$ we take $\ft{i}{2} \sigma$,
$\lambda$ being the Gell-Mann matrices (see appendix \ref{conventions})
and $\sigma$ being the Pauli matrices.
The generator of $U(1)$ is given by 
\begin{eqnarray}
T_8= \ft{i}{2} \lambda_8 \,.
\end{eqnarray}
The $SU(2)$ is diagonally embedded in $SU(3)\times SU(2)$
with generators,
\begin{eqnarray}
T_H = \ft{i}{2} \left(
\lambda_a + \sigma_a
\right)  , \qquad  a=1,2,3, 
\qquad H=9,10,11 \,.
\end{eqnarray}
where $H=9$ corresponds to $a=1$ and so on.
We will call it $SU(2)^{diag}$.
For the remaining coset generators we have
\begin{eqnarray}
T_\alpha = \ft{i}{2} \left( \lambda_1-\sigma_1, \lambda_2-\sigma_2,
            \lambda_3-\sigma_3,
             \lambda_4, \lambda_5, \lambda_6, \lambda_7 \right)
\label{cosetgenerators}
\end{eqnarray}
\par
We need the $SO(7)$ covariant derivative on the coset space.
It is defined as
\begin{eqnarray}
{\cal D} = d + B^{\alpha \beta} \, \left(T^{SO(7)}\right)_{\alpha\beta}
\label{covdergeometry}
\end{eqnarray}
where the one-form $B^{\alpha\beta}$  is the connection of the
coset space and $T^{SO(7)}$ are the generators of $SO(7)$,
be it of the vector representation or the spinor representation,
\begin{eqnarray}
\begin{array}{|c|c|}
\hline
& \cr
\,\, scalar \,\, irrep \,\, &
(T^{SO(7)})=0 \cr
& \cr
\hline
& \cr 
vector \,\, irrep &
(T^{SO(7)}_{\alpha\beta})_\gamma{}^\delta = \eta_{\gamma\epsilon} 
                 \delta^\epsilon_{[\alpha} \delta_{\beta]}^\delta \cr 
& \cr
\hline
& \cr
spinor \,\, irrep &
(T^{SO(7)}_{\alpha\beta}) = \ft14 \tau_{[\alpha} \tau_{\beta]} \cr
& \cr
\hline
\end{array}
\end{eqnarray}
where the matrices $\tau$ of the $SO(7)$ Clifford algebra are 
given in the appendix \ref{SO7conventions} and $\eta$ is the
metric of appendix \ref{conventions}.
\par
As already explained, there is some freedom in the choice of the
vielbeins, not all of them leading to ${\cal N}=3$.
\par
To see how this goes,
let us recall that on the coset there is
the invariant
\begin{equation}
L^{-1} d\, L =  e^\alpha T_\alpha + \omega^H T_H  \,,
\label{L-dL}
\end{equation}
where $L$ is coset representative.  
$H$ is the index running on $H$ and $\alpha$
is the index running on $SO(7)$, see appendix \ref{conventions}
for conventions. The fields $e^\alpha$
and $\omega^H$ are the $G/H$ vielbein and the
$H$ connection.
Using the coset vielbein $e^\alpha$
we define the coset connection
$B^\alpha{}_\beta$,
\begin{eqnarray}
d e^\alpha + B^\alpha{}_\beta \wedge e^\beta=0 \,.
\end{eqnarray}
The vielbein is specified up to $7$ rescalings of the coset directions,
see \cite{castdauriafre, noi321, cast}.
We take this into account by introducing the seven parameters $r_\alpha$
\begin{eqnarray}
e^\alpha \rightarrow r_\alpha e^\alpha \,.
\end{eqnarray}
Then the form of the connection is obtained by solving the
Maurer-Cartan equation, i.e. applying the external derivative 
$d$ to (\ref{L-dL}) and using $d^2=0$,
\begin{eqnarray}
B^\alpha{}_\beta = -\ft12 
         \Big(\frac{r_\beta r_\gamma}{r_\alpha} C_{\beta\gamma}{}^{\alpha}
       - \frac{r_\alpha r_\gamma}{r_\beta} \eta^{\epsilon\alpha}\eta_{\phi\beta}
          C_{\epsilon\gamma}{}^\phi
       - \frac{r_\alpha r_\beta}{r_\gamma} 
         \eta^{\alpha\phi}\eta_{\gamma\delta} C_{\phi\beta}{}^{\gamma}\big)
         \, e^\gamma
         - \frac{r_\beta}{r_\alpha} C_{\beta H}{}^{\alpha} \omega^H
\end{eqnarray}
where the constants $C_{\alpha\beta}{}^\gamma$ are the structure
constants of $G$.
We take the embedding of $H$ into $SO(7)$  as follows,
\begin{equation}
T_H = (T_H)^{\alpha\beta} (T^{SO(7)})_{\alpha\beta} 
\,.
\label{embedding}
\end{equation}
\par
Now we specify the rescalings $r_\alpha$. As is known \cite{noi321},
only for some well-chosen values of these parameters does
the manifold become an Einstein manifold. 
Moreover, as is clear from
\cite{castrom, cast} not all of the valuable choices for these
rescalings necessarily lead to the contemplated number of supersymmetries.
To see which of them do, we look at the curvature.
The curvature on the coset space is defined by
\begin{eqnarray}
R^\alpha{}_\beta = d \, B^\alpha{}_\beta + B^\alpha{}_\gamma \wedge
B^\gamma{}_\beta \,.
\end{eqnarray}
Then we need the following rescalings\footnote{
In fact there is the more general class of rescalings
\begin{eqnarray}
&& r_a = \pm 2 e \,, \nonumber \\
&& r_A = \pm 4 \sqrt{2} 
\end{eqnarray}
or
\begin{eqnarray}
&& r_a = \pm \ft{10}{3} e \,, \nonumber \\
&& r_A = \pm \ft{4 \sqrt{10}}{3} \, e \,, 
\end{eqnarray}
that yield the Einstein curvature (\ref{Einstein}). 
However not all of them necessarily lead to an ${\cal N}=3$
theory. As we will show in the last section of this paper,
the rescaling (\ref{rescalings}) that we adopt 
yields ${\cal N}=3$ supersymmetry. 
We refer the reader for a detailed study of these rescalings to
a future publication \cite{castn010}.
},
\begin{eqnarray}
&& r_a = - 2 e \,, \nonumber \\
&& r_A = 4 \sqrt{2} \, e 
\label{rescalings}
\end{eqnarray}
in order to have Ricci curvature
\begin{eqnarray}
R_{\alpha\beta} = 12 \, e^2 \, \eta_{\alpha\beta}  \,.
\label{Einstein}
\end{eqnarray}
Following the conventions of the papers \cite{noi321}
we put $e=1$. Then we will be able to apply
the mass relations and mass formulas that were obtained 
in these papers.
\par 
To conclude this section we give the explicit form of the embedding
of the $SU(2)^{diag}$ in $SO(7)$ according to eq. (\ref{embedding}). 
For the vector we get
\begin{eqnarray}
T_9 =
\left(
\matrix{ 0 & 0 & 0 & 0 & 0 & 0 & 0 \cr 0 & 0 & 1 & 0 & 0 & 0 & 0 \cr 0 & 
    -1 & 0 & 0 & 0 & 0 & 0 \cr 0 & 0 & 0 & 0 & 0 & 0 & {\frac{1}{2}} \cr 0 & 0 & 0 & 0 & 0 & -{\frac{1}
     {2}} & 0 \cr 0 & 0 & 0 & 0 & {\frac{1}{2}} & 0 & 0 \cr 0 & 0 & 0 & -{\frac{1}{2}} & 0 & 0 & 0 \cr  }
\right) \nonumber \\
T_{10} =
\left(
\matrix{ 0 & 0 & -1 & 0 & 0 & 0 & 0 \cr 0 & 0 & 0 & 0 & 0 & 0 & 0 \cr 1 & 0 & 0 & 0 & 0 & 0 & 0 \cr 0 & 0 & 0 & 0 & 0 & {
     \frac{1}{2}} & 0 \cr 0 & 0 & 0 & 0 & 0 & 0 & {\frac{1}{2}} \cr 0 & 0 & 0 & -{\frac{1}
     {2}} & 0 & 0 & 0 \cr 0 & 0 & 0 & 0 & -{\frac{1}{2}} & 0 & 0 \cr  }
\right)
\nonumber \\
T_{11} =
\left(
\matrix{ 0 & 1 & 0 & 0 & 0 & 0 & 0 \cr -1 & 0 & 0 & 0 & 0 & 0 & 0 \cr 0 & 0 & 0 & 0 & 0 & 0 & 0 \cr 0 & 0 & 0 & 0 & {\frac{1}
    {2}} & 0 & 0 \cr 0 & 0 & 0 & -{\frac{1}{2}} & 0 & 0 & 0 \cr 0 & 0 & 0 & 0 & 0 & 0 & -{\frac{1}
     {2}} \cr 0 & 0 & 0 & 0 & 0 & {\frac{1}{2}} & 0 \cr  } 
\right)\,,
\end{eqnarray}
and,
\begin{eqnarray}
T_8 = 
\left(
\matrix{ 0 & 0 & 0 & 0 & 0 & 0 & 0 \cr 0 & 0 & 0 & 0 & 0 & 0 & 0 \cr 0 & 0 & 0 & 0 & 0 & 0 & 0 \cr 0 & 0 & 0 & 0 & {\frac
      {{\sqrt{3}}}{2}} & 0 & 0 \cr 0 & 0 & 0 & {\frac{-{\sqrt{3}}}{2}} & 0 & 0 & 0 \cr 0 & 0 & 0 & 0 & 0 & 0 & {\frac{{
        \sqrt{3}}}{2}} \cr 0 & 0 & 0 & 0 & 0 & {\frac{-{\sqrt{3}}}{2}} & 0 \cr  }
\right)
\label{vecembed}
\end{eqnarray}
For the matrices 
$T_H^{\alpha\beta}$ the order of the indices
is $1,2,3,4,5,6,7$.
For the spinor representation we have
\begin{eqnarray}
&& T_9 =
\left(
\matrix{ 0 & 0 & 0 & 0 & 0 & 0 & 0 & 0 \cr 
         0 & 0 & {\frac{i}{2}} & 0 & 0 & 0 & 0 & 0 \cr 
         0 & {\frac{i}{2}} & 0 & 0 & 0 & 0 & 0 & 0 \cr 
         0 & 0 & 0 & 0 & {\frac{i}{2}} & 0 & 0 & 0 \cr 
         0 & 0 & 0 & {\frac{i}{2}} & 0 & 0 & 0 & 0 \cr 
         0 & 0 & 0 & 0 & 0 & 0 & i & 0 \cr 
         0 & 0 & 0 & 0 & 0 & {\frac{i}{2}} & 0 & {\frac{i}{2}} \cr 
         0 & 0 & 0 & 0 & 0 & 0 & i & 0 \cr  }
\right)
\nonumber \\
&& T_{10}=
\left(
\matrix{ & 0 & 0 & 0 & 0 & 0 & 0 & 0 \cr 
         0 & 0 & {\frac{1}{2}} & 0 & 0 & 0 & 0 & 0 \cr 
         0 & -{\frac{1}{2}} & 0 & 0 & 0 & 0 & 0 & 0 \cr 
         0 & 0 & 0 & 0 & {\frac{1}{2}} & 0 & 0 & 0 \cr 
         0 & 0 & 0 & -{\frac{1}{2}} & 0 & 0 & 0 & 0 \cr 
         0 & 0 & 0 & 0 & 0 & 0 & 1 & 0 \cr 
         0 & 0 & 0 & 0 & 0 & -{\frac{1}{2}} & 0 & {\frac{1}{2}} \cr 
         0 & 0 & 0 & 0 & 0 & 0 & -1 & 0 \cr  }
\right)
\nonumber \\
&& T_{11} =
\left(
\matrix{ 0 & 0 & 0 & 0 & 0 & 0 & 0 & 0 \cr 
         0 & {\frac{i}{2}} & 0 & 0 & 0 & 0 & 0 & 0 \cr 
         0 & 0 & -{\frac{i}{2}} & 0 & 0 & 0 & 0 & 0 \cr 
         0 & 0 & 0 & {\frac{i}{2}} & 0 & 0 & 0 & 0 \cr 
         0 & 0 & 0 & 0 & -{\frac{i}{2}} & 0 & 0 & 0 \cr 
         0 & 0 & 0 & 0 & 0 & i & 0 & 0 \cr 
         0 & 0 & 0 & 0 & 0 & 0 & 0 & 0 \cr 
         0 & 0 & 0 & 0 & 0 & 0 & 0 & -i \cr  
}
\right)
\label{spinorembedsu2}
\end{eqnarray}
\begin{eqnarray}
T_8 =
\left(
\matrix{ 0 & 0 & 0 & 0 & 0 & 0 & 0 & 0 \cr 
         0 & -{\frac{i}{2}}\,{\sqrt{3}} & 0 & 0 & 0 & 0 & 0 & 0 \cr 
         0 & 0 & -{\frac{i}{2}}\,{\sqrt{3}} & 0 & 0 & 0 & 0 & 0 \cr 
         0 & 0 & 0 & {\frac{i}{2}}\,{\sqrt{3}} & 0 & 0 & 0 & 0 \cr 
         0 & 0 & 0 & 0 & {\frac{i}{2}}\,{\sqrt{3}} & 0 & 0 & 0 \cr 
         0 & 0 & 0 & 0 & 0 & 0 & 0 & 0 \cr 
         0 & 0 & 0 & 0 & 0 & 0 & 0 & 0 \cr 
         0 & 0 & 0 & 0 & 0 & 0 & 0 & 0 \cr  }
\right)
\label{spinorembed8}
\end{eqnarray}
\section{Harmonic analysis}
In this section we provide the main ingredients of harmonic analysis
that we need for the calculation of the masses of
the zero-form, the one-form, the two-form and 
and the spinor. The technique of harmonic analysis has been well 
established, see papers 
\cite{castdauriafre, noi321, cast},
so we restrict ourselves to the essentials. We focus on its application
to our coset $N^{010}$. 
\par
The main idea is that functions on a coset space $G/H$ can be expanded
in terms of the components of the operators of the irreducible 
representations of $G$. In particular, a complete set of functions
$\Phi^{(\alpha)}_\Lambda (L(y))$
on the coset $G/H$ is given 
that transform in a irreducible representation of $H$ as follows,
\begin{eqnarray}
\Phi^{(\alpha)}_\Lambda (L(y) h)=
\Phi^{(\alpha)}_\Pi (L(y)) \, ( D^{(\alpha)} )^{\Pi}{}_\Lambda \,,
\end{eqnarray}
where
$(\alpha)$ indicates the $H$-irreps and the $\Lambda$ its indices.
These functions depend on the coordinates of the coset space via the coset 
representative $L(y)$. 
Then for such functions the expansion is given by,
\begin{eqnarray}
\Phi^{(\alpha)}_\Lambda (L(y)) = \sum_{(\mu)}  c_{\cal G}^{(\mu)}
\, (D^{(\mu)})^{\cal G}{}_\Lambda (L(y)) \,,
\label{cosetexpansion}
\end{eqnarray}
where ${\cal G}$ are the contracted indices of $G$ and
$\bf (\mu)$ label all 
the irreducible representations of $G$ that contain
the irreducible representation $\bf (\alpha)$ under reduction to $H$
as follows\footnote{
In fact $\bf (\alpha)$ may appear multiple times in the reduction,
say $m$ times.
In that case it should be understood that the above
sum contains $\bf (\mu)$ $m$ times. For a more clear 
treatment of this see \cite{castdauriafre, noi321}}
\begin{eqnarray}
{\bf (\mu)} \quad &\rightarrow& \quad \dots + {\bf (\alpha)} + \dots \,,
\nonumber \\
&H&
\label{reduction}
\end{eqnarray}
\par
The crucial step now is to use the fact that the covariant derivative
(\ref{covdergeometry}) can be expressed  in terms of
the coset generators plus some additional discrete operators.
Indeed, there is no need to express this covariant derivative
as a differential operator.
This is a generic feature of coset manifolds 
\cite{castdauriafre, noi321, cast}. It is extremely useful for our
purposes since evaluating the mass terms in the field
equation of eleven-dimensional supergravity can then be done
without solving differential equations.
\par
To see this, it is sufficient to realize that the
harmonic expansion ultimately is an expansion in 
$L(y)^{-1}$. Then one uses (\ref{L-dL}) to express 
the derivatives in terms of the vielbein.
Then on a field $Y$ on the coset space
that sits in some representation of
$SO(7)$ the covariant derivative becomes
\begin{equation}
{\cal D}_{\alpha} Y = - r_\alpha T_\alpha Y + 
                    \frac{r_\alpha r_\beta}{r_\gamma} \,
                    C_{\alpha \beta}{}^\gamma (T^{SO(7)})_{\gamma}{}^{\beta} Y
                   +\ft12 \, \eta_{\alpha\delta}\,\frac{r_\beta r_\gamma}{r_\delta}\, 
                    C_{\beta\gamma}{}^{\delta}
                    (T^{SO(7)})^{\beta\gamma} Y
\label{covder}
\end{equation}
\par
We now show how this reduction is done in
the case where the coset is $N^{010}$. In particular we
show how the representations of $SU(3)\times SU(2)$ reduce to 
representations of $SU(2)^{diag} \times U(1)$.
We also give the constraint that is 
to be imposed  in order to ensure the right $U(1)$
weight of the harmonic.
\par 
Let us  look at the index structure of the objects
$(D^{(\mu)})_{{\cal G} \Lambda} (L(y))$ in the expansion
 (\ref{cosetexpansion}).
Let us start with the indices $\cal G$. Clearly,
they have the following structure,
\begin{eqnarray}
{\cal G} &=&
\begin{array}{l}
\begin{array}{|c|c|}
\hline
           k_1 \hskip .03cm \dots \hskip .03cm k_{M_2} & l_1 \dots l_{M_1}
 \\
\hline
\end{array}\\
\begin{array}{|c|}
           n_1  \dots  n_{M_2}
 \\
\hline
\end{array}
\end{array}
\otimes
\begin{array}{l}
\begin{array}{|c|}
\hline
            m_1  \cdots  m_{2J}
\nonumber\\
\hline
\end{array}
\end{array}
\nonumber \\
&&
\hskip 0.2cm 
\underbrace{\hskip 2.1 cm}_{M_2}
\underbrace{\hskip 1.8 cm}_{M_1}
\hskip .9cm
\underbrace{\hskip 2.0 cm}_{2J}
\label{SU3Youngtableau}
\end{eqnarray}
being the product of a generic $SU(3)$ Young tableau
with a generic $SU(2)$ Young tableau.
The indices run through
\begin{eqnarray}
&& k_1, \dots, k_{M_2}, l_1,\dots, l_{M_1}, n_1, \dots,
n_{M_2}=1,2,3
\nonumber \\
&& m_1, \dots, m_{2J}=1,2
\end{eqnarray}
Let us now see whether the representation $(\mu)$
contains a given representation $(\alpha)$ of $SU(2)^{diag}$,
to clarify (\ref{reduction}). 
Hence we look at the indices $\Lambda$ in
$(D^{(\mu)})_{{\cal G} \Lambda} (L(y))$. 
\par
By making use of the symmetries of the Young tableaux
one can, for fixed values,
 arrange the indices of $(\mu)$ as follows,
\begin{eqnarray}
{\cal G} &=&
\begin{array}{l}
\begin{array}{|c|c|c|c|}
\hline
         \1bar & k_1 \dots k_{2 p} & l_1 \dots l_{2 q} & \3bar
 \\
\hline
\end{array}\\
\begin{array}{|c|c|}
           \2bar & 3 \hskip .26cm \dots \hskip .26cm 3
 \\
\hline
\end{array}
\end{array}
\otimes
\begin{array}{l}
\begin{array}{|c|}
\hline
            m_1  \cdots  m_{2J}
\nonumber\\
\hline
\end{array}
\end{array}
\nonumber \\
&&
\hskip 0.8cm 
\underbrace{\hskip 1.9 cm}_{2 p}
\underbrace{\hskip 1.7 cm}_{2 q}
\hskip 1.5cm
\underbrace{\hskip 2.0 cm}_{2J}
\nonumber \\
&&
\hskip 0.2 cm 
\underbrace{\hskip 2.5 cm}_{M_2}
\underbrace{\hskip 2.3 cm}_{M_1}
\label{rearrangedyoung}
\end{eqnarray}
where we use the following short-hand notation 
\begin{eqnarray}
\begin{array}{l}
\begin{array}{|c|}
\hline
i_1  \cdots  i_P          
 \\
\hline
\end{array}
\end{array}
\equiv
\begin{array}{l}
\begin{array}{|c|c|c|}
\hline
i_1 & \cdots & i_P          
 \\
\hline
\end{array}
\end{array}
\end{eqnarray}
and,
\begin{eqnarray}
\begin{array}{l}
\begin{array}{|c|}
\hline
         \1bar 
 \\
\hline
\end{array}
\end{array}
\equiv
\begin{array}{l}
\begin{array}{|c|}
\hline
1 \dots 1 
 \\
\hline
\end{array}
\end{array}
\,, \quad
\begin{array}{l}
\begin{array}{|c|}
\hline
         \2bar 
 \\
\hline
\end{array}
\end{array}
\equiv 
\begin{array}{l}
\begin{array}{|c|}
\hline
2 \dots 2
 \\
\hline
\end{array}
\end{array}
\,, \quad
\begin{array}{l}
\begin{array}{|c|}
\hline
         \3bar 
 \\
\hline
\end{array}
\end{array}
\equiv
\begin{array}{l}
\begin{array}{|c|}
\hline
3 \dots 3 
 \\
\hline
\end{array}
\end{array}
\,. \quad 
\end{eqnarray}
We will use this notation henceforth.
The indices $k_1, \dots, k_{2 p}, l_1, \dots, l_{2 q}$ in the above
Young tableau now get the 
values $1,2$ only. The parameters $p$ and $q$ get half-integer
values.
This yields a 
\begin{eqnarray}
{\bf p \otimes q \otimes  J}
\end{eqnarray}
$SU(2)^{diag}$-representation 
in the reduction (\ref{reduction}).
From this representation one can extract the 
irreducible representations by contracting pairs of the
indices $k,l,m$ with the $\epsilon$-symbol. Note however that it is not
necessary to contract pairs of a $k$ with an $l$ since then one would 
be over counting, as can be seen upon using the cyclic identity,
\begin{eqnarray}
\begin{array}{l}
\begin{array}{|c|c|}
\hline
            [i & j]  
 \\
\hline
\end{array}\\
\begin{array}{|c|}
            3  \hskip .05cm
\\
\hline
\end{array}
\end{array}
=
\ft12
\epsilon_{ij}
\begin{array}{l}
\begin{array}{|c|c|}
\hline
            1 & 3  
 \\
\hline
\end{array}\\
\begin{array}{|c|}
            2
\\
\hline
\end{array}
\end{array}
\label{simplecyclic}
\end{eqnarray}
Thus in the Young tableau (\ref{rearrangedyoung}) one takes only those
$SU(2)^{diag}$-irrepses that are obtained by contracting pairs of
$k$'s and $m$'s and pairs of $l$'s and $m$'s. The remaining indices are then 
completely symmetrized.
\par
We specify the most generic class
of Young tableaux that we will 
need for the harmonic analysis on $N^{010}$ and 
introduce the following notation for it,
\footnote{We use the notation,
\begin{equation}
\epsilon_{IJ} \equiv \epsilon_{i_1 j_1} \dots \epsilon_{i_{2J} j_{2J}}
\end{equation}
}
\begin{eqnarray}
&&  \Gamma^{\left(s,t,u\right)}_{k_1\dots k_{2 s},l_1\dots l_{2 t},m_1\dots m_{2 u}}
\left(M_1,M_2,J; p \right) \quad =
\nonumber \\
\nonumber \\
\nonumber \\
&& \epsilon_{IJ} 
\begin{array}{l}
\begin{array}{|c|c|c|c|c|c|}
\hline
            \1bar & k_1 \dots k_{2 s} &  i_1  \cdots  i_{2(p-s)} &
            i_{2(p-s)+1} \cdots i_{2(J-u)} & l_1 \dots l_{2 t} & 
                \3bar  
 \\
\hline
\end{array}\\
\begin{array}{|c|c|c|}
            \2bar & 3 \hskip .23cm \cdots \hskip .23cm 3 & 
              3 \hskip .49cm \cdots  \hskip .49cm 3 \\
\hline
\end{array}
\end{array}
\nonumber\\
&& \hskip 1.35cm 
\underbrace{\hskip 4.18 cm}_{2 p}
\underbrace{\hskip 3.5 cm}_{2(J-u-p+s)}
\nonumber \\
&& \hskip .75cm 
\underbrace{\hskip 4.8 cm}_{M_2}
\underbrace{\hskip 5.8 cm}_{M_1}
\nonumber \\
\nonumber \\
\nonumber \\
&& 
\hskip 7cm
\otimes
\begin{array}{l}
\begin{array}{|c|c|}
\hline
            j_1  \cdots  j_{2(J-u)} & m_1 \dots m_{2 u} 
\nonumber\\
\hline
\end{array}
\end{array}
\nonumber\\
&& 
\hskip 7cm
\hskip .5 cm 
\underbrace{\hskip 4.7 cm}_{2J}
\nonumber \\
\label{components}
\end{eqnarray}
where we have indicated the number of boxes under the Young tableaux.
In the above notation for the Young tableaux (\ref{components})
we do not a priori assume symmetrization of  the indices 
$k_1 \dots k_{2 s} l_1 \dots l_{2 t}$, although that is what we need 
in the harmonic expansion. It will be useful for our
purposes to consider the more general Young tableaux of (\ref{components}). 
It is also clear that in this notation,
\begin{eqnarray}
&2s \leq 2 p\leq M_2\,,&
\label{bound1}
\\
&2t \leq 2(J-u-p+s+t)\leq M_1\,,& 
\label{bound2}
\end{eqnarray}
is assumed without notice. Otherwise the Young tableaux simply
do not exist.
The only free $SU(2)^{diag}$-indices in the above Young tableaux
are the indices $k_1 \dots k_{2 s}$, $l_1 \dots l_{2 t}$ and 
$m_1 \dots m_{2 u}$,
since the indices $i_1 \dots i_{2(J-u)}$ are contracted with the
indices $j_1 \dots j_{2(J-u)}$ by means of the $\epsilon$'s. 
This is the most generic way finding the embedding of  a 
\begin{eqnarray}
{\bf s \otimes t \otimes u}
 \label{sxtxu}
\end{eqnarray} 
representation of 
$SU(2)^{diag}$ in a $SU(3)\times SU(2)$-representation. We will
refer to the three spins as the $\bf s$-spin, the $\bf t$-spin and the
$\bf u$-spin.
It is now clear that all the irreducible representations of $SU(2)^{diag}$
under reduction of $G$ to $H$, can be obtained be taking the
the $\bf s\oplus t \oplus u$ irreps from the decomposition of 
(\ref{sxtxu}).
\par
The Young tableaux in expression (\ref{components}) also transforms 
under $U(1)$, i.e. for the generator $\ft{i}{2} \l_8$ it has weight,
\begin{eqnarray}
i\sqrt{3}\left(
\ft{1}{3} (M_2-M_1)+ J - 2 p + (s+t-u)
\right)\,.
\end{eqnarray}
This weight has to be matched with the $U(1)$ weight 
\begin{eqnarray}
\ft{i}{2}\sqrt{3} \Delta
\label{U1weight}
\end{eqnarray} 
that is known
from the transformation of the harmonic as under (\ref{vecembed})
and (\ref{spinorembed8}). 
Hence the components of the $SU(2)^{diag}$-representation
that appear in the reduction $G\rightarrow H$ will be in
a given $U(1)$-representation if $p$ is constrained in terms
of the $G$-representation labels $M_1, M_2, J$ and $\Delta$
\begin{eqnarray}
p = \ft12 \left( J + \ft13 \left( M_2 - M_1 \right) 
+ (s+t-u) - \ft12 \Delta \right) \,.
\label{constraintp}
\end{eqnarray}
When we will refer to the $U(1)$
weight, we will refer to the number $\Delta$ 
henceforth. It is important to realize
here that this constraint not only constraints $p$
in terms of $M_1, M_2, J$ but it also
implies that the difference $M_2-M_1$ has to be a multiple of three.
Moreover as we will see later on 
$J$ has to be a positive integer. The parameter $p$ is half-integer.
\par
Now we know the embedding of the representations of
$SU(2)^{diag}\times U(1)$ in 
$SU(3)\times SU(2)$ with a given $SU(2)^{diag}$-spin and $U(1)$-weight.  
\par
In general 
all these Young tableaux are not independent. This can easily be
seen using the cyclic identity,
\begin{eqnarray}
\Gamma^{\left(s,t,u\right)}_{k_1\dots k_{2s},l_1\dots l_{2t},m_1\dots m_{2u}} 
\left(M_1,M_2,J;p \right)
&=&
\nonumber \\
& &
\Gamma^{\left(s,t+1/2,u-1/2\right)}_{k_1\dots k_{2s},l_1\dots l_{2t}
  m_{2u},m_1\dots m_{2u-1}} \left(M_1,M_2,J;p+\ft12 \right)
\nonumber \\
&&  
- \,
\Gamma^{\left(s+1/2,t,u-1/2\right)}_{k_1\dots k_{2s} m_{2u},l_1\dots l_{2t},
m_1\dots m_{2u-1}}
\left(M_1,M_2,J;p +\ft12  \right)
\nonumber \\
\label{cyclicid}
\end{eqnarray}
which is merely a generalization of the identity (\ref{simplecyclic}).
This identity allows us to reduce the number of Young tableaux that we are 
using  in most of the  cases. For instance, if we are considering 
$SU(3)$ irreducible representations that are big enough, i.e. the values
$M_1$ and $M_2$ exceed the numbers,
\begin{eqnarray}
M_2 \geq 2p+1, \qquad  M_1 \geq 2(J-u-p+s+t)+1
\end{eqnarray}
then one might make the harmonic expansion in components of the form 
$\Gamma^{(s^\prime,t^\prime,u-1/2)}$ only. 
An example where this is not possible is  the 
$\bf 0 \otimes 0 \otimes 1$ representation of 
$SU(2)^{diag}$
embedded in an $SU(3)\times SU(2)$ with 
\begin{eqnarray}
M_1=M_2=0\,, \qquad J=1 \,.
\label{nonrotcase}
\end{eqnarray}
However, there is a subset of embeddings where one can restrict oneself to the 
following components,
\pagebreak 
\begin{eqnarray}
&&\Gamma^{\left(s,t \right)}_{k_1\dots k_{2 s},l_1\dots l_{2 t}}
\left(M_1,M_2,J;p \right) \quad =
\nonumber \\
\nonumber \\
\nonumber \\
&\epsilon_{IJ}& \,
\begin{array}{l}
\begin{array}{|c|c|c|c|c|c|}
\hline
            \1bar & k_1 \dots k_{2s} &  i_1  \cdots  i_{2(p-s)} &
            i_{2(p-s)+1} \cdots i_{2J} & l_1 \dots l_{2t} & 
                3 \dots 3 
 \\
\hline
\end{array}\\
\begin{array}{|c|c|c|}
            \2bar & 3 \hskip .23cm \cdots \hskip .23cm 3 & 
                   3 \hskip .25cm \cdots  \hskip .43cm 3 \hskip .31cm \\
\hline
\end{array}
\end{array}
\otimes
\begin{array}{l}
\begin{array}{|c|}
\hline
            j_1  \cdots  j_{2J} 
\nonumber \\
\hline
\end{array}
\end{array}
\nonumber\\
&&
\hskip 0.9 cm
\underbrace{\hskip 1.8 cm}_{2s}
\underbrace{\hskip 2.3 cm}_{2(p-s)}
\underbrace{\hskip 2.9 cm}_{2(J-p+s)}
\underbrace{\hskip 1.6 cm}_{2t}
\nonumber\\
&& 
\hskip .3cm \underbrace{\hskip 4.8 cm}_{M_2}
\underbrace{\hskip 6 cm}_{M_1}
\hskip .7 cm 
\underbrace{\hskip 1.8 cm}_{2J}
\nonumber \\
\label{simplecomponents}
\end{eqnarray}
where $u=0$ is understood in the notation of $\Gamma$.
In this paper, for the calculation of the eigenvalues of the
mass operators we will always assume that $M_1$ and $M_2$
are ``big enough'' in order to justify an expansion in the
components (\ref{simplecomponents}) only. In other
words, here we calculate the mass matrix for those $SU(3)\times SU(2)$
representations where the harmonics sit in the $\bf s \oplus t$
irreducible representation of the decomposition of the
$\bf s$-spin times the $\bf t$-spin. Afterwards we will
argue the eigenvalues thus obtained, seen as functions of the
labels $M_1, M_2, J$, are also the functions for the eigenvalues
to be found in the cases where this assumption fails.
To illustrate this we will work out (\ref{nonrotcase}).
Yet a detailed study of the existing eigenvalues of the mass operators
in the cases of short multiplets will become quite a subtle book-keeping
exercise. Fortunately it can be implemented in some computer programs.
All this will be explained in the last section of this paper.
\par
We now apply this to Kaluza Klein on 
\begin{eqnarray}
AdS_4\times N^{010}\,.
\label{AdSN010}
\end{eqnarray} 
We write the 
coordinates of $AdS_4$ as $x$ and the coordinates of $N^{010}$ as $y$.
A field $\Phi$ on (\ref{AdSN010}) sits in a representation of $SO(1,3)$ as
well as in a representation of $SO(7)$, generically being some
multiple tensor product
of the vector representation and the spinor representation. As will be
exemplified further on, these $SO(7)$ representations decompose in
representations $\bf (\alpha)$ of $H$,
\begin{eqnarray}
SO(7) \rightarrow {\bf (\alpha_1)} \oplus \dots \oplus {\bf (\alpha_n)} \,,
\end{eqnarray}
hence the fields decompose in $SU(2)^{diag}$ fragments,
\begin{eqnarray}
\Phi \rightarrow
\left(
\matrix{
\Phi^{(\alpha_1)} \cr
\vdots \cr
\Phi^{(\alpha_n)}
}
\right)
\label{su2decomposition}
\end{eqnarray}
Then, using eq. (\ref{cosetexpansion}) the fragments $\Phi^{(\alpha_n)}$
can be expanded as
\begin{eqnarray}
\Phi^{(\alpha)}_\Lambda (x,y) = \sum_{(\mu)}  \Phi(x)^{(\mu)} \cdot
 (D^{(\mu)})_\Lambda (y) \,,
\label{KKharmonicexpansion}
\end{eqnarray}
The dot indicates the contraction of the index $\cal G$ and
\begin{eqnarray}
\Lambda =
&& \Gamma^{\left(s,t\right)}_{(k_1\dots k_s,l_1\dots l_t)}
\left(M_1,M_2,J;p \right) \,,
\end{eqnarray}
which is symmetrized in all indices $k_1\dots k_s,l_1\dots l_t$.
The linearized field equations on the fragments
of the $11$-dimensional theory split as follows,
\begin{eqnarray}
\left( \Box_x + \Box_y \right) \Phi^{(\alpha)}_\Lambda (x,y) = 0
\label{2boxes}
\end{eqnarray}
then $\Box_y (Y^{(\mu)})_\Lambda (y)$ can be evaluated explicitly
and provide the mass operators of the four-dimensional theory.
The evaluation
of these mass matrices and their eigenvalues is the subject
of the following sections.
\par
To see how the covariant derivative (\ref{covder})
works on the components of the Young 
tableaux (\ref{components}), one first straightforwardly derives the following 
formulae,
\begin{eqnarray}
&&
\left(\l-\tau \right)_a 
\Gamma^{\left(s,t\right)}_{k_1\dots k_{2s},l_1\dots l_{2t}}(p) =
\nonumber \\
&& \hskip 1.7cm \sum_{\mu=1}^{2s} (\l^a){}_{k_\mu}{}^{ m} \,
\Gamma^{\left(s-1/2,t\right)}_{k_1\dots  m_\mu \dots 
k_{2s},l_1\dots l_{2t}}(p-\ft12)
+ \sum_{\nu=1}^{2t} (\l^a){}_{l_\nu}{}^{ n} \,
\Gamma^{\left(s,t-1/2\right)}_{k_1\dots k_{2s} 
,l_1\dots n_\nu \dots l_{2t}}(p)
\nonumber \\
&& \hskip 1.7cm
+4\left( p-s \right) (\l^a)_{i}{}^m \epsilon^{i n}
\left(
\Gamma^{\left(s+1/2,t+1/2\right)}_{m k_1 \dots k_{2s},
l_1 \dots l_{2t} n}(p+\ft12)
 - 
\Gamma^{\left(s+1,t\right)}_{m n k_1 \dots k_{2s},
l_1 \dots l_{2t}}(p+\ft12)
\right)
\nonumber \\
&& \hskip 1.7cm
+4\left( J-p+s \right)  (\l^a)_{i}{}^m \epsilon^{i n}
\left(
\Gamma^{\left(s,t+1\right)}_{k_1 \dots k_{2s},
l_1 \dots l_{2t} m n}(p+\ft12)
- 
\Gamma^{\left(s+1/2,t+1/2\right)}_{n k_1 \dots k_{2s},
l_1 \dots l_{2t} m}(p+\ft12)
\right)
\,,
\nonumber \\
\label{lmta}
\end{eqnarray} 
and
\begin{eqnarray}
&&
\l_A \,\, 
\Gamma^{\left(s,t\right)}_{k_1\dots k_{2s},l_1\dots l_{2t}}(p) =
\nonumber \\
&&
\hskip 2.5 cm
-(M_2-2p) \, \epsilon^{mn} (\l^A)_m{}^3 
\Gamma^{\left(s+1/2,t\right)}_{n k_1\dots k_{2s},l_1\dots l_{2t}}(p+\ft12)
\nonumber \\
&&
\hskip 2.5 cm
+\sum_{\mu=1}^{2s} \epsilon_{k_\mu m} (\l^A)_3{}^m
\Gamma^{\left(s-1/2,t\right)}_{k_1\dots \hat k_\mu \dots k_{2s},
l_1\dots l_{2t}}(p-\ft12)
\nonumber \\
&&
\hskip 2.5 cm
+ 2 (s-p)(\l^A)_3{}^m \left( 
\Gamma^{\left(s+1/2,t\right)}_{m k_1\dots k_{2s}, 
l_1\dots l_{2t}}(p)
-  \Gamma^{\left(s,t+1/2\right)}_{k_1\dots k_{2s}, 
l_1\dots l_{2t} m}(p) \right)
\nonumber \\
&&
\hskip 2.5 cm
-2(J-p+s) \epsilon^{mn} (\l^A)_m{}^3
\left( 
\Gamma^{\left(s+1/2,t\right)}_{n k_1\dots k_{2s}, 
l_1\dots l_{2t}}(p+\ft12) 
-
\Gamma^{\left(s,t+1/2\right)}_{k_1\dots k_{2s}, 
l_1\dots l_{2t} n}(p+\ft12) \right)
\nonumber \\
&&
\hskip 2.5 cm
+\sum_{\nu=1}^{2t} (\l^A)_{l_\nu}{}^3
\Gamma^{\left(s,t-1/2\right)}_{k_1\dots k_{2s}, 
l_1\dots \hat l_{\nu} \dots l_{2t} }(p)
\nonumber \\
&&
\hskip 2.5 cm
+(M_1 - 2J+2p-2s-2t) (\l^A)_3{}^m
\Gamma^{\left(s,t+1/2\right)}_{k_1\dots k_{2s}, 
l_1\dots l_{2t} m}(p)
\label{lA}
\end{eqnarray}
where $m_\mu$ means that $k_\mu$ is replaced by $m$ 
and  $n_\nu$ means that $l_\nu$ is replaced by $n$. 
The hats on the indices $\hat k$ and $\hat l$ indicate that
these indices are deleted.
We have suppressed  the labels $M_1, M_2, J$, since they do not change 
under the transformations. In fact, they do not change under
the covariant derivative and the evaluation of the
mass operators can be done for fixed $M_1, M_2, J$ in the
harmonic expansion (\ref{KKharmonicexpansion}). 
For the derivation of the above expressions
we used the cyclic identity (\ref{cyclicid}).
Mind that in the Young tableaux
$\Gamma^{\left(s,t\right)}_{k_1\dots k_{2s},l_1\dots l_{2t}}(p)$
we have not symmetrized the indices $k_1\dots k_{2s},l_1\dots l_{2t}$.
Recall that in the notation (\ref{simplecomponents})
there is no symmetrization in these indices.
\par
Looking at the expression for the covariant derivative 
in eq. (\ref{covder}), one sees that the generators $T_\alpha$
are precisely given by the above formulae (\ref{lmta}) and (\ref{lA}) 
multiplied with $\ft{i}{2}$.
Moreover, these formulae can be programmed on a computer and 
in this way the covariant derivative is easily calculated. 
\section{The $0$-form, the $1$-form, the $2$-form and the spinor}
In the previous section we explained how harmonic analysis is 
done on the coset $N^{010}=SU(3)\times SU(2)/SU(2)\times U(1)$ when the 
$SU(2)$ is diagonally embedded in $SU(3)$. In this section we apply
this to the fields of eleven-dimensional supergravity.
\par
The fields of eleven-dimensional supergravity 
are expanded around the Freund-Rubin background.
Their fluctuations
are the metric $h_{MN}(x,y)$, the three-form $a_{MNP}(x,y)$ 
and the spinor $\psi_M(x,y)$. 
They can be expanded in the following harmonics \cite{castdauriafre, noi321}:
the scalar $Y(y)$, the transverse vector $Y^\alpha(y)$,
the transverse 2-form $Y^{\alpha\beta}(y)$,
the transverse 3-form $Y^{\alpha\beta\gamma}(y)$,
the symmetric transverse traceless tensor $Y^{\{\alpha\beta\}}(y)$,
the spinor $\Xi(y)$ and the irreducible vector spinor $\Xi_\alpha(y)$.
We list them in table \ref{eoms} together with their
$SO(7)$ irreducible representation and their invariant field equations
written in terms of the covariant derivatives of the coset (\ref{covder}).
\begin{table}
\begin{tabular}{|c|c|c|}
\hline
& & \cr 
harmonic &
SO(7)-irrep &
{\rm field \,\, equation} \cr
& & \cr 
\hline
\hline
& & \cr 
$Y$ & $\bf 1 = (0,0,0)$ &
$ {\cal D}^\alpha {\cal D}_\alpha Y = M_{(0)^3} Y $ \cr
& & \cr 
\hline
& & \cr 
$Y^\alpha$ & $\bf 7= (1,0,0)$ &
$2 {\cal D}^\alpha {\cal D}_{[\alpha} Y_{\beta]} = M_{(1)(0)^2} Y^\alpha$ \cr
& & \cr 
\hline
& & \cr 
$Y^{\alpha\beta}$ & $\bf 21 = (1,1,0)$ &
$ 3 {\cal D}^\gamma {\cal D}_{[\gamma} Y_{\alpha\beta]} =
M_{(1)^2 (0)} Y^{\alpha\beta} $ \cr
& & \cr 
\hline
& & \cr 
$Y^{\alpha\beta\gamma}$ & $\bf 35 = (1,1,1)$ &
$\ft{1}{24} \varepsilon_{\alpha\beta\gamma}{}^{\mu\nu\rho\sigma}
{\cal D}_\mu Y_{\nu\rho\sigma} =
M_{(1)^3} Y^{\alpha\beta\gamma}$ \cr
& & \cr 
\hline
& & \cr 
$Y^{(\alpha\beta)}$ &
${\bf 27 = (2,0,0)}$ &
$\left(
(\Box + 56) \delta^{\alpha\beta}_{\gamma\delta}
- R^{\alpha\beta}{}_{\gamma\delta}
\right) 
Y^{(\gamma\delta)} =
M_{(2)^2 (0)} Y^{(\alpha\beta)}$ \cr
& & \cr 
\hline
& & \cr 
$\Xi$ & ${\bf (\ft12,\ft12,\ft12)}$ &
$\tau^\mu \left({\cal D}_\mu - \tau_\mu \right) \Xi
= M_{(0)^3} \Xi$ \cr
& & \cr 
\hline
& & \cr 
$ \Xi_\alpha $  & $ {\bf (\ft32,\ft12,\ft12)}$ &
$\tau_\alpha{}^{\beta\gamma} \nabla_\beta \Xi_\gamma
- \ft57  \tau_\alpha \tau^{\beta\gamma} \nabla_\beta \Xi_\gamma
= M_{(\ft32)(\ft12)^2} \Xi_\alpha$
\cr
& & \cr 
\hline
\end{tabular}
\caption{The content of harmonics of eleven-dimensional supergravity:
\newline
their name as in \cite{noi321}, 
their $SO(7)$ irreducible representations and their 
\newline
field equations.}
\label{eoms}
\end{table}
The following fields are transverse,
\begin{eqnarray}
&&{\cal D}^\alpha Y_\alpha = 0 \,,
\nonumber \\
&&{\cal D}^\alpha Y_{[\alpha\beta]} = 0 \,,
\nonumber \\
&&{\cal D}^\alpha Y_{[\alpha\beta\gamma]} = 0 \,,
\nonumber \\
&&{\cal D}^\alpha Y_{(\alpha\beta)} = 0 \,,
\nonumber \\
&&{\cal D}^\alpha \Xi_\alpha =0 \,.
\end{eqnarray}
Since we are dealing with ${\cal N}=3$ supersymmetry it will suffice
to know the masses for the zero-form, the one-form, the two-form and
the spinor only. The rest of the multiplet spectrum can then be determined
using supersymmetry.
\par
The decomposition of the $SO(7)$ vector representation
into irreducible representations of $SU(2)^{diag}$
goes as follows
\begin{eqnarray}
{\bf (0,0,0)} &\rightarrow& {\bf (1,0)} \,,
\nonumber \\
{\bf (1,0,0)} &\rightarrow& {\bf (3,0)} \oplus {\bf (2,1)} \oplus 
{\bf (\bar 2,-1)} \,,
\nonumber \\
{\bf (1,1,0)} &\rightarrow& {\bf (4,1)} \oplus {\bf (\bar 4,-1)}
\oplus {\bf (3,0)} \oplus {\bf (3,0)} \oplus 
\nonumber \\
& & {\bf (2,-1)}
\oplus {\bf (\bar 2,1)} \oplus {\bf (1,2)} \oplus
{\bf (\bar 1, -2)} \oplus {\bf (1,0)} \,,
\label{fdec}
\end{eqnarray}
where the irreducible representations of $SU(2)^{diag} \times U(1)$
are written as $\bf \left(2(s+t)+1,\Delta\right)$,
$\Delta$ being the $U(1)$ weight (\ref{U1weight}).
One may check now that
 from the formula (\ref{constraintp}) it follows indeed that $J$ has 
to be an integer number.
Accordingly, an $SO(7)$ one-form and
an $SO(7)$ two-form can be expressed in terms
of the following $SU(2)^{diag}$ fragments
\begin{eqnarray}
\Phi^\alpha \rightarrow
\left(
\matrix{
{\cal C}^{(1)}_i\cr
{\cal C}^{(1)}{}^i \cr
{\cal R}^{(1)}_{ij}
}
\right) \,,
\nonumber \\
\nonumber \\
\nonumber \\
\Phi^{\alpha\beta}
\rightarrow
\left(
\matrix{
{\cal R}^{(2)} \cr
{\cal C}^{(2)} \cr
{\cal C}^{(2)}{}^* \cr
{\cal C}^{(2)}_j \cr
{\cal C}^{(2)}{}^j \cr
{\cal R}^{(2)}_{ij} \cr
{\cal O}^{(2)}_{ij} \cr
{\cal C}^{(2)}_{ijk} \cr
{\cal C}^{(2)}{}^{ijk}
}
\right)
\label{decompfragments}
\end{eqnarray}
where the fields ${\cal C}^{i_1 \dots i_n}$
are the complex conjugate fields of ${\cal C}_{i_1 \dots i_n}$
and the symmetrized indices indicate in what representation of
$SU(2)^{diag}$ they sit. The superscript $(1)$ and $(2)$ indicate 
that these are the fragments of the one-form and the two-form
respectively. The fields ${\cal R}$ are pseudo real, i.e.
\begin{eqnarray}
{\cal R}^{i_1 \dots i_n} =
\epsilon^{i_1 j_1} \dots \epsilon^{i_n j_n} {\cal R}_{j_1 \dots j_n}
\,.
\end{eqnarray}
For the choice of the coset generators (\ref{cosetgenerators}) it is useful
to introduce the indices (\ref{indexsplit}) and split 
\begin{eqnarray}
\Phi^\alpha =\{\Phi^a, \Phi^A\} \,,
\nonumber \\
\Phi^{\alpha\beta} = \{\Phi^{ab}, \Phi^{aA}, \Phi^{AB}\} \,.
\end{eqnarray}
then one can do the decomposition (\ref{decompfragments}) 
concretely by using the components of the $\lambda$ matrices,
\begin{eqnarray}
&&\Phi = {\cal R}^{(0)} \,,
\nonumber \\
\nonumber \\
&& \Phi^a = i (\lambda^a)_i{}^j\, \epsilon^{ik}\, {\cal R}^{(1)}_{jk} \nonumber\\
&& \Phi^A = (\lambda^A)_3{}^i \, {\cal C}^{(1)}_i  
         + (\lambda^A)_i{}^3 \,  {\cal C}^{(1)}{}^i \,,
\nonumber \\
\nonumber \\
&& \Phi^{ab} = i \, \varepsilon^{abc} (\lambda^c)_{i}{}^{j}
\epsilon^{ik} {\cal R}^{(2)}_{jk}
\,, \nonumber \\
&& \Phi^{aA} =
(\lambda^{a})_i{}^j
(\lambda^{A})_k{}^3 \epsilon^{ik}
{\cal C}^{(2)}_j
+
(\lambda^a)_j{}^i
(\lambda^A)_3{}^k \epsilon_{ik}
{\cal C}^{(2)}{}^j
\nonumber \\
&& \hskip 1.2cm +
(\lambda^a)_i{}^j
(\lambda^A)_3{}^k
\epsilon^{il}
{\cal C}^{(2)}_{jkl}
+
(\lambda^a)_j{}^i
(\lambda^A)_k{}^3
\epsilon_{ik}
{\cal C}^{(2)}{}^{ijk}
\,, \nonumber \\
&& \Phi^{AB} = 
(\lambda^A)_3{}^i
(\lambda^B)_3{}^j
\epsilon_{ij} {\cal C}^{(2)}
+
(\lambda^A)_i{}^3
(\lambda^B)_j{}^3
\epsilon^{ij} {\cal C}^{(2)}{}^*
\nonumber \\
&& \hskip 1.2cm +
i (\lambda^{[A})_i{}^3
( \lambda^{B]})_3{}^i
{\cal R}^{(2)}
+
(\lambda^{[A})_i{}^3
(\lambda^{B]})_3{}^j
\epsilon^{ik}
{\cal O}^{(2)}_{kj}
\,.
\label{su2decomp}
\end{eqnarray}
This shows how the decomposition (\ref{reduction}) is done.
The numbers $\Delta$ in (\ref{fdec})  are obtained by applying
the matrix $T_8$ in (\ref{vecembed})
on the components of $\Phi^\alpha$ and $\Phi^{\alpha\beta}$.
The scalar $\Phi$ does not transform under $T_8$ and hence $\Delta=0$.
One may also verify that the above decomposition (\ref{su2decomp})
is in agreement with the $SU(2)^{diag}$ matrices (\ref{vecembed}).
The expansion in the harmonics $D$ then becomes\footnote{
We expand the complex conjugates in the
conjugate representations. Upon doing so there appear some irrelevant
signs in the front of the harmonics. 
We have absorbed these signs in the $x$ fields.},
\begin{eqnarray}
{\cal R}^{(0)} &=& h^{(0,0)}(x) \cdot D^{(0,0)(0)}(y) \,,
\nonumber \\
\nonumber \\
{\cal C}_i^{(1)} &=& AW_c^{(1/2,0)}(x) \cdot D^{(1/2,0)(1)}_i(y)   +
                     AW_c^{(0,1/2)}(x) \cdot D^{(0,1/2)(1)}_i(y) \,, 
\nonumber \\
{\cal C}^{(1)}{}^i &=& \epsilon^{ij}\, \tilde{AW}_c^{(1/2,0)}(x) 
                       \cdot D^{(1/2,0)(-1)}_j(y)   +
                       \epsilon^{ij}\, \tilde{AW}_c^{(0,1/2)}(x) 
                       \cdot D^{(0,1/2)(-1)}_j(y) \,, 
\nonumber \\
{\cal R}^{(1)}_{ij} &=& AW_r^{(1,0)}(x) \cdot D^{(1,0)(0)}_{ij}(y)   +
                        AW_r^{(1/2,1/2)}(x) \cdot D^{(1/2,1/2)(0)}_{ij}(y) +
\nonumber  \\
                    & &  AW_r^{(0,1)}(x) \cdot D^{(0,1)(0)}_{ij}(y) \,,
\nonumber \\
\nonumber \\
{\cal R}^{(2)} &=& Z_r^{(0,0)}(x) \cdot D^{(0,0)(0)}(y) \,,
\nonumber \\
{\cal C}^{(2)} &=& Z_c^{(0,0)}(x) \cdot D^{(0,0)(2)}(y) \,,
\nonumber \\
{\cal C}^{(2)}{}^* &=& \tilde{Z}_c^{(0,0)}(x) \cdot D^{(0,0)(-2)}(y) \,,
\nonumber \\
{\cal C}^{(2)}_i &=& Z_c^{(1/2,0)}(x) \cdot D_i^{(1/2,0)(-1)}(y) + 
                     Z_c^{(0,1/2)}(x) \cdot D_i^{(0,1/2)(-1)}(y) \,,
\nonumber \\
{\cal C}^{(2)}{}^i &=& \epsilon^{ij}\, \tilde{Z}_c^{(1/2,0)}(x) 
                       \cdot D_j^{(1/2,0)(1)}(y) +
                       \epsilon^{ij}\, \tilde{Z}_c^{(0,1/2)}(x) 
                       \cdot D_j^{(0,1/2)(1)}(y)  \,,
\nonumber \\
{\cal R}^{(2)}_{ij} &=& Z_r^{(1,0)}(x) \cdot D^{(1,0)(0)}_{ij}(y)   +
                        Z_r^{(1/2,1/2)}(x) \cdot D^{(1/2,1/2)(0)}_{ij}(y) +
\nonumber  \\
                    & &  Z_r^{(0,1)}(x) \cdot D^{(0,1)(0)}_{ij}(y) \,,
\nonumber \\
{\cal O}^{(2)}_{ij} &=& Z_o^{(1,0)}(x) \cdot D^{(1,0)(0)}_{ij}(y)   +
                        Z_o^{(1/2,1/2)}(x) \cdot D^{(1/2,1/2)(0)}_{ij}(y) +
\nonumber  \\
                    & &  Z_o^{(0,1)}(x) \cdot D^{(0,1)(0)}_{ij}(y) \,,
\nonumber \\        
{\cal C}^{(2)}_{ijk} &=& Z_c^{(3/2,0)}(x) \cdot D^{(3/2,0)(1)}_{ijk} (y) +
                         Z_c^{(1,1/2)}(x) \cdot D^{(1,1/2)(1)}_{ijk} (y) +
\nonumber \\
                     & & Z_c^{(1/2,1)}(x) \cdot D^{(1/2,1)(1)}_{ijk} (y) +
                         Z_c^{(0,3/2)}(x) \cdot D^{(0,3/2)(1)}_{ijk} (y)\,,
\nonumber \\
{\cal C}^{(2)}{}^{ijk} &=& 
                \epsilon^{im}\epsilon^{jn}\epsilon^{kp}\, 
              \tilde{Z}_c^{(3/2,0)}(x) \cdot D^{(3/2,0)(-1)}_{mnp} (y) +
                \epsilon^{im}\epsilon^{jn}\epsilon^{kp}\, 
              \tilde{Z}_c^{(1,1/2)}(x) \cdot D^{(1,1/2)(-1)}_{mnp} (y) +
\nonumber \\
            & &  \epsilon^{im}\epsilon^{jn}\epsilon^{kp}\, 
              \tilde{Z}_c^{(1/2,1)}(x) \cdot D^{(1/2,1)(-1)}_{mnp} (y) +
                \epsilon^{im}\epsilon^{jn}\epsilon^{kp}\, 
              \tilde{Z}_c^{(0,3/2)}(x) \cdot D^{(0,3/2)(-1)}_{mnp} (y) \,.
\nonumber \\
\label{harmexp}
\end{eqnarray}
where the index structure on the harmonics
$D^{(s,t)(\Delta)}_{i_1 \dots i_{2(s+t)}}$ is
given by the {\it symmetrized} Young tableau
\begin{eqnarray}
\Gamma^{(s,t)}_{(i_1 \dots i_{2s}, i_{2s+1} \dots i_{2(s+t)})}(M_1,M_2,J;p)
\label{repsD}
\end{eqnarray}
with $p$ constrained as in (\ref{constraintp}).
We see that we have summed in the harmonic expansion a multiple of times
over the same irreducible representations of $G$  when a given
irreducible representation of $SU(2)^{diag}$ appears multiple times
in the reduction $G\rightarrow H$.
The names of the $x$-fields refer to names that were used in
the papers \cite{noi321}. 
For instance the $x$-field $h^{(0,0)}$ gives the masses
for the spin-$2$ fields $h_{\mu\nu}$, the $x$-fields $AW$ will
contribute to the vectors $A$ and $W$, and $Z$ will contribute
to the vector $Z$.
So now, for a given representation
of $G$, labeled by $M_1,M_2,J$ we have the following basis of
four-dimensional fields
\begin{eqnarray}
h\equiv
\left(
\matrix{
h^{(0,0)} \ cr
}
\right) (x) \,, \qquad
AW\equiv
\left(
\matrix{
AW_r^{(0,1)}  \cr
AW_r^{(1/2,1/2)}  \cr
AW_r^{(1,0)}  \cr
AW_c^{(0,1/2)}  \cr
AW_c^{(1/2,0)}  \cr
\tilde{AW}_c^{(0,1/2)}  \cr
\tilde{AW}_c^{(1/2,0)}  
}
\right)(x)  \,, \qquad
Z\equiv
\left(
\matrix{
Z_c^{(0,3/2)}\cr
Z_c^{(1/2,1)}\cr
Z_c^{(1,1/2)}\cr
Z_c^{(3/2,0)}\cr
\tilde{Z}_c^{(0,3/2)}\cr
\tilde{Z}_c^{(1/2,1)}\cr
\tilde{Z}_c^{(1,1/2)}\cr
\tilde{Z}_c^{(3/2,0)}\cr
Z_o^{(0,1)} \cr
Z_o^{(1/2,1/2)} \cr
Z_o^{(1,0)} \cr
Z_r^{(0,1)} \cr
Z_r^{(1/2,1/2)} \cr
Z_r^{(1,0)} \cr
Z_c^{(0,1/2)} \cr
Z_c^{(1/2,0)} \cr
\tilde{Z}_c^{(0,1/2)} \cr
\tilde{Z}_c^{(1/2,0)} \cr
Z_c^{(0,0)} \cr
\tilde{Z}_c^{(0,0)} \cr
Z_r^{(0,0)} \cr
}
\right)(x) \,.
\label{formsbasis}
\end{eqnarray}
In order to determine the mass spectrum we have to apply the
invariant Laplace Beltrami equations of table \ref{eoms}
to the harmonics
in (\ref{harmexp}) which will then determine the masses of the
four dimensional $x$ fields by means of the eq. (\ref{2boxes}).
Clearly, since we have a $1$-dimensional, a $7$-dimensional
and a $21$-dimensional basis in (\ref{formsbasis}) for the
zero-form, the one-form and the two-form respectively the
matrices $M_{(0)^3}$, $M_{(1)(0)^2}$ and $M_{(1)^2(0)}$
will be $1 \times 1$, $7 \times 7$ and $21 \times 21$.
Notice that for the two-form we will already need to handle
a $21\times 21$ matrix. Had we not restricted ourselves to
representations made with $\bf s$-spin and $\bf t$-spin only, 
then we would have had to handle a $41 \times 41$ matrix for the
two-form. Moreover we would be over counting in the harmonic
expansion.
We then calculate the eigenvalues of these operators and the
masses are then found by using the mass formulas of 
\cite{noi321}.
\par
In order to obtain the complete multiplet spectrum by the
method of paper \cite{m111}, 
it is most convenient to have the eigenvalues
of the spinor also. The spinor $\Xi$ is also decomposed in
irreducible representations of $SU(2)^{diag}$ as follows,
\begin{eqnarray}
{\bf (\ft12,\ft12,\ft12)} \rightarrow
{\bf (3,0)} \oplus {\bf (2,-1)} \oplus {\bf (2,1)} \oplus {\bf (1,0)}
\label{decompspinor}
\end{eqnarray}
Concretely, when using the $SO(7)$ $\tau$-matrices for the Clifford
algebra as we are advocating in the appendix, then one can see
from (\ref{spinorembedsu2}) that its components are 
\begin{eqnarray}
\Xi =
\left(
\matrix{
\phi \cr
\xi_i \cr
\zeta_i \cr
\psi_{ij}
}
\right)
\label{spinordecompfrags}
\end{eqnarray}
Moreover
the Majorana condition
\begin{eqnarray}
C \, \Xi^* = \Xi  \,,
\end{eqnarray}
with the conjugation matrix as in the appendix,
is translated into (pseudo) reality conditions on its components,
\begin{eqnarray}
 \phi^* &=& \phi \,, \nonumber \\
 \xi^i &=& \epsilon^{ij} \, \xi_j \,, \nonumber \\
 \zeta^i &=& - \epsilon^{ij} \,\zeta_j \,, \nonumber \\
 \psi^{ij} &=& \epsilon^{ik}\epsilon^{jl} \,\psi_{kl} \,. 
\end{eqnarray}
The numbers $\Delta$ are read off from (\ref{spinorembed8}).
then similarly as for the forms, the harmonic expansion
acquires the form,
\begin{eqnarray}
\phi &=& \chi^{(0,0)}(x) \cdot D^{(0,0)(0)} (y) \,,
\nonumber \\
\xi_i &=& \chi_\xi^{(1/2,0)}(x) \cdot D_i^{(1/2,0)(-1)} (y) 
          +  \chi_\xi^{(0,1/2)}(x) \cdot D_i^{(0,1/2)(-1)} (y) \,,
\nonumber \\
\zeta_i &=& \chi_\zeta^{(1/2,0)}(x) \cdot D_i^{(1/2,0)(1)} (y) 
          +  \chi_\zeta^{(0,1/2)}(x) \cdot D_i^{(0,1/2)(1)} (y) \,,
\nonumber \\
\psi_{ij} &=& \chi^{(1,0)} (x) \cdot D_{ij}^{(1,0)(0)} +
              \chi^{(1/2,1/2)} (x) \cdot D_{ij}^{(1/2,1/2)(0)} +
              \chi^{(0,1)} (x) \cdot D_{ij}^{(0,1)(0)} 
\,.
\label{actdecompspinor}
\end{eqnarray}
So we conclude that for the $x$-fields of the spinor
we need the basis
\begin{eqnarray}
\chi \equiv
\left(
\matrix{
\chi^{(0,1)}  \cr
\chi^{(1/2,1/2)}  \cr
\chi^{(1,0)}  \cr
\chi_\xi^{(0,1/2)} \cr
\chi_\xi^{(1/2,0)} \cr
\chi_\zeta^{(0,1/2)} \cr
\chi_\zeta^{(1/2,0)} \cr
\chi^{(0,0)}
}
\right)(x)
\end{eqnarray}
for which we should find an $8\times 8$ matrix $M_{(2)^2(0)}$.
\par
At this stage we have all the ingredients to calculate the 
matrices $M_{(0)^3}$, $M_{(0)^2(1)}$, $M_{(0)(1)^2}$ and $M_{(1/2)^3}$.
As explained before, it is possible to restrict oneself on the
terms in the expansion for a given irreducible representation of $G$.
Such representation is characterized by the labels $M_1, M_2, J$,
see (\ref{SU3Youngtableau}).
To do the calculation, we need to perform the following steps:
\begin{enumerate}
\item{Invert (\ref{su2decomp}).
We know how the mass operators work
on the the fields $\Phi$. They are given in table \ref{eoms}. To see 
how they work on the fragments we invert (\ref{su2decomp}).
}
\item{For all of the fragments in (\ref{decompfragments})
and (\ref{spinordecompfrags}) draw the Young tableaux of
$D^{(s,t)(\Delta)}$. Check whether it exists. Solve the
constraint (\ref{constraintp}) and then check the bounds
(\ref{bound1}) and (\ref{bound2}).
}
\item{Apply the operators of table \ref{eoms}, by evaluating
the covariant derivative (\ref{covder}) on the Young tableaux
of the fragments. Recall that these are not differential operators.
All one has to do is to evaluate the operators $T_\alpha$
and do the multiplication with the structure constants.
To this end one can use the formulae (\ref{lmta}) and (\ref{lA}). 
}
\end{enumerate}
All the above steps can be programmed in say Mathematica. 
We assume that the bounds (\ref{bound1}) and (\ref{bound2})
are satisfied and calculate the four contemplated operators.
Because of this assumption these
matrices are the matrices for the long ${\cal N}=3$
multiplets. Indeed, since all the possible fragments are present
that implies that all the field components of the multiplets
are present. 
We list the matrices in appendix \ref{massmatrix}.
\par
We finish this section by presenting the eigenvectors of the operators
$M_{(0)^3}$, $M_{(0)^2(1)}$, $M_{(0)(1)^2}$ and $M_{(1/2)^3}$,
which is what we ultimately need for the masses.
We use the notation
\begin{eqnarray}
H_0 \equiv
\frac{16}{3}\left(  2 (M_1^2 + M_2^2 + M_1 M_2 + 3 M_1  + 3 M_2 )
                   -3 J (J+1)
 \right)\,.
\end{eqnarray}
We write the subscript $0$ to remember that this is the eigenvalue
of the zero-form operator.
We denote the eigenvalues by $\lambda^{(f)}_\mu$, where 
$f=0,1,2,s$ refers to the zero-form, the one-form, the two-form
and the spinor
respectively and $\mu$ enumerates its different eigenvalues.
Then, 
\begin{itemize}
\item{the zero-form eigenvalues:
\begin{eqnarray}
\lambda^{(0)}=H_0 \,.
\label{zerofeig}
\end{eqnarray}
}
\item{the one-form eigenvalues:
\begin{eqnarray}
\lambda_1^{(1)}&=& H_0 - 32 J - 8 - 4 \sqrt{H_0 - 32 J + 4}
\,, \label{onefeig1}
\\
\lambda_2^{(1)}&=& H_0 - 32 J - 8 + 4 \sqrt{H_0 - 32 J + 4}
\,,\\
\lambda_3^{(1)}&=& H_0 + 24 - 4 \sqrt{H_0 + 36 }
\,,\\
\lambda_4^{(1)}&=& H_0 + 24 + 4 \sqrt{H_0 + 36 }
\,,\\
\lambda_5^{(1)}&=& H_0 + 32 J + 24 - 4 \sqrt{H_0 + 32 J + 36}
\,,\\
\lambda_6^{(1)}&=& H_0 + 32 J + 24 + 4 \sqrt{H_0 + 32 J + 36}
\,,\\
\lambda_7^{(1)}&=& H_0
\,. \label{onefeig7}
\end{eqnarray}
}
\item{the two-form eigenvalues:
\begin{eqnarray}
\lambda_1^{(2)} &=& H_0 + 64 J
\,, \label{twofeig1}
\\
\lambda_2^{(2)} &=& H_0 + 32 J + 32
\,, \\
\lambda_3^{(2)} &=& H_0 + 32 J + 32
\,, \\
\lambda_4^{(2)} &=& H_0 + 32
\,, \\
\lambda_5^{(2)} &=& H_0 + 32
\,, \\
\lambda_6^{(2)} &=& H_0 + 32
\,, \\
\lambda_7^{(2)} &=& H_0 - 32 J 
\,, \\
\lambda_8^{(2)} &=& H_0 - 32 J 
\,, \\
\lambda_9^{(2)} &=& H_0 - 64 J - 64 
\,, \\
\lambda_{10}^{(2)} &=& H_0 + 48 + 8 \sqrt{H_0 + 36}
\,, \\
\lambda_{11}^{(2)} &=& H_0 + 48 - 8 \sqrt{H_0 + 36}
\,, \\
\lambda_{12}^{(2)} &=& H_0 + 32 J + 48 + 8 \sqrt{H_0 + 32 J + 36} 
\,, \\
\lambda_{13}^{(2)} &=& H_0 + 32 J + 48 - 8 \sqrt{H_0 + 32 J + 36} 
\,, \\
\lambda_{14}^{(2)} &=& H_0 - 32 J + 16 + 8 \sqrt{H_0 - 32 J + 4}
\,, \\
\lambda_{15}^{(2)} &=& H_0 - 32 J + 16 - 8 \sqrt{H_0 - 32 J + 4}
\,, \\
\lambda_{16}^{(2)} &=& H_0 - 32 J - 8 - 4 \sqrt{H_0 - 32 J + 4}
\,, \\
\lambda_{17}^{(2)} &=& H_0 - 32 J - 8 + 4 \sqrt{H_0 - 32 J + 4}
\,, \\
\lambda_{18}^{(2)} &=& H_0 + 24 - 4 \sqrt{H_0 + 36 }
\,, \\
\lambda_{19}^{(2)} &=& H_0 + 24 + 4 \sqrt{H_0 + 36 }
\,, \\
\lambda_{20}^{(2)} &=& H_0 + 32 J + 24 - 4 \sqrt{H_0 + 32 J + 36}
\,, \\
\lambda_{21}^{(2)} &=& H_0 + 32 J + 24 + 4 \sqrt{H_0 + 32 J + 36}
\,.
\label{eigenstwo}
\end{eqnarray}
}
\item{the spinor eigenvalues:
\begin{eqnarray}
\lambda_1^{(s)}&=& -6 - \sqrt{H_0 - 32 J + 4}
\,, 
\label{speig1} \\
\lambda_2^{(s)}&=& -6 + \sqrt{H_0 - 32 J + 4}
\,, \\
\lambda_3^{(s)}&=& -6 - \sqrt{H_0 + 36}
\,, \\
\lambda_4^{(s)}&=& -6 + \sqrt{H_0 + 36}
\,, \\
\lambda_5^{(s)}&=& -6 - \sqrt{H_0 + 32 J + 36}
\,, \\
\lambda_6^{(s)}&=& -6 + \sqrt{H_0 + 32 J + 36}
\,, \\
\lambda_7^{(s)}&=& - 10 - \sqrt{H_0 + 36}
\,, \\
\lambda_8^{(s)}&=& - 10 + \sqrt{H_0 + 36}
\,.
\label{speig8}
\end{eqnarray}
}
\end{itemize}
The eigenvalues $\lambda_7^{(1)}$ and $\lambda_{16}^{(2)}, \dots,
\lambda_{21}^{(2)}$ are the unphysical longitudinal eigenvalues.
For instance  $\lambda_7^{(1)}$ is the longitudinal one-form
mode that comes from the zero-form and 
$\lambda_{16}^{(2)}, \dots, \lambda_{21}^{(2)}$ are the
longitudinal two-form modes that come from the one-form.
For an explanation, see \cite{castdauriafre, noi321, m111}.
\section{Shortening: the mechanism}
The mass matrices that we have calculated and presented in
appendix \ref{massmatrix}
are the mass matrices for the long ${\cal N}=3$ multiplets.
In this section we clarify the mechanism by which for certain
representations of $G$, some of the components of the multiplet
decouple, hence multiplet shortening. We first describe the general 
mechanism and we will illustrate this then by means of some 
concrete examples: the graviton multiplet, the Killing vector multiplet 
and the Betti multiplet.
Will see that the graviton multiplet has three gravitini in the
fundamental of $SO(3)$ and this will prove that the  theory
has indeed ${\cal N}=3$. Moreover, the case of the graviton multiplet 
will illustrate how to calculate the masses for representations
with non-zero $\bf u$-spin.
\par
To understand shortening we recall that the labels $M_1,M_2,J,p,s,t,u$
of a given Young tableau are subject to the bounds (\ref{bound1}) and
(\ref{bound2}) and to the constraint (\ref{constraintp}).
The constraint ensures the right $U(1)$-weight for the $G$-irrep.
The bounds ensure the existence of a Young tableau that corresponds to them.
The components of a multiplet that sits in 
a  $G$-irrep labeled by $M_1, M_2, J$ 
are provided by the terms in the harmonic expansion with these labels.
There will be a priori a multiple of them for different values of $s,t,u$.
However, a given term in the harmonic expansion with particular labels $s,t,u$ 
will only be there if the bounds are satisfied. If not, it will
mean that its corresponding component in one of the multiplets
decouples. Thus it is useful to reformulate the bounds (\ref{bound1})
and (\ref{bound2}) in such a way as to make the above mechanism more
transparent.
\par
To this end,
remark that for a given $s,t,u$ and $\Delta$, one can express the bounds
 (\ref{bound1}) and (\ref{bound2}) as bounds on $J$
in terms of $M_1, M_2$. This is useful
to understand shortening  since what we are after is 
to find a saturation of the unitary bound to get shortening. 
We have summarized this in table \ref{t1} and in table \ref{t2},
for all the values $s,t,u$ and $\Delta$ that appear in the
decompositions (\ref{fdec}) and (\ref{decompspinor}).
\begin{table}
\begin{tabular}{|c||c|c|c|c|c|}
\hline
 & & & & & \cr
$J \geq $ & $\frac{M_2-M_1}{3}-2$  & $ \frac{M_2-M_1}{3}-1 
$ & $ \frac{M_2-M_1}{3} $ & $ \frac{M_2-M_1}{3}+1 $ & $ \frac{M_2-M_1}{3}+2 $\cr
 & & & & & \cr
\hline
\hline
 & & & & & \cr
$\frac{M_1-M_2}{3}-2 $ &  &  &  &  & $ {\bf -1}^{(0,\ft32,0)}$ \cr
 & & & & & \cr
\hline
 & & & & & \cr
 & & & & ${\bf -2}^{(0,0,0)}$ & \cr
 & & & & ${\bf -1}^{(0,\ft12,0)}$ & \cr
$\frac{M_1-M_2}{3}-1$ 
& & & & ${\bf -1}^{(\ft12,1,0)}$ & ${\bf -1}^{(0,1,\ft12)}$ \cr
 & & & & ${\bf 0}^{(0,1,0)}$ & \cr
 & & & & ${\bf 1}^{(0,\ft32,0)}$ & \cr
 & & & & & \cr
\hline
 & & & & & \cr
 & & & ${\bf -1}^{(1,\ft12,0)}$ & & \cr
 & & & ${\bf -1}^{(\ft12,0,0)}$ & ${\bf -1}^{(0,0,\ft12)}$ & \cr
 & & & ${\bf 0}^{(0,0,0)}$ & ${\bf -1}^{(\ft12,\ft12,\ft12)}$ & \cr
$\frac{M_1-M_2}{3}$ 
& & & ${\bf 0}^{(\ft12,\ft12,0)}$ & ${\bf 0}^{(0,\ft12,\ft12)}$ 
& ${\bf -1}^{(0,\ft12,1)}$ \cr
 & & & ${\bf 1}^{(0,\ft12,0)}$ & ${\bf 1}^{(0,1,\ft12)}$ & \cr
 & & & ${\bf 1}^{(\ft12, 1, 0)}$ &  & \cr
 & & & & & \cr
\hline
 & & & & & \cr
 & & ${\bf -1}^{(\ft32,0,0)}$ &  &  & \cr
 & & ${\bf 0}^{(1,0,0)}$ & ${\bf -1}^{(1,0,\ft12)}$ & ${\bf -1}^{(\ft12,0,1)}$ & \cr
$\frac{M_1-M_2}{3}+1$ & & ${\bf 1}^{(\ft12,0,0)}$ & ${\bf 0}^{(\ft12,0,\ft12)}$ 
& ${\bf 0}^{(0,0,1)}$ & ${\bf -1}^{(0,0,\ft32)}$ \cr
 & & ${\bf 1}^{(1,\ft12,0)}$ & ${\bf 1}^{(0,0,\ft12)}$ & 
 ${\bf 1}^{(0,\ft12,1)}$ & \cr
 & & ${\bf 2}^{(0,0,0)}$ & ${\bf 1}^{(\ft12,\ft12,\ft12)}$ &  & \cr
 & & & & & \cr
\hline
 & & & & & \cr
$\frac{M_1-M_2}{3}+2$ & ${\bf 1}^{(\ft32,0,0)}$ & ${\bf 1}^{(1,0,\ft12)}$ &
${\bf 1}^{(\ft12,0,1)}$ & ${\bf 1}^{(0,0,\ft32)}$ & \cr
 & & & & & \cr
\hline
\end{tabular}
\caption{
the two lower bounds as in (\ref{bound1}) and (\ref{bound2}).
The rows represent 
\newline
the lower bound in (\ref{bound1}) and the
columns represent the lower bound in 
(\ref{bound2}).}
\label{t1}
\end{table}
\begin{table}
\begin{tabular}{|c||c|c|c|c|c|}
\hline
 & & & & & \cr
$J \leq$ & $\frac{2M_1+M_2}{3} -2$ & $\frac{2M_1+M_2}{3} -1$
 & $\frac{2M_1+M_2}{3}$ 
& $\frac{2M_1+M_2}{3} +1$ & $\frac{2M_1+M_2}{3} +2$ \cr
 & & & & & \cr
\hline
\hline
 & & & & & \cr
 & & ${\bf -1}^{(\ft32,0,0)}$ & & & \cr
$\frac{2M_2+M_1}{3}-2$ & & ${\bf -1}^{(1,\ft12,0)}$ & & & \cr
 & & ${\bf -1}^{(\ft12,1,0)}$ & & & \cr
 & & ${\bf -1}^{(0,\ft32,0)}$ & & & \cr
 & & & & & \cr
\hline
 & & & & & \cr
 & ${\bf 1}^{(\ft32,0,0)}$ & & ${\bf -1}^{(1,0,\ft12)}$ & & \cr
 & ${\bf 1}^{(1,\ft12,0)}$ & ${\bf 0}^{(1,0,0)}$ & ${\bf -1}^{(\ft12,\ft12,\ft12)}$
& & \cr
$\frac{2M_2+M_1}{3}-1$ & ${\bf 1}^{(\ft12,1,0)}$ & ${\bf 0}^{(\ft12,\ft12,0)}$ 
& ${\bf -1}^{(\ft12,0,0)}$ & ${\bf -2}^{(0,0,0)}$ & \cr
 & ${\bf 1}^{(0,\ft32,0)}$ & ${\bf 0}^{(0,1,0)}$ & ${\bf -1}^{(0,1,\ft12)}$ & & \cr
 & & & ${\bf -1}^{(0,\ft12,0)}$ & & \cr
 & & & & & \cr
\hline
 & & & & & \cr
 & & ${\bf 1}^{(1,0,\ft12)}$ & & & \cr
 & & ${\bf 1}^{(\ft12,\ft12,\ft12)}$ & ${\bf 0}^{(\ft12,0,\ft12)}$ 
& ${\bf -1}^{(\ft12,0,1)}$ & \cr
$\frac{2M_2+M_1}{3}$ & & ${\bf 1}^{(\ft12,0,0)}$ & ${\bf 0}^{(0,0,0)}$ &
${\bf -1}^{(0,\ft12,1)}$ & \cr
 & & ${\bf 1}^{(0,1,\ft12)}$ & ${\bf 0}^{(0,\ft12,\ft12)}$ &
${\bf -1}^{(0,0,\ft12)}$ & \cr
 & & ${\bf 1}^{(0,\ft12,0)}$ & & & \cr
 & & & & & \cr
\hline
 & & & & & \cr
 & & & ${\bf 1}^{(\ft12,0,1)}$ & & \cr
$\frac{2M_2+M_1}{3}+1$  & & ${\bf 2}^{(0,0,0)}$ & ${\bf 1}^{(0,\ft12,1)}$ & 
  ${\bf 0}^{(0,0,1)}$ & ${\bf -1}^{(0,0,\ft32)}$ \cr
 & & & ${\bf 1}^{(0,0,\ft12)}$ & & \cr
 & & & & & \cr
\hline
 & & & & & \cr
$\frac{2M_2+M_1}{3}+2$  & & & & ${\bf 1}^{(0,0,\ft32)}$ & \cr
 & & & & & \cr
\hline
\end{tabular}
\caption{the two upper bounds as in (\ref{bound1}) and (\ref{bound2}).
The rows represent 
\newline
the upper bound in (\ref{bound1}) and the
columns represent the upper bound in (\ref{bound2}).}
\label{t2}
\end{table}
From these tables one can read off the bounds
on $J$. These bounds
only depend on the numbers $s,t,u$ and $\Delta$. 
The notation for the entries should be understood as
\begin{eqnarray}
\bf \Delta^{(s,t,u)} \,.
\end{eqnarray}
To see which of the fragments survive the bounds we now apply the
two table \ref{t1} and \ref{t2} as if they were two sieves.
We now illustrate how to use these table by some concrete examples,
\begin{enumerate}  
\item{Let us consider the $G$-irrep with}
\begin{eqnarray}
M_1=M_2=0\,, \qquad J=0\,.
\label{exampleM10M20J0}
\end{eqnarray}
Sifting the fragments with table \ref{t1} we see that none of the
fragments of the decomposition survive the bounds except for 
${\bf 1}^{(\ft12,1,0)} \,, {\bf 1}^{(0,\ft12,0)} \,, 
{\bf 0}^{(\ft12,\ft12,0)} \,, {\bf 0}^{(0,0,0)} \,, 
{\bf -1}^{(\ft12,0,0)}, {\bf -1}^{(1,\ft12,0)}$.
Then sifting the fragments with table \ref{t2}, we see the of these
fragments only 
\begin{eqnarray}
{\bf 0}^{(0,0,0)}
\end{eqnarray}
survives. This implies that in the harmonic expansion  (\ref{harmexp})
only the harmonic $D^{(0,0)(0)}$ appears.
\item{Let us consider 
\begin{eqnarray}
M_1=M_2=0\,, \qquad J=1\,.
\label{exampleM10M20J1}
\end{eqnarray}
Then from table \ref{t2} we see that only 
${\bf 1}^{(0,0,\ft32)}, {\bf 0}^{(0,0,1)}, {\bf -1}^{(0,0,\ft32)}$
satisfy the upper bounds. 
From table \ref{t1} we see that of these only 
\begin{eqnarray}
{\bf 0}^{(0,0,1)}
\label{fragsm1m20j1}
\end{eqnarray}
satisfies the lower bounds.
}
\item{Let us also consider the $G$-irrep with
\begin{eqnarray}
M_1=M_2=1\,, \qquad J=0\,.
\label{exampleM11M21J0}
\end{eqnarray}
Then we read off from table \ref{t1} and table \ref{t2} that
only
\begin{eqnarray}
{\bf 0}^{(\ft12,\ft12,0)} \,,
{\bf 1}^{(0,\ft12,0)} \,, 
{\bf -1}^{(\ft12,0,0)} \,, 
{\bf 0}^{(0,0,0)}
\label{fragsm1m21j0} 
\end{eqnarray}
satisfy the upper and the lower bounds.
}
\end{enumerate}
This illustrates how, given a $G$-irrep, the fragments
in the $G\rightarrow H$
that survive the bounds can be obtained. 
In this way, one can study shortening for any such $G$-irrep given.
Yet, this is not 
sufficient to study shortening for the complete spectrum.
To this end we need to identify the series of $G$-irrepses
that have the same content. Before doing so, let us 
restrict our attention to some of the above examples.
\par
By means of the first example 
we show that the spectrum that we have obtained now
is really the complete spectrum in spite of the fact that we have only
been considering harmonics with $s$ and $t$ indices. This we argued
was valid for certain $SU(3)\times SU(2)$ representations where
$M_1$ and $M_2$ are big enough. Here we show that the matrices
that we have calculated in the previous section can actually be used
to obtain the eigenvalues for all the other cases where $M_1$ and $M_2$
are not big enough. 
We will do this by means of a concrete and particular 
relevant example: {\it the massless graviton multiplet}. 
Moreover, the subsequent material
will also serve as a proof of the fact that the compactification on
$AdS\times N^{010}$ formulated as $SU(3)\times SU(2)/SU(2)\times U(1)$
and with the rescalings chosen as in (\ref{rescalings})
has indeed ${\cal N}=3$ supersymmetry.
\par
Clearly, the massless graviton multiplet
contains the graviton field which is to sit in the
representation $M_1=M_2=J=0$. Applying the mass formula 
\begin{eqnarray}
m_h^2 = M_{(0)^2}\,,
\end{eqnarray}
we find that it has mass zero.
The gravitini and the graviphoton correspond to case we already
mentioned, $M_1=M_2=0$ and  $J=1$,
\begin{eqnarray}
1 \otimes 
\begin{array}{l}
\begin{array}{|c|c|}
\hline
             &  
\\
\hline
\end{array}
\end{array}
\,.
\label{Jis1}
\end{eqnarray}
This is a case where $M_1$ and $M_2$ are not big enough to 
express everything in terms of $SU(3)\times SU(2)$ Young
tableaux with $\bf s$-spin and $\bf t$-spin only, 
as we have assumed in all the
previous discussion. We illustrate that the eigenvalues of the previous
section can still be used to derive the masses of the graviphoton and the
gravitino.
\par
Let us first consider the one-form. We show that it contains the
graviphoton in its harmonic expansion.
It sits in in the fundamental of $SO(3)$.
Using the information in (\ref{fragsm1m20j1}) we see
that in the reduction $G\rightarrow H$, the representations
$\bf (2,0)$ and $\bf (\bar 2, 1)$ do not appear.
Indeed, one can not put one $SU(2)$ index $i$ in
(\ref{Jis1}).
The fragment $\bf (3,1)$ appears as (\ref{fragsm1m20j1}),
namely for the fragment ${\cal R}^{(1)}_{ij}$ there is
$
1 \otimes 
\begin{array}{l}
\begin{array}{|c|c|}
\hline
          i   &  j
\\
\hline
\end{array}
\end{array}
$. 
This can be seen as the Young tableau
\begin{eqnarray}
\Gamma^{\left(0,0,1 \right)}_{ij} (0,0,1;0)
\label{Yij}
\end{eqnarray}
where $p=1$ is given by (\ref{constraintp}). Let us now formally apply the
generalized cyclic identity (\ref{cyclicid}) to this.
We write,
\begin{eqnarray}
  \Gamma^{\left(1,0\right)}_{ij} (0,0,1;1)
- 2 \, \Gamma^{\left(1/2,1/2 \right)}_{(i, j)} (0,0,1;1)
+ \Gamma^{\left(0,1 \right)}_{ij} (0,0,1;1)
\,.
\label{2cyclics}
\end{eqnarray}
Clearly none of the above terms exists as a Young tableaux,
whereas (\ref{Yij}) does. In order to obtain the mass matrices
for the one-form we should in principle derive the formulae of the type
(\ref{lmta})  and  (\ref{lA}) for the case $u \neq 0$. 
We argue now that such work can be avoided by using the information
that we have already obtained.
\par
Since all these 
formulas are linear one does not have to derive these formulae
on the Young tableau (\ref{Yij}), but one introduces the 
objects $\Gamma$ even if they do not correspond to an irreducible representation 
of $SU(3) \times SU(2)$ with the definition that they transform
as in the transformation rules (\ref{lmta}) and (\ref{lA}).
Then one uses the identity between (\ref{Yij}) and (\ref{2cyclics})
and applies $\lambda_a$ and $\lambda_A$ to (\ref{2cyclics}).
Even though the separate terms in the calculation do not represent
irreducible representations of $SU(3) \times SU(2)$ the sum of all
of them on the other hand will. Concretely, following this line
of argument, the harmonic expansion for the 1-form can be written
as,
\begin{eqnarray}
{\cal R}_{ij}^{(1)} = AW^{(0,0,1)} \cdot
\left(
- D_{ij}^{(1,0)}
+ 2 \, D_{ij}^{(1/2,1/2)}
- D_{ij}^{(0,1)}
\right)
\label{harmexp1f}
\end{eqnarray} 
where the $D_{ij}^{\left(s,t\right)}$ are defined 
similarly as the $\Gamma$-objects upon applying the cyclic identity
two times as in (\ref{2cyclics}).
\par
In order to exhibit the results for the matrices that we have already 
obtained in the previous section, we consider $M_{(0)^2 (1)}$ for
\begin{eqnarray}
M_1=M_2=0\,, \qquad J=1 \,.
\label{M1=M2=0J=1}
\end{eqnarray} 
Since the fragments ${\cal C}^{(1)}_i$ and
${\cal C}^{(1)i}$ do not get a contribution, we have to consider 
$M_{(0)^2 (1)}$ on the basis
\begin{eqnarray}
AW=
\left(
\matrix{
AW_r^{(0,1)}  \cr
AW_r^{(1/2,1/2)}  \cr
AW_r^{(1,0)}  \cr
}
\right) 
\end{eqnarray}
only.
Hence, it is a $3\times 3$ matrix
\begin{eqnarray}
M_{(0)^2(1)} =
\left(
\matrix{ 0 & -24 & 0 \cr -96 & -48 & -96 \cr 0 & -24 & 0 \cr  }
\right)
\label{M1}
\end{eqnarray}
It is useful to realize that we can make change of basis in
the harmonic expansion (\ref{harmexp}) 
$AW^\prime = \left( {P^{-1}}\right)^T \, AW$,
\begin{eqnarray}
{\cal R}_{ij}^{(1)} =
AW^T  \, P^{-1} \cdot  P \, D_{ij} \,.
\label{harmexpgravphot}
\end{eqnarray}
where $P$ is a $3 \times 3$ invertible matrix. 
Let us take 
\begin{eqnarray}
P = \left(
\matrix{ -1 & 2 & -1 \cr 1 & 0 & -1 \cr -1 & -4 & -1 \cr  }
\right)\,, 
\label{P}
\end{eqnarray}
Then the first term in the harmonic expansion (\ref{harmexpgravphot})
gets the form (\ref{harmexp1f}). The matrix (\ref{M1}) in the
basis $AW^\prime$ gets the form,
\begin{eqnarray}
M^\prime_{(0)^2 (1)} = \left( {P^{-1}}\right)^T M_{(0)^2 (1)} P^T=
\left(
\matrix{ 48 & 0 & 0 \cr 0 & 0 & 0 \cr 0 & 0 & -96 
  \cr  }
\right)
\label{PMP}
\end{eqnarray}
So we see that the eigenvalue of $M_{(0)^2 (1)}$ for component
$AW^{(0,0,1)}$ in (\ref{harmexp1f}) of the graviphoton is given by the
first entry of the above matrix, 
\begin{eqnarray}
M_{(0)^2 (1)}=48\,. 
\end{eqnarray}
Mind that the matrix $P$ diagonalizes
the matrix $M_{(0)^2 (1)}$.
\par
As is known from \cite{noi321}, 
the mass of the vector $A$ is given by the following
mass formula,
\begin{eqnarray}
m_A^2 = M_{(0)^2 (1)} + 48 - 12 \sqrt{M_{(0)^2 (1)}+16} \,.
\label{mA}
\end{eqnarray}
So we find a massless vector indeed.
\par
So far we have found only the vectors and the graviton of the
massless graviton multiplet. Still it is crucial to 
check that we have the right number of gravitini in this multiplet.
That will prove that we have indeed ${\cal N}=3$.
Hence, let us now consider the spinor. We are interested in the representation
(\ref{Jis1}). Again there is only a contribution for the fragment
$\bf (3,0)$ in the decomposition (\ref{decompspinor}) and the treatment 
is the same as for the $\bf (3,1)$ of the one-form. Indeed,
there only appears the fragment $\psi_{ij}$ in the expansion,
\begin{eqnarray}
\psi_{ij} = \chi^{(0,0,1)} (x) \cdot \left( 
             -D_{ij}^{(1,0)} +
             2 D_{ij}^{(1/2,1/2)} 
             - D_{ij}^{(0,1)}
            \right) \,.
\end{eqnarray}
Consequently, from the $8\times 8$ matrix $M_{(1/2)^3}$ there is
only the upper $3\times 3$ matrix 
\begin{eqnarray}
M_{(1/2)^3}=
\left(
\matrix{ -4 & -2 & 0 \cr -8 & -8 & -8 \cr 0 & -2 & -4 \cr  }
\right)
\end{eqnarray}
relevant for $M_1=M_2=0$ and $J=1$. We can diagonalize this matrix
similarly by means of the matrix $P$ in (\ref{P}) and we get,
\begin{eqnarray}
M^\prime_{(1/2)^3}=
\left(
\matrix{ 0 & 0 & 0 \cr 0 & -4 & 0 \cr 0 & 0 & -12 \cr  }
\right)
\,.
\end{eqnarray}
Only the first zero entry of this matrix is physically relevant. Using the
fact,
\begin{eqnarray}
m_\chi = M_{(1/2)^3} \,,
\end{eqnarray}
we conclude that there are three massless spin-$\ft32$ fields in the 
fundamental of $SO(3)$. They provide the gravitini of the 
${\cal N}=3$ massless graviton multiplet.
\par
So we have found the massless graviton, gravitini and graviphoton of
the ${\cal N}=3$ $\left(2,3(\ft32),3(1),\ft12 \right)$ graviton multiplet.
\par
To complete the treatment of (\ref{M1=M2=0J=1}), we also consider the
eigenvalues of the two-form for this representation of $G$. As in the 
case of the one-form, the only fragments that contribute are the
ones with $\bf u$-spin $1$, i.e.  for the representation  (\ref{M1=M2=0J=1})
the $21 \times 21$ matrix works on the basis,
\begin{eqnarray}
\left(
\matrix{
Z_o^{(0,1)} \cr
Z_o^{(1/2,1/2)} \cr
Z_o^{(1,0)} \cr
Z_r^{(0,1)} \cr
Z_r^{(1/2,1/2)} \cr
Z_r^{(1,0)} \cr
}
\right)
\end{eqnarray}
only. Hence, it gets the form of a $6\times 6$ matrix,
\begin{eqnarray}
M_{(0)(1)^2}=
\left(
\matrix{ 32 & -16 & 0 & -32 & 0 & 0 \cr -64 & 0 & -64 & 0 & -32 & 0 \cr 0 & -16 & 32 & 0 & 0 & -32 \cr -16 & 0 & 0 & 32 & 
    -24 & 0 \cr 0 & -16 & 0 & -96 & -16 & -96 \cr 0 & 0 & -16 & 0 & -24 & 32 \cr  }
\right)
\end{eqnarray}
Following the same line of reasoning as  for the one-form and the
spinor, we introduce the diagonalization matrix
\begin{eqnarray}
P=
\left(
\matrix{ 2 & -4 & 2 & 1 & -2 & 1 \cr -1 & 2 & -1 & 1 & -2 & 1 \cr -{\sqrt{2}} & 0 & {\sqrt{2}} & -1 & 0 & 1 \cr {
     \sqrt{2}} & 0 & -{\sqrt{2}} & -1 & 0 & 1 \cr -1 + {\sqrt{3}} & 4\,\left( -1 + {\sqrt{3}} \right)  & -1 + 
   {\sqrt{3}} & 1 & 4 & 1 \cr -1 - {\sqrt{3}} & -4\,\left( 1 + {\sqrt{3}} \right)  & -1 - {\sqrt{3}} & 1 & 4 & 1 \cr  }
\right)
\end{eqnarray}
Thus diagonalizing we find,
\begin{eqnarray}
& M^\prime_{(0)(1)^2} = \left( {P^{-1}}\right)^T M_{(0)(1)^2} P^T =&
\nonumber \\
&
\left(
\matrix{ 48 & 0 & 0 & 0 & 0 & 0 \cr 0 & 96 & 0 & 0 & 0 & 0 \cr 0 & 0 & -16\,
   \left( -2 + {\sqrt{2}} \right)  & 0 & 0 & 0 \cr 0 & 0 & 0 & 16\,
   \left( 2 + {\sqrt{2}} \right)  & 0 & 0 \cr 0 & 0 & 0 & 0 & -16\,
   \left( 3 + {\sqrt{3}} \right)  & 0 \cr 0 & 0 & 0 & 0 & 0 & 16\,\left( -3 + {\sqrt{3}} \right)  \cr  }
\right)
&
\label{diagm2f}
\end{eqnarray}
From this matrix we have to discard the eigenvalues 
\begin{eqnarray}
-16\, \left( -2 + {\sqrt{2}} \right)\,,
16\,  \left( 2 + {\sqrt{2}} \right)\,, 
-16\, \left( 3 + {\sqrt{3}} \right)\,, 
16\,  \left( -3 + {\sqrt{3}} \right)
\end{eqnarray} 
as being unphysical. Only the first two eigenvectors in $P$
correspond to the two ${\bf 0}^{(0,0,1)}$ that we 
have found in the decomposition $G \rightarrow H$ in
(\ref{fdec}).
So we see that the first eigenvector (corresponding to the 
first row of the matrix $P$) is the longitudinal unphysical
mode that corresponds to the one-form eigenvector that
we obtained above. It has the same eigenvector $M_{(0)(1)^2}=48$.
Furthermore, there is the eigenvalue $M_{(0)(1)^2}=96$ whose 
eigenvector is an extra massive vector $Z$ with mass 
\begin{eqnarray}
m_Z^2=M_{(0)(1)^2}=96\,.
\end{eqnarray}
As will be explained in \cite{n010} this vector  is part
of a massive gravitino multiplet.
\par 
To come back to our example $M_1=M_2=J=0$, let us also consider 
the two-form for this case. This yields the
vector field of the massless Betti multiplet. Indeed,
Then the operator $M_{(0)(1)^2}$ acts on the
component $Z_r^{(0,0)}$ only in which case is has eigenvalue zero.
So we find  the vector of the massless Betti multiplet.
\par
At this stage we have elucidated the examples (\ref{exampleM10M20J0})
and (\ref{exampleM10M20J1}). This yielded the massless graviton multiplet
and the massless Betti multiplet. It is also instructive to consider 
example (\ref{exampleM11M21J0}) 
a bit more closely, since it is in this irreducible representation of
$G$ that we find the massless vector multiplet which contains the 
Killing vector of the $SU(3)$ isometry. 
\par
To this end we have to consider the one-form.
Acting on the fragments with (\ref{fragsm1m21j0}) it
it becomes the $3 \times 3$ matrix,
\begin{eqnarray}
\left(
\matrix{ 80 & -16\,i\,{\sqrt{2}} & -16\,i\,{\sqrt{2}} \cr 24\,i\,{\sqrt{2}} & 96 & 0 \cr 24\,i\,
   {\sqrt{2}} & 0 & 96 \cr  }
\right)
\,,
\end{eqnarray}
which has eigenvalues 
\begin{eqnarray}
M_{(0)^2 (1)}= 48, 96, 128 \,.
\end{eqnarray}
The first eigenvalue $48$ according to (\ref{mA}), 
gives rise to a massless vector $A$. This is the 
Killing vector of the $SU(3)$ isometry.   
\section{The series}
In the previous section we explained how the masses
of the ${\cal N}=3$ theory can be calculated
from the matrices of  appendix \ref{massmatrix}
when $M_1, M_2,J$ are given.
Yet for a systematic study of the complete spectrum
we need to do this for all the irreducible representations
 of $G$.
In order to do so it is useful
to arrange the representations that are allowed by the
bounds (\ref{bound1}) and (\ref{bound2}) and the constraint 
(\ref{constraintp}),
into series that give rise to the same content of fragments.
This can be done most easily by using the tables
\ref{t1} and \ref{t2}.
\par
First of all, notice that it is sufficient to consider the
cases where 
\begin{eqnarray}
M_2 \geq M_1 \,.
\end{eqnarray}
Indeed, the irreducible representations with
$M_2 < M_1$ are the conjugate representations
of $SU(3)$ and hence correspond to the complex conjugate
multiplets in the ${\cal N}=3$ theory. Reflecting on the 
two tables \ref{t1} and \ref{t2} we conclude that 
we should make a distinction among the
following series $R$ and $E_n$,
\begin{eqnarray}
  \label{eq:R}
&R&\,: \left( M_1\geq 4 \right)\, ; \qquad 
  \cases{
    J \geq \ft13\,(M_2 - M_1) + 2 \cr
    J \leq \ft13\,(2 M_1 + M_2) - 2
    }
\,, \\
  \label{eq:E1}
&E_1&\,: \left( M_1\geq 3 \right)\,; \qquad
  \cases{
    J \geq \ft13\,(M_2 - M_1) + 2 \cr
    J = \ft13\,(2 M_1 + M_2) - 1
    }
\,, \\
  \label{eq:E2}
&E_2&\,: \left( M_1\geq 2 \right)\,; \qquad
  \cases{
    J \geq \ft13\,(M_2 - M_1) + 2 \cr
    J = \ft13\,(2 M_1 + M_2)
    }
\,, \\
  \label{eq:E3}
&E_3&\,: \left( M_1\geq 1 \right)\,; \qquad
  \cases{
    J \geq \ft13\,(M_2 - M_1) + 2 \cr
    J = \ft13\,(2 M_1 + M_2) + 1
    }
\,, \\
  \label{eq:E4}
&E_4&\,: \left( M_1\geq 0 \right)\,; \qquad
  \cases{
    J \geq \ft13\,(M_2 - M_1) + 2 \cr
    J = \ft13\,(2 M_1 + M_2) + 2 
    }
\,, \\
  \label{eq:E5}
&E_5&\,: \left( M_1\geq 3 \right)\,; \qquad
  \cases{
    J = \ft13\,(M_2 - M_1) + 1 \cr
    J \leq \ft13\,(2 M_1 + M_2) - 2
    }
\,, \\
  \label{eq:E6}
&E_6&\,: \left( M_1 =  2 \right)\,; \qquad
  \cases{
    J =  \ft13\,(M_2 - M_1) + 1 \cr
    J = \ft13\,(2 M_1 + M_2) - 1
    }
\,, \\
  \label{eq:E7}
&E_7&\,: \left( M_1 =  1 \right)\,; \qquad
  \cases{
    J =  \ft13\,(M_2 - M_1) + 1 \cr
    J = \ft13\,(2 M_1 + M_2) 
    }
\,, \\
  \label{eq:E8}
&E_8&\,: \left( M_1 =  0 \right)\,; \qquad
  \cases{
    J =  \ft13\,(M_2 - M_1) + 1 \cr
    J = \ft13\,(2 M_1 + M_2) + 1 
    }
\,, \\
  \label{eq:E9}
&E_9&\,: \left( M_1 \geq  2 \right)\,; \qquad
  \cases{
    J =  \ft13\,(M_2 - M_1)  \cr
    J \leq \ft13\,(2 M_1 + M_2) - 2 
    }
\,, \\
  \label{eq:E10}
&E_{10}&\,: \left( M_1 = 1  \right)\,; \qquad
  \cases{
    J =  \ft13\,(M_2 - M_1)  \cr
    J = \ft13\,(2 M_1 + M_2) - 1
    }
\,, \\
  \label{eq:E11}
&E_{11}&\,: \left( M_1 = 0  \right)\,; \qquad
  \cases{
    J =  \ft13\,(M_2 - M_1)  \cr
    J = \ft13\,(2 M_1 + M_2)
    }
\,, \\
  \label{eq:E12}
&E_{12}&\,: \left( M_1 \geq 1  \right)\,; \qquad
  \cases{
    J =  \ft13\,(M_2 - M_1) - 1 \cr
    J \leq \ft13\,(2 M_1 + M_2) - 2
    }
\,, \\
  \label{eq:E13}
&E_{13}&\,: \left( M_1 = 0  \right)\,; \qquad
  \cases{
    J =  \ft13\,(M_2 - M_1) - 1 \cr
    J = \ft13\,(2 M_1 + M_2) - 1
    }
\,, \\
  \label{eq:E14}
&E_{14}&\,: \left( M_1 \geq 0  \right)\,; \qquad
  \cases{
    J =  \ft13\,(M_2 - M_1) - 2 \cr
    J \leq \ft13\,(2 M_1 + M_2) - 2
    }
\,.
\end{eqnarray}
The series $R$ is the regular series and contains the long multiplets.
For the calculation of the masses of the components, one takes the
matrices in the appendix \ref{massmatrix}. 
The series $E$ are the exceptional series.
One may verify that in these exceptional series 
some columns of the tables \ref{t1} and \ref{t2} 
are absent. In order to see which rows survive the bounds 
it is useful to make a distinction between the cases
\begin{eqnarray}
  \label{eq:m2ism1plus3j}
  M_2 = M_1 + 3 j \,,
\end{eqnarray}
for $j$ a non-negative integer. Hence, the series
(\ref{eq:E1}) ... (\ref{eq:E14}) have a different content
depending on the parameter $j$. We will denote them
as $E_n^j$. One may verify that it is sufficient to
make a distinction between 
\begin{eqnarray}
j=0 \,, \qquad j=1 \,, \qquad j \geq 2 \,.
\end{eqnarray}
We write $E^0_n, E^1_n$ and $E^{\geq}_n$.
\par
In order to determine the content of the series one 
can take a representative set of number $M_1, M_2, J$ 
for that series and do the following:
\begin{enumerate}
\item{Use table \ref{t1} and table \ref{t2} to determine
which of the fragments are present. Call this set $\cal F$.
}
\item{\label{eliminatesuperfluous}
Consider each of the fragments $\bf \Delta^{(s,t,u)}$
in $\cal F$ with non-zero 
$\bf u$-spin. If {\it both} the fragments 
$\bf \Delta^{(s,t+1/2,u-1/2)}$ and $\bf \Delta^{(s+1/2,t,u-1/2)}$ 
are present as well then eliminate 
$\bf \Delta^{(s,t,u)}$. This has to be done because of the
cyclic identity (\ref{cyclicid}). Otherwise we are overly expanding.
Call the remaining set of fragments ${\cal F}_c$.
}
\item{
Consider the fragments of ${\cal F}_c$
with non-zero $\bf u$-spin that are left.
They can not be eliminated. In fact they span a basis 
for the harmonics. However, in the spirit of the previous section,
they have to be expressed in
terms of the objects $\Gamma$ (that do not correspond to Young tableaux).
All of this has to be done
as explained in the example of (\ref{exampleM10M20J1}).
We recall that this is to avoid unnecessary work. 
This determines an unnormalized row vector 
of the conjugation matrix $P$.
See for example (\ref{2cyclics}) and (\ref{harmexp1f}), 
where the row vector is
\begin{eqnarray}
\left(
\matrix{-1 & 2 & -1}
\right)
\,.
\end{eqnarray} 
}
\item{
For all the remaining fragments with zero $\bf u$-spin, add the unit vector
to $P$. 
}
\item{\label{fillP}
Fill the rest of $P$ in such a way as to get an invertible matrix.
The choice of the vectors is completely free as long as the matrix $P$
becomes invertible.
}
\item{Apply $P$ as in (\ref{PMP}).  
}
\item{Delete the unphysical row and columns of the resulting matrix.
The rows and columns that have to be deleted are the ones for which
we had to insert the vectors in step \ref{fillP}.
}
\item{Calculate the eigenvalues of this matrix. 
They have to be in the lists
(\ref{zerofeig}), (\ref{onefeig1})...(\ref{onefeig7}), 
(\ref{twofeig1})...(\ref{eigenstwo}) and (\ref{speig1})...(\ref{speig8}).
Since we are doing this for specific values of $M_1, M_2, J$ we have
to plug in these values also in these lists.
}
\end{enumerate}
Remark that after step \ref{eliminatesuperfluous}, 
we have found a complete basis of fragments.
Given some values for $M_1, M_2, J$, all the above steps 
can be implemented in a Mathematica program. So it is sufficient to
choose one representative $M_1, M_2, J$ for the series in 
(\ref{eq:E1})...(\ref{eq:E14}) and
let the computer perform these steps.
\par
We now list the results of this. For the zero-form, the one-form, 
the two-form and the spinor we list which of the eigenvalues are present
and we give the masses of the fields (see \cite{castdauriafre, noi321, m111}
for conventions concerning names)
\begin{eqnarray}
h, \chi, A, W, Z, \lambda_L, \Sigma, S 
\end{eqnarray}
that can be calculated from them. 
\subsection{The zero-form}
The eigenvalue $\lambda^{(0)}$ of the zero-form (\ref{zerofeig}) is 
{\it only} present in the series 
\begin{eqnarray}
&& R\,, 
\nonumber \\
&& E_1^0\,, E_2^0\,, E_5^0\,, E_6^0\,, E_7^0\,, E_9^0\,, E_{10}^0\,, 
E_{11}^0\,,
\nonumber \\
&& E_1^1\,, E_2^1\,, E_5^1\,, E_6^1\,, E_7^1\,, E_9^1\,, E_{10}^1\,,
E_{11}^1\,,
\nonumber \\
&& E_1^{\geq}\,, E_2^{\geq}\,, E_5^{\geq}\,, E_6^{\geq}\,, 
E_7^{\geq}\,, E_9^{\geq}\,, E_{10}^{\geq}\,, E_{11}^{\geq}
\,.
\label{zerofseries}
\end{eqnarray}
From the zero-form eigenvalue $\lambda^{(0)}$
we can obtain the mass of the spin-$2$
field $h$ and the spin-$0$ fields $\Sigma$ and $S$ 
\cite{castdauriafre, noi321, m111}:
\begin{eqnarray}
m_h^2 &=& \lambda^{(0)} \,, \nonumber \\
m_\Sigma^2 &=& \lambda^{(0)} + 176 + 24 \, \sqrt{\lambda^{(0)} + 36} \,, 
\nonumber\\
m_S^2 &=& \lambda^{(0)} + 176 - 24 \, \sqrt{\lambda^{(0)} + 36} \,.
\label{hSigmsSmasses}
\end{eqnarray}
So for each $G$-irrep in the series
(\ref{zerofseries}) there is a field $h$, a fields $\Sigma$ and
a field $S$ with masses as given in (\ref{hSigmsSmasses}).
\subsection{The one-form}
We list which of the eigenvalues are present for the one-form.
We use the notation of (\ref{onefeig1})...(\ref{onefeig7}). In the 
following tables we suppress the superscript $(1)$.
\par
\noindent
For the series $R$ all seven eigenvalues are present:
\begin{eqnarray}
\begin{array}{|c||c|}
\hline
R
&
{{\lambda }_1},{{\lambda }_2},{{\lambda }_3},{{\lambda }_4},{{\lambda }_5},{{
      \lambda }_6},{{\lambda }_7}
\cr
\hline
\end{array}
\end{eqnarray}
For the exceptional series with $j=0$ we have,
\begin{eqnarray}
\begin{array}{|c||c|}
\hline
E_1^0
&
{{\lambda }_1},{{\lambda }_2},{{\lambda }_3},{{\lambda }_4},{{\lambda }_5},{{
      \lambda }_6},{{\lambda }_7}
\cr
\hline
E_2^0
&
{{\lambda }_4},{{\lambda }_5},{{\lambda }_6},{{\lambda }_7}
\cr
\hline
E_3^0
&
{{\lambda }_6}
\cr
\hline
E_4^0
&
none
\cr
\hline
E_5^0
&
{{\lambda }_1},{{\lambda }_2},{{\lambda }_3},{{\lambda }_4},{{\lambda }_5},{{
      \lambda }_6},{{\lambda }_7}
\cr
\hline
E_6^0
&
{{\lambda }_1},{{\lambda }_2},{{\lambda }_3},{{\lambda }_4},{{\lambda }_5},{{
      \lambda }_6},{{\lambda }_7}
\cr
\hline
E_7^0
&
{{\lambda }_4},{{\lambda }_5},{{\lambda }_6},{{\lambda }_7}
\cr
\hline
E_8^0
&
{{\lambda }_6}
\cr
\hline
E_9^0
&
{{\lambda }_1},{{\lambda }_2},{{\lambda }_7}
\cr
\hline
E_{10}^2
&
{{\lambda }_1},{{\lambda }_2},{{\lambda }_7}
\cr
\hline
E_{11}^0
&
none
\cr
\hline
E_{12}^0
&
empty
\cr
\hline
E_{13}^0
&
empty
\cr
\hline
E_{14}^0
&
empty
\cr
\hline
\end{array}
\end{eqnarray}
Mind that there are no values for $M_1, M_2, J$ that match the series
$E_{12}^0, E_{13}^0, E_{14}^0$. Indeed, if $j=0$ then (\ref{eq:E12}),
(\ref{eq:E13}) and (\ref{eq:E14}) would imply that $J=-1$ or $J=-2$. 
We have indicated this in the table with $empty$. For the series $E_4^0$ 
there exist values of $M_1, M_2, J$. However, no eigenvalues $\lambda^{(1)}$
are present. We have indicated that with $none$.
\par
\noindent
For the exceptional series with $j=1$:
\begin{eqnarray}
\begin{array}{|c||c|}
\hline
E_1^1
&
{{\lambda }_1},{{\lambda }_2},{{\lambda }_3},{{\lambda }_4},{{\lambda }_5},{{
      \lambda }_6},{{\lambda }_7}
\cr
\hline
E_2^1
&
{{\lambda }_3},{{\lambda }_4},{{\lambda }_5},{{\lambda }_6},{{\lambda }_7}
\cr
\hline
E_3^1
&
{{\lambda }_5},{{\lambda }_6}
\cr
\hline
E_4^1
&
none
\cr
\hline
E_5^1
&
{{\lambda }_1},{{\lambda }_2},{{\lambda }_3},{{\lambda }_4},{{\lambda }_5},{{
      \lambda }_6},{{\lambda }_7}
\cr
\hline
E_6^1
&
{{\lambda }_1},{{\lambda }_2},{{\lambda }_3},{{\lambda }_4},{{\lambda }_5},{{
      \lambda }_6},{{\lambda }_7}
\cr
\hline
E_7^1
&
{{\lambda }_3},{{\lambda }_4},{{\lambda }_5},{{\lambda }_6},{{\lambda }_7}
\cr
\hline
E_8^1
&
{{\lambda }_5},{{\lambda }_6}
\cr
\hline
E_9^1
&
{{\lambda }_1},{{\lambda }_2},{{\lambda }_3},{{\lambda }_4},{{\lambda }_7}
\cr
\hline
E_{10}^1
&
{{\lambda }_1},{{\lambda }_2},{{\lambda }_3},{{\lambda }_4},{{\lambda }_7}
\cr
\hline
E_{11}^1
&
{{\lambda }_3},{{\lambda }_4},{{\lambda }_7}
\cr
\hline
E_{12}^1
&
{{\lambda }_1},{{\lambda }_2}
\cr
\hline
E_{13}^1
&
{{\lambda }_1},{{\lambda }_2}
\cr
\hline
E_{14}^1
&
empty
\cr
\hline
\end{array}
\end{eqnarray}
\par
\noindent
For the exceptional series with $j\geq 2$:
\begin{eqnarray}
\begin{array}{|c||c|}
\hline
E_1^{\geq}
&
{{\lambda }_1},{{\lambda }_2},{{\lambda }_3},{{\lambda }_4},{{\lambda }_5},{{
      \lambda }_6},{{\lambda }_7}
\cr
\hline
E_2^{\geq}
&
{{\lambda }_3},{{\lambda }_4},{{\lambda }_5},{{\lambda }_6},{{\lambda }_7}
\cr
\hline
E_3^{\geq}
&
{{\lambda }_5},{{\lambda }_6}
\cr
\hline
E_4^{\geq}
&
none
\cr
\hline
E_5^{\geq}
&
{{\lambda }_1},{{\lambda }_2},{{\lambda }_3},{{\lambda }_4},{{\lambda }_5},{{
      \lambda }_6},{{\lambda }_7}
\cr
\hline
E_6^{\geq}
&
{{\lambda }_1},{{\lambda }_2},{{\lambda }_3},{{\lambda }_4},{{\lambda }_5},{{
      \lambda }_6},{{\lambda }_7}
\cr
\hline
E_7^{\geq}
&
{{\lambda }_3},{{\lambda }_4},{{\lambda }_5},{{\lambda }_6},{{\lambda }_7}
\cr
\hline
E_8^{\geq}
&
 {{\lambda }_5},{{\lambda }_6}
\cr
\hline
E_9^{\geq}
&
{{\lambda }_1},{{\lambda }_2},{{\lambda }_3},{{\lambda }_4},{{\lambda }_7}
\cr
\hline
E_{10}^{\geq}
&
{{\lambda }_1},{{\lambda }_2},{{\lambda }_3},{{\lambda }_4},{{\lambda }_7}
\cr
\hline
E_{11}^{\geq}
&
{{\lambda }_3},{{\lambda }_4},{{\lambda }_7}
\cr
\hline
E_{12}^{\geq}
&
{{\lambda }_1},{{\lambda }_2}
\cr
\hline
E_{13}^{\geq}
&
{{\lambda }_1},{{\lambda }_2}
\cr
\hline
E_{14}^{\geq}
&
none
\cr
\hline
\end{array}
\end{eqnarray}
From the one-form eigenvalue $\lambda^{(1)}$ we can obtain
the masses of the vectors $A$ and $W$,
\begin{eqnarray}
m_A^2 &=& \lambda^{(1)} + 48 - 12 \, \sqrt{\lambda^{(1)} + 16}
\,, \nonumber \\
m_W^2 &=& \lambda^{(1)} + 48 + 12 \, \sqrt{\lambda^{(1)} + 16}
\,.
\label{massesAW}
\end{eqnarray}
So for each entry $\lambda$ in the above tables there is both a vector
$A$ and a vector $W$ whose mass can be obtained from $\lambda$
by the formulas (\ref{massesAW}). 
\subsection{The two-form} 
We list which of the eigenvalues are present for the two-form.
We use the notation of (\ref{twofeig1})...(\ref{eigenstwo}). 
Mind that there are the multiplicities of the eigenvalues
\begin{eqnarray}
&& \lambda_2^{(2)} = \lambda_3^{(2)} 
\nonumber \\
&& \lambda_4^{(2)} = \lambda_5^{(2)} = \lambda_6^{(2)}
\nonumber \\
&& \lambda_7^{(2)} = \lambda_8^{(2)}
\label{multiplicities}
\end{eqnarray}
In the following tables we will suppress the superscript
$(2)$. 
\par
\noindent
The series $R$ contains all eigenvalues:
\begin{eqnarray}
\begin{array}{|c||c|}
\hline
R
&
 {{\lambda }_1},{{\lambda }_2},{{\lambda }_3},{{\lambda }_4},{{
            \lambda }_5},{{\lambda }_6},{{\lambda }_7},{{\lambda }_8},{{
            \lambda }_9},{{\lambda }_{10}},{{\lambda }_{11}},{{\lambda }_{
            12}},{{\lambda }_{13}},{{\lambda }_{14}},{{\lambda }_{15}},{{
            \lambda }_{16}},{{\lambda }_{17}},{{\lambda }_{18}},{{\lambda }_{
            19}},{{\lambda }_{20}},{{\lambda }_{21}}
\cr
\hline
\end{array}
\nonumber \\
\end{eqnarray}
\par
\noindent
The exceptional series $E$ with $j=0$ contain:
\begin{eqnarray}
\begin{array}{|c||c|}
\hline
E_1^0 
&
{{\lambda }_1},{{\lambda }_2},{{\lambda }_3},{{\lambda }_4},{{\lambda }_5},{{
      \lambda }_6},{{\lambda }_7},{{\lambda }_8},{{\lambda }_{10}},{{
      \lambda }_{11}},{{\lambda }_{12}},{{\lambda }_{13}},{{\lambda }_{14}},{{
      \lambda }_{16}},{{\lambda }_{17}},{{\lambda }_{18}},{{\lambda }_{19}},{{
      \lambda }_{20}},{{\lambda }_{21}}
\cr
\hline
E_2^0
&
{{\lambda }_1},{{\lambda }_2},{{\lambda }_3},{{\lambda }_4},
       {{\lambda }_{5,14}},{{
      \lambda }_{10}},{{\lambda }_{12}},{{\lambda }_{13}},{{\lambda }_{19}},{{
      \lambda }_{20}},{{\lambda }_{21}}
\cr
\hline
E_3^0
&
{{\lambda }_1},{{\lambda }_{2,10}},{{\lambda }_{12}},{{\lambda }_{21}}
\cr
\hline
E_4^0
&
none
\cr
\hline
E_5^0
&
{{\lambda }_{1,2}},
{{\lambda }_4},{{\lambda }_5},{{\lambda }_6},{{\lambda }_7},{{
      \lambda }_8},{{\lambda }_9},{{\lambda }_{10}},{{\lambda }_{11}},{{
      \lambda }_{12}},{{\lambda }_{13}},{{\lambda }_{14}},{{\lambda }_{15}},{{
      \lambda }_{16}},{{\lambda }_{17}},{{\lambda }_{18}},{{\lambda }_{19}},{{
      \lambda }_{20}},{{\lambda }_{21}}
\cr
\hline
E_6^0
&
 {{\lambda }_{1,2}},
{{\lambda }_4},{{\lambda }_5},{{\lambda }_6},{{\lambda }_7},{{
      \lambda }_8},{{\lambda }_{10}},{{\lambda }_{11}},{{\lambda }_{12}},{{
      \lambda }_{13}},{{\lambda }_{14}},{{\lambda }_{16}},{{\lambda }_{17}},{{
      \lambda }_{18}},{{\lambda }_{19}},{{\lambda }_{20}},{{\lambda }_{21}}
\cr
\hline
E_7^0
&
{{\lambda }_{1,2,19}},{{\lambda }_{1,3,19}},
{{\lambda }_4},{{\lambda }_{5,14}},{{\lambda }_{
      10}},{{\lambda }_{12}},{{\lambda }_{13}},{{\lambda }_{20}},{{\lambda }_{
      21}}
\cr
\hline
E_8^0
&
{{\lambda }_{12}},{{\lambda }_{21}}
\cr
\hline
E_9^0
&
{{\lambda }_{1,7}},{{\lambda }_{2,4}},
{{\lambda }_9},{{\lambda }_{14}},{{\lambda }_{
      15}},{{\lambda }_{16}},{{\lambda }_{17}} 
\cr
\hline
E_{10}^0
&
{{\lambda }_{1,7}},{{\lambda }_{2,4}},{{\lambda }_{3,5}},{{\lambda }_{14}},{{\lambda }_{
      16}}
\cr
\hline
E_{11}^0
&
 {{\lambda }_{1,7,11,13,15,17,18,20}}
\cr
\hline
E_{12}^0
&
empty
\cr
\hline
E_{13}^0
&
empty
\cr
\hline
E_{14}^0
&
empty
\cr
\hline
\end{array}
\nonumber \\
\end{eqnarray}
\par
\noindent
The exceptional series $E$ with $j=1$ contain:
\begin{eqnarray}
\begin{array}{|c||c|}
\hline
E_1^1
&
{{\lambda }_1},{{\lambda }_2},{{\lambda }_3},{{\lambda }_4},{{\lambda }_5},{{
      \lambda }_6},{{\lambda }_7},{{\lambda }_8},{{\lambda }_{10}},{{
      \lambda }_{11}},{{\lambda }_{12}},{{\lambda }_{13}},{{\lambda }_{14}},{{
      \lambda }_{15}},{{\lambda }_{16}},{{\lambda }_{17}},{{\lambda }_{18}},{{
      \lambda }_{19}},{{\lambda }_{20}},{{\lambda }_{21}}
\cr
\hline
E_2^1
&
{{\lambda }_1},{{\lambda }_2},{{\lambda }_3},{{\lambda }_4},{{\lambda }_5},{{
      \lambda }_6},{{\lambda }_{10}},{{\lambda }_{11}},{{\lambda }_{12}},{{
      \lambda }_{13}},{{\lambda }_{18}},{{\lambda }_{19}},{{\lambda }_{20}},{{
      \lambda }_{21}}
\cr
\hline
E_3^1
&
{{\lambda }_1},{{\lambda }_2},{{\lambda }_3},{{\lambda }_{12}},{{\lambda }_{
      20}},{{\lambda }_{21}}
\cr
\hline
E_4^1
&
{{\lambda }_1}
\cr
\hline
E_5^1
&
{{\lambda }_2},{{\lambda }_3},{{\lambda }_4},{{\lambda }_5},{{\lambda }_6},{{
      \lambda }_7},{{\lambda }_8},{{\lambda }_9},{{\lambda }_{10}},{{
      \lambda }_{11}},{{\lambda }_{12}},{{\lambda }_{13}},{{\lambda }_{14}},{{
      \lambda }_{15}},{{\lambda }_{16}},{{\lambda }_{17}},{{\lambda }_{18}},{{
      \lambda }_{19}},{{\lambda }_{20}},{{\lambda }_{21}}
\cr
\hline
E_6^1
&
{{\lambda }_{1,14}},
{{\lambda }_2},{{\lambda }_3},{{\lambda }_4},{{\lambda }_5},{{
      \lambda }_6},{{\lambda }_7},{{\lambda }_8},{{\lambda }_{10}},{{
      \lambda }_{11}},{{\lambda }_{12}},{{\lambda }_{13}},{{\lambda }_{15}},{{
      \lambda }_{16}},{{\lambda }_{17}},{{\lambda }_{18}},{{\lambda }_{19}},{{
      \lambda }_{20}},{{\lambda }_{21}}
\cr
\hline
E_7^1
&
{{\lambda }_2},{{\lambda }_3},{{\lambda }_4},{{\lambda }_5},{{\lambda }_6},{{
      \lambda }_{10}},{{\lambda }_{11}},{{\lambda }_{12}},{{\lambda }_{13}},{{
      \lambda }_{18}},{{\lambda }_{19}},{{\lambda }_{20}},{{\lambda }_{21}}
\cr
\hline
E_8^1
&
{{\lambda }_2},{{\lambda }_3},{{\lambda }_{4,20}},
{{\lambda }_{12}},{{\lambda }_{
      21}}
\cr
\hline
E_9^1
&
{{\lambda }_4},{{\lambda }_5},{{\lambda }_6},{{\lambda }_7},{{\lambda }_8},{{
      \lambda }_9},{{\lambda }_{10}},{{\lambda }_{11}},{{\lambda }_{14}},{{
      \lambda }_{15}},{{\lambda }_{16}},{{\lambda }_{17}},{{\lambda }_{18}},{{
      \lambda }_{19}}
\cr
\hline
E_{10}^1
&
{{\lambda }_4},{{\lambda }_5},{{\lambda }_6},{{\lambda }_7},{{\lambda }_8},{{
      \lambda }_{10}},{{\lambda }_{11}},{{\lambda }_{14}},{{\lambda }_{15}},{{
      \lambda }_{16}},{{\lambda }_{17}},{{\lambda }_{18}},{{\lambda }_{19}}
\cr
\hline
E_{11}^1
&
{{\lambda }_4},{{\lambda }_5},{{\lambda }_6},{{\lambda }_{7,18}},{{\lambda }_{
      10}},{{\lambda }_{11}},{{\lambda }_{19}}
\cr
\hline
E_{12}^1
&
{{\lambda }_{1,7}},{{\lambda }_9},{{\lambda }_{14}},{{\lambda }_{15}},{{
      \lambda }_{16}},{{\lambda }_{17}}
\cr
\hline
E_{13}^1
&
{{\lambda }_{1,7}},{{\lambda }_{9,16}},{{\lambda }_{14}},{{\lambda }_{15}},{{
      \lambda }_{17}}
\cr
\hline
E_{14}^1
&
empty
\cr
\hline
\end{array}
\nonumber \\
\end{eqnarray}
\par
\noindent
The exceptional series $E$ with $j\geq 2$ contain:
\begin{eqnarray}
\begin{array}{|c||c|}
\hline
E_1^{\geq}
&
{{\lambda }_1},{{\lambda }_2},{{\lambda }_3},{{\lambda }_4},{{\lambda }_5},{{
      \lambda }_6},{{\lambda }_7},{{\lambda }_8},{{\lambda }_{10}},{{
      \lambda }_{11}},{{\lambda }_{12}},{{\lambda }_{13}},{{\lambda }_{14}},{{
      \lambda }_{15}},{{\lambda }_{16}},{{\lambda }_{17}},{{\lambda }_{18}},{{
      \lambda }_{19}},{{\lambda }_{20}},{{\lambda }_{21}}
\cr
\hline
E_2^{\geq}
&
{{\lambda }_1},{{\lambda }_2},{{\lambda }_3},{{\lambda }_4},{{\lambda }_5},{{
      \lambda }_6},{{\lambda }_{10}},{{\lambda }_{11}},{{\lambda }_{12}},{{
      \lambda }_{13}},{{\lambda }_{18}},{{\lambda }_{19}},{{\lambda }_{20}},{{
      \lambda }_{21}}
\cr
\hline
E_3^{\geq}
&
{{\lambda }_1},{{\lambda }_2},{{\lambda }_3},{{\lambda }_{12}},{{\lambda }_{
      13}},{{\lambda }_{20}},{{\lambda }_{21}}
\cr
\hline
E_4^{\geq}
&
 {{\lambda }_1}
\cr
\hline
E_5^{\geq}
&
{{\lambda }_2},{{\lambda }_3},{{\lambda }_4},{{\lambda }_5},{{\lambda }_6},{{
      \lambda }_7},{{\lambda }_8},{{\lambda }_9},{{\lambda }_{10}},{{
      \lambda }_{11}},{{\lambda }_{12}},{{\lambda }_{13}},{{\lambda }_{14}},{{
      \lambda }_{15}},{{\lambda }_{16}},{{\lambda }_{17}},{{\lambda }_{18}},{{
      \lambda }_{19}},{{\lambda }_{20}},{{\lambda }_{21}}
\cr
\hline
E_6^{\geq}
&
{{\lambda }_2},{{\lambda }_3},{{\lambda }_4},{{\lambda }_5},{{\lambda }_6},{{
      \lambda }_7},{{\lambda }_8},{{\lambda }_{10}},{{\lambda }_{11}},{{
      \lambda }_{12}},{{\lambda }_{13}},{{\lambda }_{14}},{{\lambda }_{15}},{{
      \lambda }_{16}},{{\lambda }_{17}},{{\lambda }_{18}},{{\lambda }_{19}},{{
      \lambda }_{20}},{{\lambda }_{21}}
\cr
\hline
E_7^{\geq}
&
{{\lambda }_2},{{\lambda }_3},{{\lambda }_4},{{\lambda }_5},{{\lambda }_6},{{
      \lambda }_{10}},{{\lambda }_{11}},{{\lambda }_{12}},{{\lambda }_{13}},{{
      \lambda }_{18}},{{\lambda }_{19}},{{\lambda }_{20}},{{\lambda }_{21}}
\cr
\hline
E_8^{\geq}
&
{{\lambda }_2},{{\lambda }_3},{{\lambda }_{12}},{{\lambda }_{13}},{{
      \lambda }_{20}},{{\lambda }_{21}}
\cr
\hline
E_9^{\geq}
&
{{\lambda }_4},{{\lambda }_5},{{\lambda }_6},{{\lambda }_7},{{\lambda }_8},{{
      \lambda }_9},{{\lambda }_{10}},{{\lambda }_{11}},{{\lambda }_{14}},{{
      \lambda }_{15}},{{\lambda }_{16}},{{\lambda }_{17}},{{\lambda }_{18}},{{
      \lambda }_{19}}
\cr
\hline
E_{10}^{\geq}
&
{{\lambda }_4},{{\lambda }_5},{{\lambda }_6},{{\lambda }_7},{{\lambda }_8},{{
      \lambda }_{10}},{{\lambda }_{11}},{{\lambda }_{14}},{{\lambda }_{15}},{{
      \lambda }_{16}},{{\lambda }_{17}},{{\lambda }_{18}},{{\lambda }_{19}}
\cr
\hline
E_{11}^{\geq}
&
{{\lambda }_4},{{\lambda }_5},{{\lambda }_6},{{\lambda }_{10}},{{\lambda }_{
      11}},{{\lambda }_{18}},{{\lambda }_{19}}
\cr
\hline
E_{12}^{\geq}
&
{{\lambda }_7},{{\lambda }_8},{{\lambda }_9},{{\lambda }_{14}},{{\lambda }_{
      15}},{{\lambda }_{16}},{{\lambda }_{17}}
\cr
\hline
E_{13}^{\geq}
&
{{\lambda }_7},{{\lambda }_8},{{\lambda }_{14}},{{\lambda }_{15}},{{
      \lambda }_{16}},{{\lambda }_{17}}
\cr
\hline
E_{14}^{\geq}
&
{{\lambda }_9}
\cr
\hline
\end{array}
\nonumber \\
\end{eqnarray}
Mind that besides the multiplicities in (\ref{multiplicities}),
there may be occasional multiplicities that arise for certain
values of $M_1, M_2, J$. If that occurs,
then we have indicated 
it in the above tables by writing the eigenvalues $\lambda$
with multiple indices. For instance, $\lambda_{1,2}$ in $E_5^0$ 
means that for the values $M_1=M_2\geq 3, J=1$ we have
$\lambda_1^{(2)} = \lambda_2^{(2)} = \lambda_3^{(2)}$
in (\ref{eigenstwo}).
\par
From this one can obtain the masses of the vector field $Z$,
\begin{eqnarray}
m_Z^2 = \lambda^{(2)} \,.
\label{massZ}
\end{eqnarray}
So for each entry in the above tables there is a vector $Z$
with mass (\ref{massZ}).
\subsection{The spinor}
We list which of the eigenvalues are present for the spinor.
We use the notation of  (\ref{speig1})...(\ref{speig8}).
In the following tables we will suppress the superscript 
$(s)$.
\par
\noindent
For the series $R$ we find all of the eight eigenvalues: 
\begin{eqnarray}
\begin{array}{|c||c|}
\hline
R 
&
{{\lambda }_1},{{\lambda }_2},{{\lambda }_3},{{\lambda }_4},{{\lambda }_5},{{
      \lambda }_6},{{\lambda }_7},{{\lambda }_8}
\cr
\hline
\end{array}
\end{eqnarray}
\par
\noindent
For the exceptional series with $j=0$:
\begin{eqnarray}
\begin{array}{|c||c|}
\hline
E_1^0 
&
{{\lambda }_1},{{\lambda }_2},{{\lambda }_3},{{\lambda }_4},{{\lambda }_5},{{
      \lambda }_6},{{\lambda }_7},{{\lambda }_8}
\cr
\hline
E_2^0
&
 {{\lambda }_3},{{\lambda }_4},{{\lambda }_5},{{\lambda }_6},{{\lambda }_7}
\cr
\hline
E_3^0
&
{{\lambda }_6}
\cr
\hline
E_4^0
&
none
\cr
\hline
E_5^0
&
{{\lambda }_1},{{\lambda }_2},{{\lambda }_3},{{\lambda }_4},{{\lambda }_5},{{
      \lambda }_6},{{\lambda }_7},{{\lambda }_8}
\cr
\hline
E_6^0
&
{{\lambda }_1},{{\lambda }_2},{{\lambda }_3},{{\lambda }_4},{{\lambda }_5},{{
      \lambda }_6},{{\lambda }_7},{{\lambda }_8}
\cr
\hline
E_7^0
&
{{\lambda }_3},{{\lambda }_4},{{\lambda }_5},{{\lambda }_6},{{\lambda }_7}
\cr
\hline
E_8^0
&
{{\lambda }_6}
\cr
\hline
E_9^0
&
 {{\lambda }_1},{{\lambda }_2},{{\lambda }_7},{{\lambda }_8}
\cr
\hline
E_{10}^0
&
{{\lambda }_1},{{\lambda }_2},{{\lambda }_7},{{\lambda }_8}
\cr
\hline
E_{11}^0
&
{{\lambda }_7}
\cr
\hline
E_{12}^0
&
empty
\cr
\hline
E_{13}^0
&
empty
\cr
\hline
E_{14}^0
&
empty
\cr
\hline
\end{array}
\end{eqnarray}
\par
\noindent
For the exceptional series with $j=1$:
\begin{eqnarray}
\begin{array}{|c||c|}
\hline
E_1^1
&
{{\lambda }_1},{{\lambda }_2},{{\lambda }_3},{{\lambda }_4},{{\lambda }_5},{{
      \lambda }_6},{{\lambda }_7},{{\lambda }_8}
\cr
\hline
E_2^1
&
{{\lambda }_3},{{\lambda }_4},{{\lambda }_5},{{\lambda }_6},{{\lambda }_7},{{
      \lambda }_8}
\cr
\hline
E_3^1
&
{{\lambda }_5},{{\lambda }_6}
\cr
\hline
E_4^1
&
none
\cr
\hline
E_5^1
&
{{\lambda }_1},{{\lambda }_2},{{\lambda }_3},{{\lambda }_4},{{\lambda }_5},{{
      \lambda }_6},{{\lambda }_7},{{\lambda }_8}
\cr
\hline
E_6^1
&
{{\lambda }_1},{{\lambda }_2},{{\lambda }_3},{{\lambda }_4},{{\lambda }_5},{{
      \lambda }_6},{{\lambda }_7},{{\lambda }_8}
\cr
\hline
E_7^1
&
{{\lambda }_3},{{\lambda }_4},{{\lambda }_5},{{\lambda }_6},{{\lambda }_7},{{
      \lambda }_8}
\cr
\hline
E_8^1
&
{{\lambda }_5},{{\lambda }_6}
\cr
\hline
E_9^1
&
{{\lambda }_1},{{\lambda }_2},{{\lambda }_3},{{\lambda }_4},{{\lambda }_7},{{
      \lambda }_8}
\cr
\hline
E_{10}^1
&
{{\lambda }_1},{{\lambda }_2},{{\lambda }_3},{{\lambda }_4},{{\lambda }_7},{{
      \lambda }_8}
\cr
\hline
E_{11}^1
&
{{\lambda }_3},{{\lambda }_4},{{\lambda }_7},{{\lambda }_8}
\cr
\hline
E_{12}^1
&
{{\lambda }_1},{{\lambda }_2}
\cr
\hline
E_{13}^1
&
{{\lambda }_1},{{\lambda }_2}
\cr
\hline
E_{14}^1
&
empty
\cr
\hline
\end{array}
\end{eqnarray}
\par
\noindent
For the exceptional series with $j\geq 2$:
\begin{eqnarray}
\begin{array}{|c||c|}
\hline
E_1^{\geq}
&
{{\lambda }_1},{{\lambda }_2},{{\lambda }_3},{{\lambda }_4},{{\lambda }_5},{{
      \lambda }_6},{{\lambda }_7},{{\lambda }_8}
\cr
\hline
E_2^{\geq}
&
{{\lambda }_3},{{\lambda }_4},{{\lambda }_5},{{\lambda }_6},{{\lambda }_7},{{
      \lambda }_8}
\cr
\hline
E_3^{\geq}
&
{{\lambda }_5},{{\lambda }_6}
\cr
\hline
E_4^{\geq}
&
none
\cr
\hline
E_5^{\geq}
&
{{\lambda }_1},{{\lambda }_2},{{\lambda }_3},{{\lambda }_4},{{\lambda }_5},{{
      \lambda }_6},{{\lambda }_7},{{\lambda }_8}
\cr
\hline
E_6^{\geq}
&
{{\lambda }_1},{{\lambda }_2},{{\lambda }_3},{{\lambda }_4},{{\lambda }_5},{{
      \lambda }_6},{{\lambda }_7},{{\lambda }_8}
\cr
\hline
E_7^{\geq}
&
{{\lambda }_3},{{\lambda }_4},{{\lambda }_5},{{\lambda }_6},{{\lambda }_7},{{
      \lambda }_8}
\cr
\hline
E_8^{\geq}
&
{{\lambda }_5},{{\lambda }_6}
\cr
\hline
E_9^{\geq}
&
{{\lambda }_1},{{\lambda }_2},{{\lambda }_3},{{\lambda }_4},{{\lambda }_7},{{
      \lambda }_8}
\cr
\hline
E_{10}^{\geq}
&
{{\lambda }_1},{{\lambda }_2},{{\lambda }_3},{{\lambda }_4},{{\lambda }_7},{{
      \lambda }_8}
\cr
\hline
E_{11}^{\geq}
&
{{\lambda }_3},{{\lambda }_4},{{\lambda }_7},{{\lambda }_8}
\cr
\hline
E_{12}^{\geq}
&
{{\lambda }_1},{{\lambda }_2}
\cr
\hline
E_{13}^{\geq}
&
{{\lambda }_1},{{\lambda }_2}
\cr
\hline
E_{14}^{\geq}
&
none
\cr
\hline
\end{array}
\end{eqnarray}
From the this one can obtain the masses of the spin-$\ft32$ field 
and the spin-$\ft12$ field $\lambda_L$,
\begin{eqnarray}
m_\chi &=& \lambda^{(s)}
\,, \nonumber \\
m_{\lambda_L} &=& -\left( \lambda^{(s)} + 16 \right)
\,.
\label{masschilL}
\end{eqnarray}
So for each entry in the above tables there is the field
$\chi$ and the field $\lambda_L$
with masses (\ref{masschilL}).
\section{Conclusions and outlook}
To conclude this paper, we have calculated the 
operators $M_{(0)^3}$, $M_{(0)^2 (1)}$, $M_{(1/2)^2}$ and $M_{(0)(1)^2}$.
For the long multiplets, they are listed in appendix \ref{massmatrix}.
We have found the massless
graviton multiplet with massless gravitini in the
fundamental of $SO(3)$, hence we have found the ${\cal N}=3$
graviton multiplet. This proves that
the result is indeed ${\cal N}=3$. We have also 
found the Betti multiplet and the massless vector 
multiplet whose vectors gauge the $SU(3)$ isometry. 
Moreover, this serves as a check on our results.
We explained that the operators
$M_{(0)^3}$, $M_{(0)^2 (1)}$, $M_{(1/2)^2}$ and $M_{(0)(1)^2}$
that are listed in the appendix contain the complete information. 
We explained the mechanism for shortening and
we showed how the masses of the representations
that contain $\bf u$-spin can be obtained.
We identified the series of $G$-irrepses with a common 
field content. We have formulated the procedure to calculate
the masses for all the shortened series. Finally we listed
the results of this.
\par
Hence we have done everything that can be done concerning the harmonic
analysis. The next step is to draw group theory into the analysis
to uncover the structure of the ${\cal N}=3$ multiplets. 
This will yield a small zoo of new ${\cal N}=3$ multiplets
beside Freedman and Nicolai's vector multiplet in table
\ref{vectormultiplet}. Then it will be easy to recognize these 
multiplets as superfields on the ${\cal N}=2$ superspace of
\cite{susp}. The invariant constraints that characterize these 
multiplets can then be read off directly.
All of this will be done in a forthcoming publication
\cite{n010}.
\par
The subsequent step is to to check the anti-de Sitter correspondence 
\begin{eqnarray}
AdS_4 \times N^{010}/CFT_3
\end{eqnarray}
\vskip 1 cm
\par
{\bf Acknowledgements}
\par
This work fits in a much wider project at the 
university of Torino which is mentored by Pietro Fr\'e.
I am indebted to him for providing this background.
I have received the essential knowledge about the geometry of 
the spaces $N^{010}$ from Leonardo Castellani. 
I also benefited from useful 
discussions with Leonardo Gualtieri and Davide Fabbri at an 
early stage of this work. 
I cordially thank all of them. 
I also thank Lorenzo Magnea for letting me terrorize
his computer.
I thank Leonardo Gualtieri for carefully reading through the
manuscript.
\newpage
\appendix
\section{The mass matrices}
\label{massmatrix}
In this appendix we list the operators 
$M_{(0)^3}$, $M_{(0)^2 (1)}$, $M_{(1/2)^2}$ and $M_{(0)(1)^2}$
for representations with $\bf s$-spin and $\bf t$-spin only,
as output of the Mathematica programs.
\par
The eigenvalue of the operator $M_{(0)^3}$ is,
\begin{eqnarray}
M_{(0)^3} = H_0 =
\frac{16}{3}\left( + 2 (M_1^2 + M_2^2 + M_1 M_2 + 3 M_1  + 3 M_2 )
                   -3 J (J+1)
 \right)
\end{eqnarray}
\par
The matrix $M_{(0)^2 (1)}$ of the one-form is
\begin{eqnarray}
&
\begin{array}{|c|c|c|}
\hline
AW_r^{(0,1)} & AW_r^{(1/2,1/2)} & AW_r^{(1,0)} \cr
\hline
   & & \cr
    32 + H_0 - 16\,M_1 + 16\,M_2 & -8\,
   \left( 3\,J + M_1 - M_2 \right)  & 0 \cr -16\,
   \left( 3 + 3\,J - M_1 + M_2 \right)  & -16 + 
   H_0 & -16\,\left( 3 + 3\,J + M_1 - 
     M_2 \right)  \cr 0 & -8\,
   \left( 3\,J - M_1 + M_2 \right)  & H_0 + 
   16\,\left( 2 + M_1 - M_2 \right)  \cr 
   {\frac{16\,i}{3}}\,{\sqrt{2}}\,
   \left( 3 + 3\,J - M_1 + M_2 \right)  & 
   {\frac{8\,i}{3}}\,{\sqrt{2}}\,
   \left( 6 + M_1 + 2\,M_2 \right)  & 0 \cr 0 & 
   {\frac{8\,i}{3}}\,{\sqrt{2}}\,
   \left( 3\,J - M_1 + M_2 \right)  & {\frac{16\,i}
     {3}}\,{\sqrt{2}}\,\left( 3 + M_1 + 2\,M_2 \right) 
    \cr {\frac{16\,i}{3}}\,{\sqrt{2}}\,
   \left( 3 + 2\,M_1 + M_2 \right)  & {\frac{8\,i}{3}}\,
   {\sqrt{2}}\,\left( 3\,J + M_1 - M_2 \right)  & 0 \cr
   0 & {\frac{8\,i}{3}}\,{\sqrt{2}}\,
   \left( 6 + 2\,M_1 + M_2 \right)  & {\frac{16\,i}
     {3}}\,{\sqrt{2}}\,\left( 3 + 3\,J + M_1 - M_2
      \right)  \cr
     & &  \cr
\hline
      \end{array}&
\nonumber \\
\nonumber \\
\nonumber \\
&  
\begin{array}{|c|c|}
\hline
   AW_c^{(0,1/2)} &    AW_c^{(1/2,0)} \cr 
\hline
 & \cr
   {\frac{16\,i}{3}}\,{\sqrt{2}}\,
   \left( 3\,J + M_1 - M_2 \right)  & 0 \cr 
   {\frac{-16\,i}{3}}\,{\sqrt{2}}\,
   \left( 2\,M_1 + M_2 \right)  & {\frac{16\,i}{3}}\,
   {\sqrt{2}}\,\left( 3 + 3\,J + M_1 - M_2 \right)  \cr
   0 & {\frac{-16\,i}{3}}\,{\sqrt{2}}\,
   \left( 3 + 2\,M_1 + M_2 \right)  \cr H_0
     - \frac{16 }{3}\left( M_1 - M_2 \right) &
   \frac{-16 }{3}\left( 3 + 3\,J + M_1 - M_2 \right)
     \cr \frac{-16}{3}\left( 3\,J - M_1 + M_2 \right)
        & H_0 + 
   \frac{16 }{3}\left( 3 + M_1 - M_2 \right) \cr 
   0 & 0 \cr 0 & 0 \cr 
 & \cr
\hline
  \end{array}
&
\nonumber \\
\nonumber \\
\nonumber \\
&
\begin{array}{|c|c|}
\hline
\tilde{AW}_c^{(0,1/2)} & \tilde{AW}_c^{(1/2,0)} \cr
\hline
 & \cr 
 {\frac{-16\,i}{3}}\,{\sqrt{2}}\,
   \left( 3 + M_1 + 2\,M_2 \right)  & 0 \cr 
   {\frac{16\,i}{3}}\,{\sqrt{2}}\,
   \left( 3 + 3\,J - M_1 + M_2 \right)  & 
   {\frac{-16\,i}{3}}\,{\sqrt{2}}\,
   \left( M_1 + 2\,M_2 \right)  \cr 0 & {\frac{16\,i}
     {3}}\,{\sqrt{2}}\,\left( 3\,J - M_1 + M_2 \right) 
    \cr 0 & 0 \cr 0 & 0 \cr H_0 - 
   \frac{16 }{3}\left( -3 + M_1 - M_2 \right) &
   \frac{-16 }{3}\left( 3\,J + M_1 - M_2 \right)
    \cr \frac{-16 }{3}\left( 3 + 3\,J - M_1 + 
        M_2 \right) & H_0 + 
   \frac{16 }{3}\left( M_1 - M_2\right) \cr 
 & \cr 
\hline
  \end{array}
&
\nonumber \\
  \label{eq:oneformmatrix}
\end{eqnarray}
\newpage
The matrix $M_{(1/2)^3}$ of the spinor is
\begin{eqnarray}
  \label{eq:spinormatrix1_3}
&  \begin{array}{|c|c|c|}
\hline
   \chi^{(0,1)}  &   \chi^{(1/2,1/2)} &   \chi^{(1,0)} \cr
\hline
 & & \cr
 {\frac{-4\,\left( 3 + M_1 - M_2 \right) }
    {3}} & {\frac{-2\,\left( 3\,J + M_1 - M_2 \right) }
    {3}} & 0 \cr {\frac{-4\,\left( 3 + 3\,J - M_1 + 
        M_2 \right) }{3}} & -8 & {\frac{-4\,
      \left( 3 + 3\,J + M_1 - M_2 \right) }{3}} \cr 0 &
   {\frac{-2\,\left( 3\,J - M_1 + M_2 \right) }
    {3}} & {\frac{4\,\left( -3 + M_1 - M_2 \right) }
    {3}} \cr {\frac{4\,{\sqrt{2}}\,
      \left( 3 + 2\,M_1 + M_2 \right) }{3}} & {
     \frac{2\,{\sqrt{2}}\,\left( 3\,J + M_1 - M_2
        \right) }{3}} & 0 \cr 0 & {\frac{2\,{\sqrt{2}}\,
      \left( 6 + 2\,M_1 + M_2 \right) }{3}} & {
     \frac{4\,{\sqrt{2}}\,\left( 3 + 3\,J + M_1 - 
        M_2 \right) }{3}} \cr {\frac{-4\,{\sqrt{2}}\,
      \left( 3 + 3\,J - M_1 + M_2 \right) }{3}} & {
     \frac{-2\,{\sqrt{2}}\,\left( 6 + M_1 + 2\,M_2
         \right) }{3}} & 0 \cr 0 & {\frac{-2\,{\sqrt{2}}\,
      \left( 3\,J - M_1 + M_2 \right) }{3}} & {
     \frac{-4\,{\sqrt{2}}\,\left( 3 + M_1 + 2\,M_2
         \right) }{3}} \cr {\frac{-2\,
      \left( 3 + 3\,J - M_1 + M_2 \right) }{3}} & {
     \frac{2\,\left( M_1 - M_2 \right) }{3}} & {
     \frac{2\,\left( 3 + 3\,J + M_1 - M_2 \right) }{3}}
    \cr 
 & & \cr
\hline
  \end{array} &
\nonumber \\
\nonumber \\
\nonumber \\
 & \begin{array}{|c|c|}
\hline
\chi_\xi^{(0,1/2)} & \chi_\xi^{(1/2,0)} \cr
\hline
 & \cr
{\frac{4\,{\sqrt{2}}\,\left( 3 + M_1 + 
        2\,M_2 \right) }{3}} & 0 \cr {\frac{-4\,{\sqrt{2}}\,
      \left( 3 + 3\,J - M_1 + M_2 \right) }{3}} & {
     \frac{4\,{\sqrt{2}}\,\left( M_1 + 2\,M_2 \right) }
    {3}} \cr 0 & {\frac{-4\,{\sqrt{2}}\,
      \left( 3\,J - M_1 + M_2 \right) }{3}} \cr {
     \frac{4\,\left( -6 + M_1 - M_2 \right) }{3}} &
   {\frac{4\,\left( 3\,J + M_1 - M_2 \right) }
    {3}} \cr {\frac{4\,\left( 3 + 3\,J - M_1 + M_2
         \right) }{3}} & {\frac{-4\,
      \left( 3 + M_1 - M_2 \right) }{3}} \cr 0 & 0 \cr 
   0 & 0 \cr {\frac{2\,{\sqrt{2}}\,
      \left( 3 + 3\,J - M_1 + M_2 \right) }{3}} & {
     \frac{2\,{\sqrt{2}}\,\left( 6 + M_1 + 2\,M_2
        \right) }{3}} \cr  
 & \cr
\hline 
  \end{array}&
\nonumber \\
\nonumber \\
\nonumber \\
&  \begin{array}{|c|c|c|}
\hline
\chi_\zeta^{(0,1/2)} & \chi_\zeta^{(1/2,0)} &\chi^{(0,0)} \cr
\hline
 & & \cr
 {\frac{4\,{\sqrt{2}}\,\left( 3\,J + M_1 - 
        M_2 \right) }{3}} & 0 & {\frac{-4\,
      \left( 3\,J + M_1 - M_2 \right) }{3}} \cr {
     \frac{-4\,{\sqrt{2}}\,\left( 2\,M_1 + M_2 \right) }
    {3}} & {\frac{4\,{\sqrt{2}}\,
      \left( 3 + 3\,J + M_1 - M_2 \right) }{3}} & {
     \frac{8\,\left( M_1 - M_2 \right) }{3}} \cr 0 &
   {\frac{-4\,{\sqrt{2}}\,\left( 3 + 2\,M_1 + M_2
        \right) }{3}} & {\frac{4\,
      \left( 3\,J - M_1 + M_2 \right) }{3}} \cr 0 & 0 &
   {\frac{-4\,{\sqrt{2}}\,\left( 3\,J + M_1 - M_2
        \right) }{3}} \cr 0 & 0 & {\frac{4\,{\sqrt{2}}\,
      \left( 2\,M_1 + M_2 \right) }{3}} \cr {\frac{4\,
      \left( -3 + M_1 - M_2 \right) }{3}} & {\frac{4\,
      \left( 3 + 3\,J + M_1 - M_2 \right) }{3}} & {
     \frac{4\,{\sqrt{2}}\,\left( M_1 + 2\,M_2 \right) }
    {3}} \cr {\frac{4\,\left( 3\,J - M_1 + M_2 \right) }
    {3}} & {\frac{-4\,\left( 6 + M_1 - M_2 \right) }
    {3}} & {\frac{-4\,{\sqrt{2}}\,
      \left( 3\,J - M_1 + M_2 \right) }{3}} \cr {
     \frac{2\,{\sqrt{2}}\,\left( 6 + 2\,M_1 + M_2
        \right) }{3}} & {\frac{2\,{\sqrt{2}}\,
      \left( 3 + 3\,J + M_1 - M_2 \right) }{3}} & 
    -16 \cr 
 & & \cr
\hline
  \end{array}
&
\end{eqnarray}
The matrix $M_{(0)(1)^2}$ of the two-form is
\begin{eqnarray}
  \label{eq:twoformmatrix1_3}
  \begin{array}{|c|c|c|}
\hline
 & & \cr 
Z_c^{(0,3/2)} & Z_c^{(1/2,1)}  & Z_c^{(1,1/2)} \cr
 & & \cr 
\hline
 & & \cr 
 H_0 - {\frac{64\,
       \left( M_1 - M_2 \right) }{3}} & {\frac{-64\,
      \left( 3\,J + M_1 - M_2 \right) }{9}} & 0 \cr
   {\frac{-64\,\left( 3 + 3\,J - M_1 + M_2 \right) }
    {3}} & H_0 - {\frac{64\,
       \left( 9 + M_1 - M_2 \right) }{9}} & {\frac{
      -128\,\left( 3 + 3\,J + M_1 - M_2 \right) }{9}}
    \cr 0 & {\frac{-128\,\left( 3\,J - M_1 + M_2 \right)
        }{9}} & {H_0 + \frac{ 
      64\,\left( -6 + M_1 - M_2 \right) }{9}} \cr 0 & 
   0 & {\frac{-64\,\left( -3 + 3\,J - M_1 + M_2 \right)
        }{9}} \cr 0 & 0 & 0 \cr 0 & 0 & 0 \cr 0 & 0 & 0 \cr 0 & 0 & 0 \cr
   {\frac{32\,{\sqrt{2}}\,\left( 6 + 2\,M_1 + M_2
        \right) }{3}} & {\frac{32\,{\sqrt{2}}\,
      \left( 3\,J + M_1 - M_2 \right) }{9}} & 0 \cr 0 &
   {\frac{64\,{\sqrt{2}}\,\left( 9 + 2\,M_1 + M_2
        \right) }{9}} & {\frac{64\,{\sqrt{2}}\,
      \left( 3 + 3\,J + M_1 - M_2 \right) }{9}} \cr 0 &
   0 & {\frac{32\,{\sqrt{2}}\,\left( 12 + 2\,M_1 + 
        M_2 \right) }{9}} \cr {\frac{-16\,{\sqrt{2}}\,
      \left( 6 + 2\,M_1 + M_2 \right) }{3}} & {
     \frac{-16\,{\sqrt{2}}\,\left( 3\,J + M_1 - M_2
         \right) }{9}} & 0 \cr 0 & {\frac{-32\,{\sqrt{2}}\,
      \left( 9 + 2\,M_1 + M_2 \right) }{9}} & {
     \frac{-32\,{\sqrt{2}}\,\left( 3 + 3\,J + M_1 - 
        M_2 \right) }{9}} \cr 0 & 0 & {\frac{-16\,{\sqrt{2}}\,
      \left( 12 + 2\,M_1 + M_2 \right) }{9}} \cr 0 & 
   0 & 0 \cr 0 & 0 & 0 \cr {\frac{-16\,
      \left( 3 + 3\,J - M_1 + M_2 \right) }{9}} & {
     \frac{32\,\left( M_1 - M_2 \right) }{27}} & {
     \frac{16\,\left( 3 + 3\,J + M_1 - M_2 \right) }{27}
    } \cr 0 & {\frac{-16\,\left( 3\,J - M_1 + M_2
        \right) }{27}} & {\frac{32\,
      \left( 3 + M_1 - M_2 \right) }{27}} \cr 0 & 0 & 
   0 \cr 0 & 0 & 0 \cr 0 & 0 & 0 \cr 
 & & \cr 
\hline
  \end{array}
\end{eqnarray}

\newpage

\begin{eqnarray}
  \label{eq:twoformmatrix4_6}
  \begin{array}{|c|c|c|}
\hline
 & & \cr 
Z_c^{(3/2,0)} & \tilde Z_c^{(0,3/2)} & \tilde Z_c^{(1/2,1)} \cr
 & & \cr 
\hline
 & & \cr
 0 & 0 & 0 \cr 0 & 0 & 0 \cr {\frac{-64\,
      \left( 6 + 3\,J + M_1 - M_2 \right) }{3}} & 0 & 
   0 \cr H_0 + {\frac{64\,
       \left( 3 + M_1 - M_2 \right) }{3}} & 0 & 0 \cr 
   0 & H_0 - {\frac{64\,
       \left( -3 + M_1 - M_2 \right) }{3}} & {\frac
      {-64\,\left( -3 + 3\,J + M_1 - M_2 \right) }{9}}
    \cr 0 & {\frac{-64\,\left( 6 + 3\,J - M_1 + M_2
         \right) }{3}} & {H_0 - \frac{ 
      64\,\left( 6 + M_1 - M_2 \right) }{9}} \cr 0 & 
   0 & {\frac{-128\,\left( 3 + 3\,J - M_1 + M_2 \right)
        }{9}} \cr 0 & 0 & 0 \cr 0 & {\frac{32\,{\sqrt{2}}\,
      \left( 6 + 3\,J - M_1 + M_2 \right) }{3}} & {
     \frac{32\,{\sqrt{2}}\,\left( M_1 + 
        2\,\left( 6 + M_2 \right)  \right) }{9}} \cr 0 & 0 & {
     \frac{64\,{\sqrt{2}}\,\left( 3 + 3\,J - M_1 + 
        M_2 \right) }{9}} \cr {\frac{32\,{\sqrt{2}}\,
      \left( 6 + 3\,J + M_1 - M_2 \right) }{3}} & 0 & 
   0 \cr 0 & {\frac{-16\,{\sqrt{2}}\,
      \left( 6 + 3\,J - M_1 + M_2 \right) }{3}} & {
     \frac{-16\,{\sqrt{2}}\,\left( M_1 + 
        2\,\left( 6 + M_2 \right)  \right) }{9}} \cr 0 & 0 & {
     \frac{-32\,{\sqrt{2}}\,\left( 3 + 3\,J - M_1 + 
        M_2 \right) }{9}} \cr {\frac{-16\,{\sqrt{2}}\,
      \left( 6 + 3\,J + M_1 - M_2 \right) }{3}} & 0 & 
   0 \cr 0 & {\frac{16\,\left( 6 + 3\,J - M_1 + M_2
         \right) }{9}} & {\frac{-32\,
      \left( -3 + M_1 - M_2 \right) }{27}} \cr 0 & 0 &
   {\frac{16\,\left( 3 + 3\,J - M_1 + M_2 \right) }
    {27}} \cr 0 & 0 & 0 \cr {\frac{16\,
      \left( 6 + 3\,J + M_1 - M_2 \right) }{9}} & 0 & 
   0 \cr 0 & 0 & 0 \cr 0 & 0 & 0 \cr 0 & 0 & 0 \cr 
 & & \cr
\hline
  \end{array}
\end{eqnarray}

\newpage

\begin{eqnarray}
  \label{eq:twoformmatrix7_9}
  \begin{array}{|c|c|c|}
\hline
 & & \cr
\tilde Z_c^{(1,1/2)} & \tilde Z_c^{(3/2,1)} & Z_o^{(0,1)} \cr
 & & \cr
\hline
 & & \cr
 0 & 0 & {\frac{8\,{\sqrt{2}}\,
      \left( M_1 + 2\,M_2 \right) }{3}} \cr 0 & 0 &
   {\frac{-8\,{\sqrt{2}}\,\left( 3 + 3\,J - M_1 + 
        M_2 \right) }{3}} \cr 0 & 0 & 0 \cr 0 & 0 & 0 \cr 0 & 0 &
   {\frac{-8\,{\sqrt{2}}\,\left( -3 + 3\,J + M_1 - 
        M_2 \right) }{3}} \cr {\frac{-128\,
      \left( 3\,J + M_1 - M_2 \right) }{9}} & 0 & {
     \frac{8\,{\sqrt{2}}\,\left( -6 + 2\,M_1 + M_2
         \right) }{3}} \cr H_0 + 
   {\frac{64\,\left( -9 + M_1 - M_2 \right) }{9}} &
   {\frac{-64\,\left( 3 + 3\,J + M_1 - M_2 \right) }
    {3}} & 0 \cr {\frac{-64\,\left( 3\,J - M_1 + 
        M_2 \right) }{9}} & H_0 + 
   {\frac{64\,\left( M_1 - M_2 \right) }{3}} & 0 \cr 
   0 & 0 & H_0 - {\frac{32\,
       \left( -6 + M_1 - M_2 \right) }{3}} \cr {
     \frac{64\,{\sqrt{2}}\,\left( 9 + M_1 + 2\,M_2
         \right) }{9}} & 0 & {\frac{-32\,
      \left( 3 + 3\,J - M_1 + M_2 \right) }{3}} \cr
   {\frac{32\,{\sqrt{2}}\,\left( 3\,J - M_1 + M_2
        \right) }{9}} & {\frac{32\,{\sqrt{2}}\,
      \left( 6 + M_1 + 2\,M_2 \right) }{3}} & 0 \cr 0 &
   0 & -16 \cr {\frac{-32\,{\sqrt{2}}\,
      \left( 9 + M_1 + 2\,M_2 \right) }{9}} & 0 & 0 \cr
   {\frac{-16\,{\sqrt{2}}\,\left( 3\,J - M_1 + M_2
         \right) }{9}} & {\frac{-16\,{\sqrt{2}}\,
      \left( 6 + M_1 + 2\,M_2 \right) }{3}} & 0 \cr
   {\frac{-16\,\left( 3\,J + M_1 - M_2 \right) }
    {27}} & 0 & {\frac{-8\,{\sqrt{2}}\,
      \left( 3 + 2\,M_1 + M_2 \right) }{9}} \cr {
     \frac{-32\,\left( M_1 - M_2 \right) }{27}} & {
     \frac{-16\,\left( 3 + 3\,J + M_1 - M_2 \right) }{9}
    } & 0 \cr 0 & 0 & {\frac{-8\,{\sqrt{2}}\,
      \left( 3 + 3\,J - M_1 + M_2 \right) }{9}} \cr 0 &
   0 & 0 \cr 0 & 0 & 0 \cr 0 & 0 & 0 \cr 0 & 0 & 0 \cr 
 & & \cr
\hline
  \end{array}
\end{eqnarray}

\newpage

\begin{eqnarray}
  \label{eq:twoformmatrix10_12}
  \begin{array}{|c|c|c|}
\hline
& & \cr
Z_o^{(1/2,1/2)} & Z_o^{(1,0)} & Z_r^{(0,1)} \cr
& & \cr
\hline
& & \cr
 0 & 0 & {\frac{-8\,{\sqrt{2}}\,
      \left( M_1 + 2\,M_2 \right) }{3}} \cr {\frac{8\,
      {\sqrt{2}}\,\left( -3 + M_1 + 2\,M_2 \right) }{3}}
   & 0 & {\frac{8\,{\sqrt{2}}\,\left( 3 + 3\,J - M_1 + 
        M_2 \right) }{3}} \cr {\frac{-8\,{\sqrt{2}}\,
      \left( 3\,J - M_1 + M_2 \right) }{3}} & {
     \frac{8\,{\sqrt{2}}\,\left( -6 + M_1 + 2\,M_2
         \right) }{3}} & 0 \cr 0 & {\frac{-8\,{\sqrt{2}}\,
      \left( -3 + 3\,J - M_1 + M_2 \right) }{3}} & 
   0 \cr 0 & 0 & {\frac{8\,{\sqrt{2}}\,
      \left( -3 + 3\,J + M_1 - M_2 \right) }{3}} \cr
   {\frac{-8\,{\sqrt{2}}\,\left( 3\,J + M_1 - M_2
        \right) }{3}} & 0 & {\frac{-8\,{\sqrt{2}}\,
      \left( -6 + 2\,M_1 + M_2 \right) }{3}} \cr {
     \frac{8\,{\sqrt{2}}\,\left( -3 + 2\,M_1 + M_2
         \right) }{3}} & {\frac{-8\,{\sqrt{2}}\,
      \left( 3 + 3\,J + M_1 - M_2 \right) }{3}} & 0 \cr
   0 & {\frac{8\,{\sqrt{2}}\,\left( 2\,M_1 + M_2 \right)
        }{3}} & 0 \cr {\frac{-16\,
      \left( 3\,J + M_1 - M_2 \right) }{3}} & 0 & 
    -32 \cr 32 + H_0 & {\frac{-32\,
      \left( 3 + 3\,J + M_1 - M_2 \right) }{3}} & 0 \cr
   {\frac{-16\,\left( 3\,J - M_1 + M_2 \right) }
    {3}} & H_0 + {\frac{32\,
       \left( 6 + M_1 - M_2 \right) }{3}} & 0 \cr 0 & 
   0 & H_0 - 16\,\left( -4 + M_1 - M_2
      \right)  \cr -16 & 0 & -16\,
   \left( 3 + 3\,J - M_1 + M_2 \right)  \cr 0 & 
    -16 & 0 \cr {\frac{-4\,{\sqrt{2}}\,
      \left( 3\,J + M_1 - M_2 \right) }{9}} & 0 & {
     \frac{32\,{\sqrt{2}}\,\left( 3 + 2\,M_1 + M_2
         \right) }{9}} \cr {\frac{-4\,{\sqrt{2}}\,
      \left( 6 + 2\,M_1 + M_2 \right) }{9}} & {
     \frac{-8\,{\sqrt{2}}\,\left( 3 + 3\,J + M_1 - 
        M_2 \right) }{9}} & 0 \cr {\frac{-4\,{\sqrt{2}}\,
      \left( 6 + M_1 + 2\,M_2 \right) }{9}} & 0 & {
     \frac{32\,{\sqrt{2}}\,\left( 3 + 3\,J - M_1 + 
        M_2 \right) }{9}} \cr {\frac{-4\,{\sqrt{2}}\,
      \left( 3\,J - M_1 + M_2 \right) }{9}} & {
     \frac{-8\,{\sqrt{2}}\,\left( 3 + M_1 + 2\,M_2
         \right) }{9}} & 0 \cr 0 & 0 & 0 \cr 0 & 0 & 0 \cr 0 & 0 & 0 \cr 
& & \cr
\hline
  \end{array}
\end{eqnarray}

\newpage

\begin{eqnarray}
  \label{eq:twoformatrix13_15}
  \begin{array}{|c|c|c|}
\hline
 & & \cr
Z_r^{(1/2,1/2)} & Z_r^{(1,0)} &Z_c^{(0,1/2)} \cr
 & & \cr
\hline
 & & \cr
 0 & 0 & 0 \cr {\frac{-8\,{\sqrt{2}}\,
      \left( -3 + M_1 + 2\,M_2 \right) }{3}} & 0 & 
   0 \cr {\frac{8\,{\sqrt{2}}\,\left( 3\,J - M_1 + 
        M_2 \right) }{3}} & {\frac{-8\,{\sqrt{2}}\,
      \left( -6 + M_1 + 2\,M_2 \right) }{3}} & 0 \cr 
   0 & {\frac{8\,{\sqrt{2}}\,\left( -3 + 3\,J - M_1 + 
        M_2 \right) }{3}} & 0 \cr 0 & 0 & {\frac{8\,
      \left( -3 + 3\,J + M_1 - M_2 \right) }{3}} \cr
   {\frac{8\,{\sqrt{2}}\,\left( 3\,J + M_1 - M_2 \right)
        }{3}} & 0 & {\frac{-16\,
      \left( -3 + M_1 - M_2 \right) }{3}} \cr {
     \frac{-8\,{\sqrt{2}}\,\left( -3 + 2\,M_1 + M_2
         \right) }{3}} & {\frac{8\,{\sqrt{2}}\,
      \left( 3 + 3\,J + M_1 - M_2 \right) }{3}} & {
     \frac{-8\,\left( 3 + 3\,J - M_1 + M_2 \right) }{3}}
    \cr 0 & {\frac{-8\,{\sqrt{2}}\,
      \left( 2\,M_1 + M_2 \right) }{3}} & 0 \cr 0 & 0 &
   {\frac{-16\,{\sqrt{2}}\,\left( 3 + M_1 + 2\,M_2
         \right) }{3}} \cr -32 & 0 & {\frac{16\,{\sqrt{2}}\,
      \left( 3 + 3\,J - M_1 + M_2 \right) }{3}} \cr 0 &
   -32 & 0 \cr -8\,\left( 3\,J + M_1 - M_2 \right)
      & 0 & {\frac{32\,{\sqrt{2}}\,
      \left( 3 + M_1 + 2\,M_2 \right) }{3}} \cr 16 + 
   H_0 & -16\,\left( 3 + 3\,J + M_1 - 
     M_2 \right)  & {\frac{-32\,{\sqrt{2}}\,
      \left( 3 + 3\,J - M_1 + M_2 \right) }{3}} \cr
   -8\,\left( 3\,J - M_1 + M_2 \right)  & H_0
     + 16\,\left( 4 + M_1 - M_2 \right)  & 0 \cr {
     \frac{16\,{\sqrt{2}}\,\left( 3\,J + M_1 - M_2
         \right) }{9}} & 0 & H_0 - 
   {\frac{16\,\left( -33 + 5\,M_1 - 5\,M_2 \right) }{9}}
    \cr {\frac{16\,{\sqrt{2}}\,\left( 6 + 2\,M_1 + 
        M_2 \right) }{9}} & {\frac{32\,{\sqrt{2}}\,
      \left( 3 + 3\,J + M_1 - M_2 \right) }{9}} & {
     \frac{-80\,\left( 3 + 3\,J - M_1 + M_2 \right) }{9}
    } \cr {\frac{16\,{\sqrt{2}}\,
      \left( 6 + M_1 + 2\,M_2 \right) }{9}} & 0 & 0 \cr
   {\frac{16\,{\sqrt{2}}\,\left( 3\,J - M_1 + M_2
        \right) }{9}} & {\frac{32\,{\sqrt{2}}\,
      \left( 3 + M_1 + 2\,M_2 \right) }{9}} & 0 \cr 0 &
   0 & 0 \cr 0 & 0 & 8\,{\sqrt{2}}\,
   \left( 2\,M_1 + M_2 \right)  \cr 0 & 0 & 8\,i\,
   {\sqrt{2}}\,\left( 3 + 3\,J - M_1 + M_2 \right)  \cr
 & & \cr
\hline
  \end{array}
\end{eqnarray}

\newpage

\begin{eqnarray}
  \label{eq:twoformmatrix16_18}
  \begin{array}{|c|c|c|}
\hline
 & & \cr
Z_c^{(1/2,0)} & \tilde Z_c^{(0,1/2)} & \tilde Z_c^{(1/2,0)} \cr
 & & \cr
\hline
 & & \cr
 0 & {\frac{-8\,\left( 3\,J + M_1 - M_2 \right)
        }{3}} & 0 \cr 0 & {\frac{16\,
      \left( M_1 - M_2 \right) }{3}} & {\frac{-8\,
      \left( 3 + 3\,J + M_1 - M_2 \right) }{3}} \cr 0 &
   {\frac{8\,\left( 3\,J - M_1 + M_2 \right) }
    {3}} & {\frac{16\,\left( 3 + M_1 - M_2 \right) }
    {3}} \cr 0 & 0 & {\frac{8\,\left( -3 + 3\,J - M_1 + 
        M_2 \right) }{3}} \cr 0 & 0 & 0 \cr {\frac{8\,
      \left( 3\,J + M_1 - M_2 \right) }{3}} & 0 & 0 \cr
   {\frac{-16\,\left( M_1 - M_2 \right) }
    {3}} & 0 & 0 \cr {\frac{-8\,
      \left( 3\,J - M_1 + M_2 \right) }{3}} & 0 & 0 \cr
   0 & {\frac{16\,{\sqrt{2}}\,\left( 3\,J + M_1 - 
        M_2 \right) }{3}} & 0 \cr {\frac{-16\,{\sqrt{2}}\,
      \left( M_1 + 2\,M_2 \right) }{3}} & {\frac{-16\,
      {\sqrt{2}}\,\left( 2\,M_1 + M_2 \right) }{3}} &
   {\frac{16\,{\sqrt{2}}\,\left( 3 + 3\,J + M_1 - 
        M_2 \right) }{3}} \cr {\frac{16\,{\sqrt{2}}\,
      \left( 3\,J - M_1 + M_2 \right) }{3}} & 0 & {
     \frac{-16\,{\sqrt{2}}\,\left( 3 + 2\,M_1 + M_2
         \right) }{3}} \cr 0 & {\frac{-32\,{\sqrt{2}}\,
      \left( 3\,J + M_1 - M_2 \right) }{3}} & 0 \cr
   {\frac{32\,{\sqrt{2}}\,\left( M_1 + 2\,M_2 \right) }
    {3}} & {\frac{32\,{\sqrt{2}}\,
      \left( 2\,M_1 + M_2 \right) }{3}} & {\frac{-32\,
      {\sqrt{2}}\,\left( 3 + 3\,J + M_1 - M_2 \right) }
      {3}} \cr {\frac{-32\,{\sqrt{2}}\,
      \left( 3\,J - M_1 + M_2 \right) }{3}} & 0 & {
     \frac{32\,{\sqrt{2}}\,\left( 3 + 2\,M_1 + M_2
         \right) }{3}} \cr {\frac{-80\,
      \left( 3\,J + M_1 - M_2 \right) }{9}} & 0 & 0 \cr
   H_0 + {\frac{16\,\left( 18 + 5\,M_1 - 
         5\,M_2 \right) }{9}} & 0 & 0 \cr 0 & H_0 - 
   {\frac{16\,\left( -18 + 5\,M_1 - 5\,M_2 \right) }{9}}
   & {\frac{-80\,\left( 3 + 3\,J + M_1 - M_2 \right) }
    {9}} \cr 0 & {\frac{-80\,\left( 3\,J - M_1 + 
        M_2 \right) }{9}} & H_0 + 
   {\frac{16\,\left( 33 + 5\,M_1 - 5\,M_2 \right) }{9}}
    \cr 0 & 8\,{\sqrt{2}}\,\left( 3\,J - M_1 + M_2
      \right)  & 8\,{\sqrt{2}}\,
   \left( M_1 + 2\,M_2 \right)  \cr 8\,{\sqrt{2}}\,
   \left( 3\,J + M_1 - M_2 \right)  & 0 & 0 \cr 8\,i\,
   {\sqrt{2}}\,\left( 6 + M_1 + 2\,M_2 \right)  & 8\,i\,
   {\sqrt{2}}\,\left( 6 + 2\,M_1 + M_2 \right)  & 8\,i\,
   {\sqrt{2}}\,\left( 3 + 3\,J + M_1 - M_2 \right)  \cr
 & & \cr
\hline
  \end{array}
\end{eqnarray}

\newpage

\begin{eqnarray}
  \label{eq:twoformmatrix19_21}
  \begin{array}{|c|c|c|}
\hline
 & & \cr
Z_c^{(0,0)} & \tilde Z_c^{(0,0)} & Z_r^{(0,0)} \cr
 & & \cr
\hline
 & & \cr
   0 & 0 & 0 \cr 0 & 0 & 0 \cr 0 & 0 & 0 \cr 0 & 0 & 0 \cr 0 & 0 & 0 \cr 0 & 
   0 & 0 \cr 0 & 0 & 0 \cr 0 & 0 & 0 \cr 0 & 0 & 0 \cr 0 & 0 & 0 \cr 0 & 0 & 
   0 \cr 0 & 0 & 0 \cr 0 & 0 & 0 \cr 0 & 0 & 0 \cr 0 & {\frac{16\,{\sqrt{2}}\,
      \left( 6 + M_1 + 2\,M_2 \right) }{3}} & 
   {\frac{8\,i}{3}}\,{\sqrt{2}}\,
   \left( 3\,J + M_1 - M_2 \right)  \cr 0 & {\frac{-16\,
      {\sqrt{2}}\,\left( 3 + 3\,J - M_1 + M_2 \right) }
      {3}} & {\frac{-8\,i}{3}}\,{\sqrt{2}}\,
   \left( 2\,M_1 + M_2 \right)  \cr {\frac{-16\,
      {\sqrt{2}}\,\left( 3 + 3\,J + M_1 - M_2 \right) }
      {3}} & 0 & {\frac{-8\,i}{3}}\,{\sqrt{2}}\,
   \left( M_1 + 2\,M_2 \right)  \cr {\frac{16\,
      {\sqrt{2}}\,\left( 6 + 2\,M_1 + M_2 \right) }{3}}
   & 0 & {\frac{8\,i}{3}}\,{\sqrt{2}}\,
   \left( 3\,J - M_1 + M_2 \right)  \cr H_0
    & 0 & 0 \cr 0 & H_0 & 0 \cr 0 & 0 & H_0 \cr 
 & & \cr
\hline
  \end{array}
\end{eqnarray}
\newpage
\section{conventions}
\label{conventions}
The indices $\alpha, \beta$ are the $SO(7)$ Lorentz 
indices, we have negative definite metric
\begin{eqnarray}
\eta_{\alpha\beta}=(-------).
\end{eqnarray}
The indices on the generators of the coset $G/H$:
\begin{eqnarray}
a,b,c &=& 1,2,3 \nonumber \\
A,B,C &=& 4,5,6,7 \nonumber \\
\label{indexsplit}
\end{eqnarray}
The indices on $H$:
\begin{eqnarray}
m,n,p,q  &=& 9,10,11 \nonumber \\
N &=& 8
\end{eqnarray}
If written on the structure constants of $SU(3)$ we use $m,n,p,q $ in stead of $a,b,c$.
\par
The generators of $SU(3)$ are given by
 $\ft{i}{2} \lambda_i$, where $\lambda_i$ are the
Gell--Mann matrices:
\begin{eqnarray}
    \lambda_1    =
    \left(\begin{array}{ccc}
          0   &  1  & 0 \\
          1   &  0   & 0 \\
          0   &  0  & 0 \\
          \end{array}
    \right)\,, \quad
    \lambda_2    =
    \left(\begin{array}{ccc}
          0   & -i   & 0 \\
          i   &  0  & 0  \\
          0   &  0  & 0  \\
          \end{array}
    \right)\,, \quad
    \lambda_3    =
    \left(\begin{array}{ccc}
          1   & 0   & 0 \\
          0   & -1  & 0 \\
          0   & 0   & 0 \\
          \end{array}
    \right)\,, 
\nonumber \\
    \lambda_4    =
    \left(\begin{array}{ccc}
           0  & 0   & 1 \\
           0  & 0   & 0 \\
           1  &  0  & 0 \\
          \end{array}
    \right)\,, \quad
    \lambda_5    =
    \left(\begin{array}{ccc}
           0  & 0   & -i \\
           0  & 0   &  0\\
           i  & 0   &  0\\
          \end{array}
    \right)\,,\quad
    \lambda_6    =
    \left(\begin{array}{ccc}
           0  & 0   & 0 \\
           0  & 0   & 1 \\
            0 & 1   & 0 \\
          \end{array}
    \right)\,, 
\nonumber \\
\quad 
    \lambda_7    =
    \left(\begin{array}{ccc}
           0  & 0   & 0 \\
           0  & 0   & -i \\
            0 & i   & 0 \\
          \end{array}
    \right)\,, \qquad
    \lambda_8    = \ft{1}{\sqrt{3}}
    \left(\begin{array}{ccc}
           1  &  0  & 0 \\
           0  &  1  & 0 \\
           0  &  0  & -2 \\
          \end{array}
    \right)\,.
\nonumber \\
\end{eqnarray}
The $SU(2)$ generators are: $\frac{i}{2} \sigma_a$
\begin{equation}
\sigma_1 =
    \left(\begin{array}{cc}
           0  &  1  \\
           1  &  0  \\
          \end{array}
    \right) \,,
\sigma_2 =
    \left(\begin{array}{cc}
           0  &  -i  \\
           i  &  0  \\
          \end{array}
    \right) \,,
\sigma_3 =
    \left(\begin{array}{cc}
           1  &  0  \\
           0  &  -1  \\
          \end{array}
    \right) \,,
\end{equation}
\par
The structure constants of $SU(3)$ are given by $f_{ijk}=f_{[ijk]}$,
$[\lambda_i, \lambda_j]=2i f_{ijk} \lambda_k$
\begin{eqnarray}
f_{123} &=& 1 \,, 
\nonumber \\
f_{147}&=& \ft12\,, \quad f_{156}=-\ft12\,, \quad f_{246} = \ft12\,,
f_{257} = \ft12 \,, \quad f_{345} = \ft12 \,, \quad f_{367} = -\ft12 \,,
\nonumber \\ 
f_{458} &=& \ft{\sqrt{3}}{2} \,, \quad f_{678} = \ft{\sqrt{3}}{2}
\end{eqnarray}

\section{the $SO(7)$ spinor representation}
\label{SO7conventions}
For our purposes it is convenient to take the following $SO(7)$ Clifford
algebra, 
\begin{small}
\begin{eqnarray}
\tau_1 =
\left(
\matrix{ 0 & 0 & 0 & 0 & 0 & {\frac{i}{2}} & 0 & {\frac{-i}
    {2}} \cr 0 & 0 & i & 0 & 0 & 0 & 0 & 0 \cr 0 & i & 0 & 0 & 0 & 0 & 0 & 0 \cr 0 & 0 & 0 & 0 & i & 0 & 0 & 0 \cr 0 & 0 & 0 & 
   i & 0 & 0 & 0 & 0 \cr i & 0 & 0 & 0 & 0 & 0 & -i & 0 \cr 0 & 0 & 0 & 0 & 0 & {\frac{-i}{2}} & 0 & {\frac{-i}{2}} \cr 
    -i & 0 & 0 & 0 & 0 & 0 & -i & 0 \cr  }
\right)\,,
\tau_2 =
\left(\matrix{ 0 & 0 & 0 & 0 & 0 & -{\frac{1}{2}} & 0 & -{\frac{1}{2}} \cr 0 & 0 & 1 & 0 & 0 & 0 & 0 & 0 \cr 0 & 
    -1 & 0 & 0 & 0 & 0 & 0 & 0 \cr 0 & 0 & 0 & 0 & 1 & 0 & 0 & 0 \cr 0 & 0 & 0 & 
    -1 & 0 & 0 & 0 & 0 \cr 1 & 0 & 0 & 0 & 0 & 0 & -1 & 0 \cr 0 & 0 & 0 & 0 & 0 & {\frac{1}{2}} & 0 & -{\frac{1}
     {2}} \cr 1 & 0 & 0 & 0 & 0 & 0 & 1 & 0 \cr  }
\right)
\nonumber \\ 
\tau_3 =
\left(\matrix{ 0 & 0 & 0 & 0 & 0 & 0 & -i & 0 \cr 0 & i & 0 & 0 & 0 & 0 & 0 & 0 \cr 0 & 0 & 
    -i & 0 & 0 & 0 & 0 & 0 \cr 0 & 0 & 0 & i & 0 & 0 & 0 & 0 \cr 0 & 0 & 0 & 0 & -i & 0 & 0 & 0 \cr 0 & 0 & 0 & 0 & 0 & 
    -i & 0 & 0 \cr -i & 0 & 0 & 0 & 0 & 0 & 0 & 0 \cr 0 & 0 & 0 & 0 & 0 & 0 & 0 & i \cr  } \right)
\tau_4 =
\left(\matrix{ 0 & 0 & {\frac{i}{2}} & {\frac{-i}{2}} & 0 & 0 & 0 & 0 \cr 0 & 0 & 0 & 0 & 0 & -i & 0 & 0 \cr i & 0 & 0 & 0 & 0 & 0 & 
    -i & 0 \cr -i & 0 & 0 & 0 & 0 & 0 & -i & 0 \cr 0 & 0 & 0 & 0 & 0 & 0 & 0 & -i \cr 0 & 
    -i & 0 & 0 & 0 & 0 & 0 & 0 \cr 0 & 0 & {\frac{-i}{2}} & {\frac{-i}{2}} & 0 & 0 & 0 & 0 \cr 0 & 0 & 0 & 0 & 
    -i & 0 & 0 & 0 \cr  }\right)
\nonumber \\
\tau_5 =
\left(\matrix{ 0 & 0 & {\frac{1}{2}} & {\frac{1}{2}} & 0 & 0 & 0 & 0 \cr 0 & 0 & 0 & 0 & 0 & 1 & 0 & 0 \cr 
    -1 & 0 & 0 & 0 & 0 & 0 & 1 & 0 \cr -1 & 0 & 0 & 0 & 0 & 0 & -1 & 0 \cr 0 & 0 & 0 & 0 & 0 & 0 & 0 & -1 \cr 0 & 
    -1 & 0 & 0 & 0 & 0 & 0 & 0 \cr 0 & 0 & -{\frac{1}{2}} & {\frac{1}
    {2}} & 0 & 0 & 0 & 0 \cr 0 & 0 & 0 & 0 & 1 & 0 & 0 & 0 \cr  }\right)
\tau_6 =
\left(\matrix{ 0 & {\frac{-i}{2}} & 0 & 0 & {\frac{-i}{2}} & 0 & 0 & 0 \cr -i & 0 & 0 & 0 & 0 & 0 & 
    -i & 0 \cr 0 & 0 & 0 & 0 & 0 & 0 & 0 & -i \cr 0 & 0 & 0 & 0 & 0 & i & 0 & 0 \cr 
    -i & 0 & 0 & 0 & 0 & 0 & i & 0 \cr 0 & 0 & 0 & i & 0 & 0 & 0 & 0 \cr 0 & {\frac{-i}{2}} & 0 & 0 & {\frac{i}
    {2}} & 0 & 0 & 0 \cr 0 & 0 & -i & 0 & 0 & 0 & 0 & 0 \cr  } \right)
\nonumber \\
\tau_7 =
\left(\matrix{ 0 & -{\frac{1}{2}} & 0 & 0 & {\frac{1}
    {2}} & 0 & 0 & 0 \cr 1 & 0 & 0 & 0 & 0 & 0 & 1 & 0 \cr 0 & 0 & 0 & 0 & 0 & 0 & 0 & 1 \cr 0 & 0 & 0 & 0 & 0 & 1 & 0 & 0 \cr 
    -1 & 0 & 0 & 0 & 0 & 0 & 1 & 0 \cr 0 & 0 & 0 & -1 & 0 & 0 & 0 & 0 \cr 0 & -{\frac{1}{2}} & 0 & 0 & -{\frac{1}
     {2}} & 0 & 0 & 0 \cr 0 & 0 & -1 & 0 & 0 & 0 & 0 & 0 \cr  }\right)
\nonumber 
\end{eqnarray}
The conjugation matrix,
\begin{eqnarray}
C=
\left(
\matrix{ 1 & 0 & 0 & 0 & 0 & 0 & 0 & 0 \cr 0 & 0 & 0 & 0 & 
    -1 & 0 & 0 & 0 \cr 0 & 0 & 0 & 1 & 0 & 0 & 0 & 0 \cr 0 & 0 & 1 & 0 & 0 & 0 & 0 & 0 \cr 0 & 
    -1 & 0 & 0 & 0 & 0 & 0 & 0 \cr 0 & 0 & 0 & 0 & 0 & 0 & 0 & 1 \cr 0 & 0 & 0 & 0 & 0 & 0 & 
    -1 & 0 \cr 0 & 0 & 0 & 0 & 0 & 1 & 0 & 0 \cr  }\right)
\end{eqnarray}
\end{small}

\end{document}